\documentclass[%
amsmath,amssymb,
aps,
twocolumn
]{revtex4-2}

\usepackage{graphicx}
\usepackage{dcolumn}
\usepackage{bm}

\usepackage{physics}
\usepackage{amsmath,amssymb}
\usepackage{geometry}
\usepackage[colorlinks=true]{hyperref}

\begin{document}
	
	\preprint{APS/123-QED}
	
	\title{Understanding the nature of the $ \mathbf{\Delta(1600)} $ resonance}
	
	\author{Liam Hockley}
	\email{liam.hockley@adelaide.edu.au}
	\author{Curtis Abell}
	\author{Derek Leinweber}
	\author{Anthony Thomas}
	\affiliation{%
		ARC Special Research Centre for the Subatomic Structure of Matter (CSSM), Department of Physics, The University of Adelaide, SA, 5005,
		Australia.
	}%
	
	
	\date{\today}
	
	\begin{abstract}
		We present a coupled-channel analysis of the $ J^P = 3/2^+\ \Delta $-baryon spectrum, based in the framework of Hamiltonian Effective Field Theory (HEFT). We construct a Hamiltonian which mixes quark model-like single-particle states and two-particle meson-baryon channels, and constrain this via experimentally measured $ \pi N \to \pi N $ scattering observables.
		In the same vein as L\"{u}scher's approach, we then connect this infinite-volume inspired Hamiltonian with finite-volume lattice QCD results. Drawing on lattice correlation-matrix eigenvectors identifying the $ 1s $ and $ 2s $ states in the finite-volume $ \Delta(3/2^+) $ spectrum, and utilising the HEFT eigenvectors describing the composition of the energy eigenstates, we resolve the structure of these states and their relation to the $ \Delta(1600) $ resonance. We find the dominant contributions to this resonance come from strong rescattering in the $ \pi N $ and $ \pi \Delta $ channels. This contrasts the long-held view of a dominant quark model-like core for the $ \Delta(1600) $. Further discussion of other contemporary lattice results for the $ \Delta $ spectrum and $ \pi N $ scattering states is also presented. 
	\end{abstract}
	
	\maketitle
	
	
	
	\section{Introduction}\label{sec:Intro}
	
	The strong nuclear force presents many unique challenges in particle physics, one of which being short-lived states known as resonances. While the lowest-lying hadrons can be arranged based on isospin, charge and strangeness symmetries in a quark model for mesons and baryons \cite{Gell-Mann:1961omu,Gell-Mann:1962yej}, the complex dynamics of resonances makes modelling and analysis of these excited states non-trivial. 
	
	The foremost example of this difficulty is the $ N^*(1440) $ Roper resonance \cite{Roper:1964zza, Krehl:1999km, Burkert:2017djo}, with similar concerns lying in the $ \Delta $ baryon spectrum's $ \Delta(1600) $ state \cite{Liu:2022ndb, Ramalho:2010cw}. Essentially, the issue with these states is that they have masses which disagree with the most simple quark model predictions without any fine-tuning. For example, a naive quark model predicts that states in the $ \Delta (J = 3/2) $ spectrum should alternate in the signs of their internal parities as one climbs in energy. However, considering the experimentally observed resonances, one immediately notices the apparent mismatch in the ordering of states since two positive parity states (the $ \Delta(1232) $ and $ \Delta(1600) $ resonances) lie below the first odd parity state, the $ \Delta(1700) $. This tension between theory and experiment represents a fundamental gap in the collective understanding of QCD. 
		
	The only currently tenable method of resolving the properties of QCD exactly in the hadronic region is lattice QCD. Lattice QCD is well established as a non-perturbative framework within which many physical quantities can be extracted such as particle masses, decay constants, form factors, etc \cite{FLAG:2021npn}. Studying scattering processes and resonances has historically proven challenging, with no way of instituting asymptotic scattering states on the finite-volume of the lattice. 
	
	A breakthrough methodology was first realised by L\"{u}scher \cite{Luscher:1985dn,Luscher:1986pf,Luscher:1990ux}, allowing for scattering properties to be extracted from finite-volume spectra. Initially derived for the case of single-channel 2-body decays, the formalism has been extended for dealing with cases such as that of multiple channels \cite{Hansen:2012tf} as well as 3-body decays \cite{Hansen:2014eka, Hansen:2019nir, Draper:2023xvu}. The development of this approach towards ever-more generalised systems is ongoing. Useful reviews of the L\"{u}scher approach and other methods for studying hadron resonances are provided in Refs. \cite{Briceno:2017max, Mai:2022eur}.
	
	While the L\"{u}scher formalism provides rigourous quantisation conditions relating finite-volume spectra to infinite-volume scattering observables, for high energy resonances, the opening of multiple relevant channels and the proximity of various 2- and 3-body thresholds rapidly complicates the calculations. More recently, Hamiltonian Effective Field Theory (HEFT) has emerged as a L\"{u}scher based formalism which seamlessly incorporates multiple scattering channels and 3-quark bare states into its description of resonances \cite{Hall:2013qba}. Using infinite-volume scattering data, one is able to place initial constraints on the Hamiltonian and check for validity through extraction of the T-matrix poles. HEFT acts as an alternative for the L\"{u}scher formalism in the sense that this infinite-volume inspired Hamiltonian can then be confined to a finite-size box and thus the energy eigenstates and eigenvalues become discretised. This allows for direct comparison with results from lattice QCD through the Hamiltonian eigenvalues, with additional structural information coming from the eigenvectors.
	
	The combination of lattice QCD and HEFT has been shown to yield insightful results for the historically enigmatic Roper resonance $ N^*(1440) $ \cite{Mahbub:2010rm, Roberts:2013ipa, Roberts:2013oea,Alexandrou:2014mka, Leinweber:2015kyz,Kiratidis:2015vpa, Kiratidis:2016hda,Liu:2016uzk,Wu:2017qve} and the $ \Lambda(1405) $ \cite{Hall:2014uca,Hall:2014gqa,Liu:2016wxq, Liu:2023xvy}, allowing for new understanding of these complicated resonances. Similarly, an initial case study of the $ \Delta(1600) $ \cite{Abell:2021awi} suggested there is more to this resonance than a predominant 3-quark radial excitation of the $ \Delta(1232) $, contrasting a mainstay in the literature as suggested, for example, by work using Dyson-Schwinger models \cite{Burkert:2017djo, Liu:2022ndb}. 
	
\begin{center}
		\begin{figure*}[]
			\includegraphics[width=0.9\linewidth]{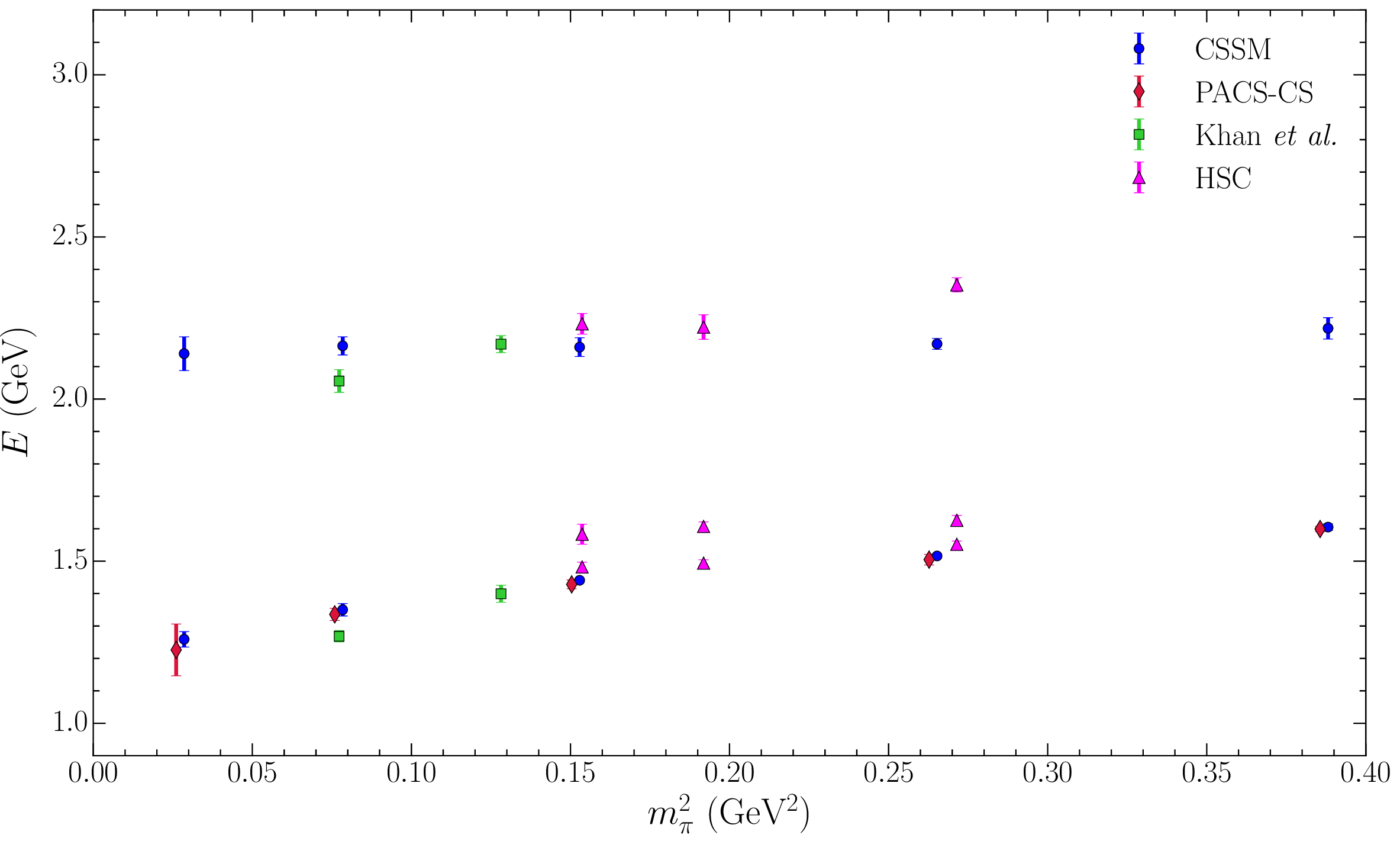}
			\caption{A summary of key lattice QCD results from Ref.~\cite{Hockley:2023yzn}. Importantly, the first excitation in the CSSM results was identified as a $ 2s $ excitation of the ground state based on nodes of the wave function used to excite the state. This state is far above the energy region one typically associates with the $ \Delta(1600) $ resonance. Results from other groups using 3-quark operators are given for comparison. Discussion of the HSC \cite{Bulava:2010yg} and  Khan {\it et al.} \cite{Khan:2020ahz} results is given in Section~\ref{sec:CompLat}.}		
			\label{fig:lattice_data}
		\end{figure*}
	\end{center}

	The current work looks to build on recent lattice QCD results for the $ \Delta $-spectrum which identify the energy of the $ 2s $ excitation of the ground state at $ \sim 2.15 $ GeV \cite{Hockley:2023yzn}. This is remarkably reminiscent of the findings for the nucleon spectrum where the first radial excitation of the ground state was found to lie at approximately $ 1.9 $ GeV, about $ 500 $ MeV above the physical Roper resonance \cite{Alexandrou:2014mka, Leinweber:2015kyz, Roberts:2013ipa, Roberts:2013oea,Kiratidis:2015vpa, Kiratidis:2016hda}. The key results of the recent lattice QCD $ \Delta $ spectrum \cite{Hockley:2023yzn} are given in Fig.~\ref{fig:lattice_data}. The current study seeks to use HEFT to understand this $ 2s $ state in the context of the $ \Delta(1600) $ resonance. 
		
	This paper is organised in the following way. Section~\ref{sec:HEFT} describes the HEFT formalism and how it can be applied for an arbitrary number of single- and two-particle basis states. We outline how one can constrain the various Hamiltonian parameters by fitting phase shifts and inelasticities to $ \pi N $ scattering data, and then extend the Hamiltonian to finite volumes at unphysical quark masses. This allows for direct contact with the lattice results of Ref.~\cite{Hockley:2023yzn} through the Hamiltonian eigenvalues, with additional information about quark-model-like and meson-baryon-like components coming from the eigenvectors. In Section~\ref{sec:LowEnergyModel} this approach is taken to investigate the low energy region around the mass of the $ \Delta(1232) $ resonance before being extended in Section~\ref{sec:FullModel} to include additional bare states and channels which couple to the $ \Delta(1600) $ resonance. We obtain a model which simultaneously describes both the infinite-volume phase shifts and inelasticities, and the finite-volume lattice results of Ref.~\cite{Hockley:2023yzn}. The key information about the structure of the states in the finite-volume spectrum comes from the Hamiltonian eigenvectors, and these are discussed in detail.
	
	We show the diverse applicability of HEFT in Section~\ref{sec:CompLat} by recomputing the finite-volume spectrum at a range of lattice sizes and comparing with recent $ \pi N $ energy level results from Refs.~\cite{Andersen:2017una,Morningstar:2021ewk,Alexandrou:2023elk}, as well as other lattice QCD results using single-hadron operators in Refs.~\cite{Bulava:2010yg,Khan:2020ahz}. This affords another way of constraining our Hamiltonian and understanding the lattice results of other groups. This also demonstrates the powerful insight offered by HEFT in guiding future lattice QCD calculations.

	\section{Hamiltonian Effective Field Theory}\label{sec:HEFT}
	Hamiltonian Effective Field Theory (HEFT) has been shown to be a useful tool in exploring the nucleon spectrum to understand the nature of the Roper resonance \cite{Liu:2015ktc,Liu:2016uzk,Liu:2016wxq,Wu:2017qve} and more recently in the $ \Delta $-baryon spectrum \cite{Abell:2021awi, Abell:2023qgj}. Here, we briefly review the formalism for constructing the Hamiltonian and constraining it using infinite-volume scattering data. We then present the finite-volume, unphysical-quark-mass extension of the model and describe how this can be used to make direct contact with results from lattice QCD.
	
	\subsection{Hamiltonian Model} \label{ssHamMod}
	In general, the Hamiltonian for an interacting system can be decomposed into two pieces
	\begin{equation}
		H = H_0 + H_I \,,
	\end{equation}
	where $ H_0 $ is the free (non-interacting) part, and $ H_I $ governs the non-trivial dynamics of the system through various interactions between its basis states. In our HEFT formalism we consider two types of basis states: single-particle bare baryon states (one can interpret these as being quark model-like states \cite{Miller:1979kg, Theberge:1980ye, Thomas:1982kv}) and two-particle meson-baryon states. The two-particle basis states are non-interacting.
	
	We introduce the general notation $ \ket{B_0} $ to describe our bare basis states, and $ \ket*{\alpha(\boldsymbol{k})} $ to denote our meson-baryon pairs in the center of mass frame with back-to-back 3-momentum $ \boldsymbol{k} $. The free Hamiltonian is then given by
	\begin{align}
		H_0 &= \sum_{B_0}\, \ket{B_0}\, m_{B_0} \, \bra{B_0} + \sum_\alpha \int d^3 k \nonumber \\
		&\times \ket*{\alpha(\boldsymbol{k})} \qty{ \sqrt{m_{\alpha_B}^2 + k^2} + \sqrt{m_{\alpha_M}^2 + k^2} }\, \bra*{\alpha(\boldsymbol{k})}
	\end{align}
	where $ m_{B_0} $ is the mass of the bare state $ B_0 $, $ m_{\alpha_B} $ is the mass of the baryon in the two-particle channel $ \alpha $, and likewise $ m_{\alpha_M} $ is the mass of the meson in channel $ \alpha $. 
	
\begin{center}
		\begin{figure}[t]
			\includegraphics[width=0.6\linewidth]{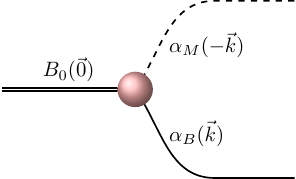}
			
			\vspace{5mm}	
			
			\includegraphics[width=0.6\linewidth]{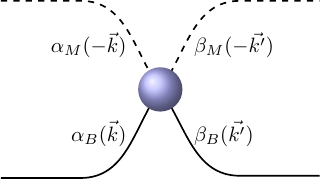}
			\caption{Interaction vertices for bare to two-particle state processes (top), and 2-to-2 processes (bottom). Here $ B_0 $ is a bare state while $ \alpha $ denotes a channel consisting of meson $ \alpha_M $ and baryon $ \alpha_B $.}		
			\label{fig:int_diags}
		\end{figure}
	\end{center}

	Including interactions in our model, we decompose the interacting Hamiltonian into two parts
	\begin{equation}
		H_I = g + v \,,
	\end{equation}
	where $ g $ describes interactions between bare states and two-particle basis states, and $ v $ describes the coupling of the channels through two-particle states interacting with other two-particle states. These are represented graphically by the processes shown in Fig.~\ref{fig:int_diags}.

	Mathematically, we describe these vertex interactions by
	\begin{equation}
		g = \sum_{\alpha,\, B_0} \int d^3{k}\, 
		\qty{ \,
		\ket*{\alpha(\boldsymbol{k})} \, G^\dagger_{\alpha, B_0}(k)\, \bra*{B_0} 
		+ h.c.\,
		}\,, \label{eq:geqn}
	\end{equation}
	where $ h.c. $ denotes the Hermitian conjugate of the first term, and
	\begin{equation}
		v = \sum_{\alpha,\beta} \int d^3{k}\, d^3{k}'\,
		\ket*{\alpha(\boldsymbol{k})}\,  V_{\alpha\beta}(k,k')\, 
		\bra*{\beta(\boldsymbol{k}')}\,. \label{eq:veqn}
	\end{equation}
	
	The momentum-dependent forms of the various $ G_{\alpha, B_0}(k) $ are motivated by $ \chi $PT. We use the form
	\begin{equation}
		\tilde{G}_{B_0, \alpha}(k) = \frac{g^{B_0}_{\alpha}}{2\pi}\, \qty( \frac{k}{f_\pi} )^{l_\alpha}\, \frac{1}{\sqrt{\omega_\pi(k)}}\,,
	\end{equation}
	where the tilde denotes an unregulated function, and we've introduced some as yet unknown coupling strength $ g^{B_0}_{\alpha} $, as well as the pion decay constant $ f_\pi = 92.4 $ MeV. The angular momentum in channel $ \alpha $ is denoted $ l_\alpha $. In general, this function will diverge as we perform integrals over all values of 3-momentum in the two-particle states, so we adopt a regularisation scheme to keep the results convergent and finite. 
	
	To regulate the so-called ``1-to-2" interactions contained in the $ g $ term of $ H_I $, we introduce a regulator 
	\begin{equation}
		u^{g}_{l_\alpha}(k,\Lambda_{B_0,\alpha}) = \frac{1}{\qty(1 + k^2/\Lambda^{2}_{B_0,\alpha})^{(l_\alpha+3)/2}}\,. \label{eq:regulator_g}
	\end{equation}
	
	The exponent of $ (l_\alpha+3)/2 $ is motivated by the overall dimensionful factor of $ k^{{l_\alpha}+2} $ in Eq.~\eqref{eq:geqn} after converting the integral to spherical coordinates.
	
	For the case of a channel which is in $ p $-wave, this has the usual form of a dipole with $ (l_\alpha+3)/2 = 2 $, while for $ f $-wave, we get $ (l_\alpha+3)/2 = 3 $, in other words, a tripole form factor. These two cases will be the relevant angular momenta for this study.
	
	We can thus compute a fully regularised and finite contribution for each interaction by taking
	\begin{equation}
		G_{B_0, \alpha}(k) = \tilde{G}_{B_0, \alpha}(k) \, u^{g}_{l_\alpha}(k,\Lambda_{B_0,\alpha})\,. \label{eq:full_g}
	\end{equation}
	
	Similarly, we define couplings for the 2-to-2 interactions by the separable potentials 
	\begin{equation}
		\tilde{V}_{\alpha \beta}(k, k') = \frac{v_{\alpha \beta}}{4\pi^2}\, \qty(\frac{k}{f_\pi} )^{\l_\alpha}\, \qty(\frac{k'}{f_\pi})^{l_\beta}\, \frac{1}{\omega_\pi(k)}\, \frac{1}{\omega_\pi(k')}\,, 
	\end{equation}
	where we've introduced a parameter $ v_{\alpha\beta} $ which determines how strongly the channels $ \alpha $ and $ \beta $ couple to one another.
	
	To regulate these interactions, we introduce regulators $ u^{v}_{l_\alpha} $. These are of the same form as Eq.~\eqref{eq:regulator_g}
	\begin{equation}
		u^{v}_{l_\alpha}(k,\Lambda^v_{\alpha}) = \frac{1}{\qty(1 + k^2/\qty(\Lambda^{v}_{\alpha})^2)^{(l_\alpha+3)/2}} \,,
	\end{equation}
	but we make the distinction here that the regulator parameters $ \Lambda^{v}_{\alpha} $ may in principle be different from the $ \Lambda_{B_0,\alpha} $.
	
	Hence, we can construct fully regularised 2-to-2 couplings by
	\begin{equation}
		V_{\alpha \beta}(k, k') = \tilde{V}_{\alpha \beta}(k, k')\, u^{v}_{l_\alpha}(k,\Lambda^v_{\alpha})\, u^{v}_{l_\beta}(k',\Lambda^v_{\beta})\,. \label{eq:full_v}
	\end{equation}
	
	In this study, we find dipole and tripole regulators to be sufficient to describe the experimental data. Previous HEFT studies have explored alternative regulators \cite{Abell:2021awi}.
	
	\subsection{Constraints from Scattering Observables}
	In order to constrain the various parameters entering into the model, we look to construct the scattering T-matrix and from this, extract channel phase shifts and inelasticities. This is achieved by first constructing a coupled-channel potential
	\begin{equation}
		V^{\text{coup}}_{\alpha\beta}(k,k',E) = \sum_{B_0} \frac{G^{B_0\dagger}_\alpha(k)G^{B_0}_\beta(k')}{E-m_{B_0}} + V_{\alpha\beta}(k,k')\,,
	\end{equation}
	and then solving a 3-dimensional reduction of the Bethe-Salpeter equations using this potential,
	\begin{align}
		T_{\alpha\beta}(k,k';E) &= V^{\text{coup}}_{\alpha\beta}(k,k';E) + \sum_\gamma \int dq\, q^2 \times \nonumber \\
		& V^{\text{coup}}_{\alpha\gamma}(k, q, E) \frac{1}{E-\omega_\gamma(q)+i\epsilon} T_{\gamma\beta}(q,k';E)\,.
	\end{align}
	
	The corresponding S-matrix elements are obtained from
	\begin{align}
		S_{\alpha \beta}(E) 
		&= \delta_{\alpha\beta}-2i\pi \sqrt{\rho_\alpha \rho_\beta} \, T_{\alpha\beta}(k_{\text{on},\alpha},\, k_{\text{on},\beta};\, E)
	\end{align}
	where the functions $ \rho_\alpha $ are defined by
	\begin{equation}
		\rho_\alpha = \frac{\sqrt{k^2_{\text{on},\, \alpha} + m^2_{\alpha_M}}\, \sqrt{k^2_{\text{on},\, \alpha} + m^2_{\alpha_B}}}{E}\, k_{\text{on},\, \alpha}\,.
	\end{equation}
	Here, $ k_{\text{on},\, \alpha} $ is the on-shell back-to-back momentum carried by the meson and baryon in the two-particle channel $ \alpha $.
	
	The diagonal elements of the S-matrix are readily related to the scattering phase shift $ \delta_\alpha $ and inelasticity $ \eta_\alpha $ by
	\begin{equation}
		S_{\alpha\alpha}(E) = \eta_\alpha\, \exp(2i\delta_{\alpha})\,.
	\end{equation}
	
	For the case of $ \pi N \to \pi N $ scattering, which is of most importance to us, these observables have been derived using partial wave analysis methods applied to experimental scattering data. The results are publicly available on the George Washington University (GWU) SAID partial wave analysis database \cite{GWU:2023ex}. Hence, one can perform fits to this data and so constrain the Hamiltonian model.
	
	An additional point of comparison between our results and those reported by experiment lies in the extraction of poles of the $ T $-matrix. These are complex energies $ E_{\text{pole}} $ for which $ T_{\alpha\beta}(k,k';E_{\text{pole}})^{-1} = 0 $. 
	
	\subsection{Finite-volume Hamiltonian} \label{subsec:FinVolHam}
	In order to compare our Hamiltonian model with lattice QCD results, we need to consider that lattice QCD results are calculated on a finite volume at unphysical quark masses. Neither of these facets has been incorporated into our infinite-volume Hamiltonian.
	
	To address the first of these, consider taking our Hamiltonian and constraining it to describe a finite-volume system with periodic boundary conditions. The basis states interact within the volume to create scattering states. The primary difference this induces is that our momenta obey the quantisation condition
	\begin{equation}
		\boldsymbol{k_n} = \frac{2\pi}{L}\, \boldsymbol{n}\,,
	\end{equation}
	where $ \boldsymbol{n} \in \mathbb{Z}_3 $ and $ L $ describes the length of a side of the cubic volume. Thus, the continuous integrals over momenta encountered throughout Section~\ref{ssHamMod} are converted to discrete sums over momentum $ 3 $-indices $ \boldsymbol{n} $ via
	\begin{equation} \label{eq:3dreduction}
		4\pi \int k^2 dk = \int d^3k \to \sum_{\boldsymbol{n}\in\mathbb{Z}^3} \qty(\frac{2\pi}{L})^3\,. 
	\end{equation}
	
	The entries of $ \boldsymbol{n} $ can take on any integer value, however when constructing the $ \pi N $ and $ \pi \Delta $ states required for studying the $ \Delta $, one ignores the pure $ S $-wave contribution ($ \boldsymbol{n} = \boldsymbol{0} $). 
	
	We can reduce this 3-dimensional sum down to a 1-dimensional sum by identifying degenerate momenta magnitudes
	\begin{equation}
		k_n = \frac{2\pi}{L} \sqrt{n}\,, \label{eq:DiscreteMom}
	\end{equation}
	where
	\begin{equation} \label{eq:neq}
		n = n_x^2 + n_y^2 + n_z^2 \,.
	\end{equation}

	The degeneracy of the magnitudes is captured by introducing the function $ C_3(n) $ which counts the number of ways to produce a given integer $ n $ from the sum of three squared integers. Some examples are $ C_3(2)~=~12 $ since there are 12 ways to sum three squared integers and obtain a result of 2 taking into account the four allowed combinations of $ \pm 1 $ and the three allowed positions for 0. As another example, $ C_3(7)~=~0 $, since $ n=7 $ is not a solution of Eq.~\eqref{eq:neq} $ \text{for any } n_i \in \mathbb{Z} $.
	
	Finally then, the $ 3 $-dimensional sums in Eq.~\eqref{eq:3dreduction} become
	\begin{equation}
		\sum_{\boldsymbol{n}\in\mathbb{Z}^3} \qty(\frac{2\pi}{L})^3 \to \sum_{n\in\mathbb{Z}} \qty(\frac{2\pi}{L})^3\, C_3(n)\,. \label{eq:cubicsymm}
	\end{equation}
	Thus we need only label our momentum states by a single index $ n $ and all 3-dimensional symmetries are carried in the degeneracy factors $ C_3(n) $.
	
	The discussion above leads to various finite-volume corrections to our couplings and potentials $ G_{B_0,\alpha} $ and $ V_{\alpha \beta} $. Before elaborating on this however, we note that our Hamiltonian is regularised by the parameters $ \Lambda_{B_0,\alpha},\ \Lambda^v_{\alpha} $ which allow for the introduction of momentum cutoffs. Considering the coupling suppression of high-momentum Hamiltonian basis states, we can impose a maximum magnitude allowed for the momentum and thus a maximum number of basis states in the Hamiltonian. Importantly then, the regulator parameters influence the size of the Hamiltonian matrix; note that the choice of $ L $ also plays a role since with larger volumes we can fit more momentum states in the box. For fixed values of $ L $, we can thus determine a maximum momentum in the box by imposing a minimum value on the regulators $ u $, for example for a generic regulator with $ l_\alpha = 1 $
	\begin{equation}
		u(k,\Lambda) = \frac{1}{\qty(1 + k^2/\Lambda^2)^{2}}\,.
	\end{equation}
	Rearranging for the maximum value of momentum $ k_\text{max} $ we have
	\begin{equation}
		k_\text{max} = \Lambda\sqrt{u^{-1/2}_{\text{min}}-1}\,.
	\end{equation}
	Setting $ u_\text{min} = 0.01 $ as in previous HEFT analyses \cite{Abell:2023qgj,Abell:2021awi,Liu:2015ktc,Liu:2016uzk,Liu:2016wxq,Wu:2017qve} this fixes a maximum momentum index by
	\begin{equation}
		n_\text{max} = \qty(\frac{k_\text{max}L}{2\pi})^2\,.
	\end{equation}
	In principle this maximum index can be computed for each channel, taking into account the different values of angular momenta, and so determine a momentum cutoff for each channel in the box. To keep the Hamiltonian matrix in a suitable form we work with a single value for $ k_\text{max} $.
	
	Thus we see explicitly that the two-particle states (of which we had an infinite number previously), are now of the form $ \ket{\alpha(k_n)} $ for some single index $ n $ which runs from 1 to $ n_{\text{max}} $. Hence we are now in a position to consider our Hamiltonian as a matrix, and solve for its eigenvalues and eigenvectors.
	
	The final consideration for the finite-volume Hamiltonian is that the coupling functions $ G_{B_0,\,\alpha} $ and $ V_{\alpha\beta} $ receive finite-volume factors associated with momentum quantisation and cubic symmetry in Eqs.~\eqref{eq:3dreduction} and \eqref{eq:cubicsymm} respectively. The finite-volume coupling functions $ \bar{G} $ and $ \bar{V} $ are
	\begin{equation} \label{fincoup}
		\bar{G}^{B_0}_\alpha (k_n) = \sqrt{\frac{C_3(n)}{4\pi}} \qty(\frac{2\pi}{L})^{3/2} G^{B_0}_{\alpha}(k_n)\,,
	\end{equation}

	\begin{equation} \label{finpot}
		\bar{V}_{\alpha\beta}(k_n,k_m) = \sqrt{\frac{C_3(n)}{4\pi}} \sqrt{\frac{C_3(m)}{4\pi}} \qty(\frac{2\pi}{L})^{3} V_{\alpha\beta}(k_n,k_m)\,.
	\end{equation}
	
	These finite-volume corrected functions replace their corresponding infinite-volume counterparts in Eqs.~\eqref{eq:geqn} and \eqref{eq:veqn} in the calculation of the Hamiltonian matrix.

	\subsection{Hamiltonian Matrix}\label{subsec:HamMat}
	Following the preceding discussion, we can simply fill in the entries of the Hamiltonian matrix, now using the finite-volume coupling functions of Eqs.~\eqref{fincoup} and \eqref{finpot}. For ease of presentation, we will only show the specific case of $ n_b = 2 $ bare basis states and $ n_c = 2 $ species of meson-baryon basis states. We adopt the labels $ \Delta_1 $ and $ \Delta_2 $ for the bare states, with the channel labels being $ \alpha \in \{ \pi N,\, \pi \Delta \} $, though the following is easily generalised to any number of channels and bare states. Recalling $ H = H_0 + H_I $, then the Hamiltonian matrix is given by

\begin{widetext}
\begin{equation}
H_0 = \text{diag}\qty(m_{\Delta_1},\, m_{\Delta_2},\, 
\omega_{\pi N}(k_1),\, \omega_{\pi \Delta}(k_1),\, 
\dots, 
\omega_{\pi N}(k_{n_\text{max}}),\, \omega_{\pi \Delta}(k_{n_\text{max}}) )	\,, \label{eq:H0Mat}
\end{equation}

\begin{equation} \label{eq:HIMat}
H_I = 
\begin{pmatrix}
	0 & 0 & 
	\bar{G}^{\Delta_1}_{\pi N}(k_1) & \bar{G}^{\Delta_1}_{\pi \Delta}(k_1) & \bar{G}^{\Delta_1}_{\pi N}(k_2) & \bar{G}^{\Delta_1}_{\pi \Delta}(k_2) & \dots & 
	\bar{G}^{\Delta_1}_{\pi N}(k_{n_\text{max}}) & \bar{G}^{\Delta_1}_{\pi \Delta}(k_{n_\text{max}})\\
	
	0 & 0 & 
	\bar{G}^{\Delta_2}_{\pi N}(k_1) & \bar{G}^{\Delta_2}_{\pi \Delta}(k_1) & \bar{G}^{\Delta_2}_{\pi N}(k_2) & \bar{G}^{\Delta_2}_{\pi \Delta}(k_2) & \dots & 
	\bar{G}^{\Delta_2}_{\pi N}(k_{n_\text{max}}) & \bar{G}^{\Delta_2}_{\pi \Delta}(k_{n_\text{max}}) \\
	
	\bar{G}^{\Delta_1}_{\pi N}(k_1) & \bar{G}^{\Delta_2}_{\pi N}(k_1) \\
	
	\bar{G}^{\Delta_1}_{\pi \Delta}(k_1) & \bar{G}^{\Delta_2}_{\pi \Delta}(k_1) \\
	
	\bar{G}^{\Delta_1}_{\pi N}(k_2) & \bar{G}^{\Delta_2}_{\pi N}(k_2) \\
	
	\bar{G}^{\Delta_1}_{\pi \Delta}(k_2) & \bar{G}^{\Delta_2}_{\pi \Delta}(k_2) & & & & \tilde{\tilde{V}} \\
	
	\vdots & \vdots \\
	
	\bar{G}^{\Delta_1}_{\pi N}(k_{n_\text{max}}) & \bar{G}^{\Delta_2}_{\pi N}(k_{n_\text{max}})\\
	
	\bar{G}^{\Delta_1}_{\pi \Delta}(k_{n_\text{max}}) & \bar{G}^{\Delta_2}_{\pi \Delta}(k_{n_\text{max}})\\
	
\end{pmatrix} \,,
\end{equation}

where the $ (n_c \times n_{\text{max}})^2 $ symmetric matrix $ \tilde{\tilde{V}} $ is given by
\begin{equation} \label{eq:VMat}
\tilde{\tilde{V}} =
\begin{pmatrix}
	\bar{V}_{\pi N, \pi N}(k_1,k_1) & \bar{V}_{\pi N, \pi \Delta}(k_1,k_1) & 
	\bar{V}_{\pi N, \pi N}(k_1,k_2) & \bar{V}_{\pi N, \pi \Delta}(k_1,k_2) & \dots & \bar{V}_{\pi N, \pi \Delta}(k_1,k_{n_\text{max}}) \\
	
	\bar{V}_{\pi \Delta, \pi N}(k_1,k_1) & \bar{V}_{\pi \Delta, \pi \Delta}(k_1,k_1) & 
	\bar{V}_{\pi \Delta, \pi N}(k_1,k_2) & \bar{V}_{\pi \Delta, \pi \Delta}(k_1,k_2) & \dots & \bar{V}_{\pi \Delta, \pi \Delta}(k_1,k_{n_\text{max}}) \\
	
	\bar{V}_{\pi N, \pi N}(k_2,k_1) & \bar{V}_{\pi N, \pi \Delta}(k_2,k_1) & 
	\bar{V}_{\pi N, \pi N}(k_2,k_2) & \bar{V}_{\pi N, \pi \Delta}(k_2,k_2) & \dots &	 \bar{V}_{\pi N, \pi \Delta}(k_2,k_{n_\text{max}}) \\
	
	\bar{V}_{\pi \Delta, \pi N}(k_2,k_1) & \bar{V}_{\pi \Delta, \pi \Delta}(k_2,k_1) & 
	\bar{V}_{\pi \Delta, \pi N}(k_2,k_2) & \bar{V}_{\pi \Delta, \pi \Delta}(k_2,k_2) & \dots & \bar{V}_{\pi \Delta, \pi \Delta}(k_2,k_{n_\text{max}}) \\
	
	\vdots & \vdots & \vdots & \vdots & \ddots & \vdots \\
	
	\bar{V}_{\pi N, \pi N}(k_{n_\text{max}},k_1) & \bar{V}_{\pi N, \pi \Delta}(k_{n_\text{max}},k_1) & 
	\bar{V}_{\pi N, \pi N}(k_{n_\text{max}},k_2) & \bar{V}_{\pi N, \pi \Delta}(k_{n_\text{max}},k_2) & \dots & \bar{V}_{\pi N, \pi \Delta}(k_{n_\text{max}},k_{n_\text{max}}) \\
	
	\bar{V}_{\pi \Delta, \pi N}(k_{n_\text{max}},k_1) & \bar{V}_{\pi \Delta, \pi \Delta}(k_{n_\text{max}},k_1) & 
	\bar{V}_{\pi \Delta, \pi N}(k_{n_\text{max}},k_2) & \bar{V}_{\pi \Delta, \pi \Delta}(k_{n_\text{max}},k_2) & \dots & \bar{V}_{\pi \Delta, \pi \Delta}(k_{n_\text{max}},k_{n_\text{max}})
	
\end{pmatrix}\,.
\end{equation}
\end{widetext}

With the matrices above defined we can solve this symmetric Hamiltonian matrix for its eigenvalues and eigenvectors. The final step required to compare with results from lattice QCD is to extend our Hamiltonian model to unphysical quark/pion masses.

\subsection{Pion Mass Dependence}\label{subsec:PionMassDep}
Thus far, all calculations have been carried out at the physical point, where the hadron masses take their measured values. As noted in the introduction, state-of-the-art lattice QCD calculations can now be carried out at the physical quark-mass point. Nevertheless, comparison with lattice QCD results computed over a broad range of quark masses introduces strong constraints on the Hamiltonian vital to obtaining the composition of states in the Hamiltonian eigenvectors. To extend our calculations at the physical point into the heavier quark-mass regime, we seek a simple interpolation of the finite-volume PACS-CS hadron masses used in Ref.~\cite{Hockley:2023yzn}. Using the linear dependence of $m_\pi^2$ on the quark mass, we find a simple linear relationship between the baryon masses and $m_\pi^2$ to be adequate in reproducing results from lattice QCD. While one anticipates important chiral curvature, this is suppressed in the finite volume and at pion masses far from the physical point where hadron masses follow a linear trend in $ m_\pi^2 $ \cite{Leinweber:2000, PACS-CS:2008bkb}. Thus, for the $ N $ and $ \Delta $ masses appearing in the two-particle basis-state channels, we consider
\begin{equation}
	m_{B}(m^2_{\pi}) = m^{*}_{B} + \alpha_{B}\, \qty(m_{\pi}^2 - m_{\pi}|_\text{phys}^2)\,, \label{eq:mass_extrap}
\end{equation}
where $ m^*_B $ is the interpolated finite-volume mass of baryon $ B $ at the physical quark/pion mass point $ m_\pi|_\text{phys} $. This and the slope parameter $ \alpha_B $ are obtained in a fit to lattice QCD results from the PACS-CS collaboration \cite{PACS-CS:2008bkb}. Note that we allow for finite-volume effects by allowing $ m^*_B \ne m_B|_\text{phys} $ on the lattice. 

One aims to repeat the finite-volume calculation at various choices of pion mass and so extend to regimes compatible with lattice QCD calculations. This builds up a finite-volume energy spectrum where each of the energy values is an eigenvalue of the Hamiltonian at the given pion mass and lattice size. In addition to interpolating the masses appearing in the two-particle channels, one also needs to extend the bare masses from the physical point. This time there is an additional complication as one needs to know which eigenstates of the HEFT spectrum should be fit to the available lattice QCD results. Thus we turn our attention to the Hamiltonian eigenvectors describing the composition of the energy eigenstates.

\subsection{Eigenvector Analysis} \label{subsec:EigvecAnalysis}
At the outset of this HEFT approach, the number of bare basis states $ n_b $ is chosen to match the number of 3-quark interpolated lattice states to which we aim to compare. For example, suppose one is comparing to only a set of ground state lattice masses, then the number of bare states is set to $ n_b = 1 $. This essentially allows for the case where the bare state corresponds to the observed lattice state. To see if there really is a correspondence between the two, we inspect the content of the energy eigenvectors.

Considering the form of the Hamiltonian in Eqs.~(\ref{eq:H0Mat})-(\ref{eq:VMat}), a general eigenvector can be decomposed in terms of bare basis states and two-particle basis states of different momenta as
\begin{equation}
	\ket{E_n} = \sum_{i=1}^{n_b} b_n^i\ket{B^i_0} + \sum_{i = 1}^{n_c} \sum_{j=1}^{n_\text{max}}\, c_{n, j}^{\alpha_i}\ket{\alpha_{i} (k_j)}\,,
\end{equation}
where, for the energy eigenstate with energy-level index $ n $, the $ b_n^i $ denote eigenvector components for each of the bare basis states $ B^i_0 $, and the $ c^{\alpha_i}_{n,j} $ are the eigenvector components for each two-particle basis state $ \alpha_i $ with discrete back-to-back momentum $ k_j $.

Continuing with the $ n_b = n_c = 2 $ case as an example, we write
\begin{align}
	\ket{E_n} 
	&= b_n^1\ket{\Delta_1} 
	+ b_n^2\ket{\Delta_2} \nonumber \\ 
	&+ \sum_{j=1}^{n_\text{max}}\, c_{n, j}^{\pi N}\ket{\pi N (k_j)}
	+ \sum_{j=1}^{n_\text{max}}\, c_{n, j}^{\pi \Delta}\ket{\pi \Delta (k_j)}\,.
\end{align}

One can then consider the overlap of these eigenvectors with each of the basis states, and in particular the bare basis states $ \ket{\Delta_1} $ and $ \ket{\Delta_2} $. Since the lattice results of Ref.~\cite{Hockley:2023yzn} were generated using 3-quark interpolators, they should have significant overlap with the bare basis states in our Hamiltonian. Thus we expect the bare basis state content to be significant for energy eigenvalues near the lattice results and we tune the parameters of the Hamiltonian to ensure this is the case. We can test if our Hamiltonian allows this by simply computing the overlap probabilities
\begin{equation}
	\abs{b^n_i}^2 = \abs{\braket{E_n}{B_i}}^2\,, \label{eq:overlap}
\end{equation}
and optimising the bare masses, $ m_{B_i} $, and the bare mass slope parameters, $ \alpha_{B_i} $, such that the states with highest overlap with the bare states are closest to the lattice QCD results. This optimisation is controlled by a $ \chi^2/{\text{dof}} $ measure. This is done with the aim of achieving a single Hamiltonian which can describe both infinite-volume scattering data, and the lattice data on finite volumes.

\subsection{Model (in)dependence in HEFT}

Understanding the model-dependent and model-independent aspects of HEFT is important. HEFT incorporates the L\"uscher formalism \cite{Wu:2014vma,Hall:2013qba}, and therefore there are aspects of the calculation that share the same level of model independence as the L\"uscher formalism itself.

\subsubsection{Model independence}

The L\"uscher formalism provides a rigorous relationship between the finite-volume energy spectrum
and the scattering amplitudes of infinite-volume experiment.  In HEFT, this relationship is
mediated by a Hamiltonian.  In the traditional approach, the parameters of the Hamiltonian are
tuned to describe lattice QCD results.  When the fit provides a high-quality description of lattice
QCD results, the associated scattering-amplitude predictions are of high quality.  The key is to
have a sufficient number of tunable parameters within the Hamiltonian to accurately describe the
lattice QCD results.

However, in the baryon sector, high-quality precise lattice QCD results are scarce and HEFT is
usually fit to experimental data first. The HEFT formalism then describes the finite-volume
dependence of the baryon spectrum, indicating where high-precision lattice QCD results will reside.
This is the approach adopted herein.  We will show high-quality fits to the experimental scattering
observables such that HEFT provides rigorous predictions of the finite-volume lattice QCD spectrum
with model independence at the level of the L\"uscher formalism.

Of course, this model independence is restricted to the case of matched quark masses in
finite-volume and infinite-volume.  The L\"uscher formalism provides no avenue for changing the
quark mass.  In other words, direct contact with lattice QCD results is only possible when the
quark masses used in the lattice QCD simulations are physical.

On the other hand, $\chi$PT is renowned for describing the quark mass dependence of hadron
properties in a model-independent manner, provided one employs the truncated expansion in the
power-counting regime, where higher-order terms not considered in the expansion are small by
definition.  Given that finite-volume HEFT reproduces finite-volume $\chi$PT in the perturbative
limit by construction \cite{Hall:2013qba,Abell:2021awi}, it is reasonable to explore the extent to
which this model independence persists in the full nonperturbative calculation of HEFT.

This has been explored in Ref.~\cite{Abell:2021awi}.  In the one-channel case where a single
particle basis state (e.g. a quark-model like $ \Delta $) couples to one two-particle channel (e.g. $\pi
N$), the independence of the results on the form of regularisation is reminiscent of that realised
in $\chi$PT.  Any change in the regulator is absorbed by the low-energy coefficients such that the
renormalised coefficients are physical, independent of the renormalisation scheme.

However, in the more complicated two-channel case with a $\pi \Delta$ channel added, the same is
not observed.  The form of the Hamiltonian becomes constrained, describing experimental data
accurately for only a limited range of parameters with specific regulator shapes. This property is
also observed in the calculations performed herein. The Hamiltonian becomes a model in this case,
with regulator-function scales and shapes governed by the experimental data.  The principles of
$ \chi $PT no longer apply in this nonperturbative calculation.  However, for
fit parameters that describe the data well, the model independence of the L\"uscher formalism
remains intact.  The Hamiltonian is only mediary.

\subsubsection{Quark mass variation}

The consideration of variation of the quark masses away from the physical point provides further
constraints on the Hamiltonian.  In particular, lattice QCD results away from the physical point
provide new constraints on the form of the Hamiltonian.  In the two-channel case, we find the
Hamiltonian becomes tightly constrained when considering experimental scattering data and lattice
QCD results together.  This was also observed in the analysis of Ref.~\cite{Abell:2021awi}.

With the Hamiltonian determined by one set of lattice results, one can then make predictions of the
finite-volume spectrum considered by other lattice groups at different volumes and different quark
masses.  This is a central aim of the current investigation where we will confront legacy results
from the HSC \cite{Bulava:2010yg}, as well as contemporary results from Khan {\it et al.} \cite{Khan:2020ahz} and the CLS Consortium \cite{Morningstar:2021ewk}, and very recent lattice QCD predictions for the $\Delta$ spectrum from the Cyprus collaboration at a physical pion mass of 139.4(1) MeV \cite{Alexandrou:2023elk}.

Our Hamiltonian will be constrained by lattice QCD results from Ref.~\cite{Hockley:2023yzn} for
the $1s$ and $2s$ radial excitations of the $\Delta$, such that the confrontation with contemporary
lattice QCD results is parameter free and predictive.

For the cases previously considered in the baryon spectrum the predictions of HEFT are in agreement with lattice QCD spectrum predictions. For example, in the $N(1/2^+)$ channel, HEFT reproduces the lattice QCD results from Lang \textit{et al.} \cite{Wu:2017qve, Lang:2016hnn}. In the $N(1/2^-)$ channel, HEFT successfully predicts spectra from the CLS consortium \cite{Abell:2023nex, Bulava:2022vpq}, the HSC \cite{Abell:2023nex, Edwards:2011jj, Edwards:2012fx} and Lang \& Verducci \cite{Abell:2023nex, Lang:2012db}.  In the odd-parity $\Lambda (1/2^-)$ spectrum, HEFT accurately predicts the finite-volume spectrum of the BaSc collaboration \cite{Liu:2023xvy, BaryonScatteringBaSc:2023ori}. In the present analysis we will show successful predictions for the finite-volume $\Delta$ spectrum of the CLS consortium \cite{Morningstar:2021ewk}. Thus one concludes that the systematic errors of the HEFT approach to quark-mass variation are small on the scale of contemporary lattice QCD uncertainties. As the Hamiltonian is constrained by model-independent scattering data and lattice QCD results, we expect this success to be realised in general.

Variation in the quark mass is conducted in the same spirit as for $\chi$PT.  The couplings are
held constant and the hadron masses participating in the theory take values as determined in
lattice QCD.  The single-particle bare basis states acquire a quark mass dependence and this is
done in the usual fashion by drawing on terms analytic in the quark mass.  In most cases, lattice
QCD results are only able to constrain a term linear in $m_\pi^2$, as is the case here, e.g. in Eq.~\eqref{eq:mass_extrap}.

The model independence associated with the movement of quark masses away from the physical point is
largely governed by the distance one chooses to move from the physical quark-mass point. The HEFT
approach is systematically improvable, reliant on high-quality lattice QCD results to constrain the
higher-order terms that one can introduce.  For example, one could include an additional analytic
$m_\pi^4$ term or higher-order interaction terms from the chiral Lagrangian.  However, this
increased level of precision is not yet demanded by current experimental measurements nor
contemporary lattice QCD results.

\subsubsection{Model dependence}

Now that the Hamiltonian has become a tightly constrained model, the eigenvectors describing the
manner in which the non-interacting basis states come together to compose the eigenstates of the
spectrum are model dependent. At the same time, there is little freedom in the model parameters of
the Hamiltonian such that the predictions of the Hamiltonian are well defined.

The information contained in the Hamiltonian eigenvectors describing the basis-state composition of
finite-volume energy eigenstates is analogous to the information contained within the eigenvectors
of lattice QCD correlation matrices describing the linear combination of interpolating fields
isolating energy eigenstates on the lattice.  These too are model dependent, governed by the nature
of the interpolating fields used to construct the correlation matrix.

What is remarkable is that, as shown in Refs.~\cite{Wu:2017qve,Abell:2023nex}, with a suitable renormalisation scheme on the lattice
(e.g. interpolators are normalised to set diagonal correlators equal to 1 at one time slice after
the source), the composition of the states drawn from the lattice correlation matrix is very
similar to the description provided by HEFT. While both eigenvector
sets are model dependent, their similarity does indeed provide some relevant insight into hadron
structure.  And because regularisation in the Hamiltonian is tightly constrained, one can begin to
separate out the contributions of bare versus two-particle channels.

\subsubsection{Summary}

In summary, there is a direct model-independent link between the scattering observables of
experiment and the finite-volume spectrum calculated in HEFT at physical quark masses.  This model
independence is founded on the L\"uscher formalism embedded within HEFT. Similarly, variations of the quark masses away from the physical quark mass have systematic uncertainties that are small relative to contemporary lattice QCD spectral uncertainties. Finally, the Hamiltonian eigenvectors
describing the basis-state composition of finite-volume energy eigenstates are model dependent.
They are analogous to the interpolator dependent eigenvectors of lattice QCD correlation matrices
describing the linear combination of interpolating fields isolating energy eigenstates on the
lattice.  The similarity displayed by these two different sets of eigenvectors suggests that they
do indeed provide insight into hadron structure.

\section{Preliminary Calculation: $ L = 3\ \text{fm} $}\label{sec:LowEnergyModel}
To begin our analysis of the $ \Delta $-baryon spectrum, we choose our basis states based on the following:
\begin{itemize}
	\item The number of single-particle basis states should match the number of 3-quark-interpolated states identified on the lattice.
	\item The meson-baryon pairs should hold significant branching fractions in the decays of the relevant $ \Delta $ resonances.
\end{itemize}

Another consideration to take into account is the limited availability of $ S $-matrix scattering data. By this, we mean that while $ \pi N \to \pi N $ scattering data is readily available to fit our infinite-volume Hamiltonian to, the same cannot be said for other relevant transitions such as $ \pi N \to \pi \Delta $ and $ \pi \Delta \to \pi \Delta $. This means that including many channels will make it difficult to constrain the Hamiltonian as the number of parameters increases rapidly with additional channels in which there is little experimental support. It is here that lattice QCD results can provide important constraints.

With these considerations in mind, we start by only looking to analyse the lowest state in the lattice spectrum, that is, the $ 1s $ state identified in Ref.~\cite{Hockley:2023yzn} as corresponding to the $ \Delta(1232) $ resonance. Hence, we take $ n_b = 1 $, with $ \Delta_1 $ denoting the single bare basis state. The most relevant channels at this energy are $ \pi N $ and $ \pi \Delta $, both in $ p $-wave, based on branching fractions in the PDG \cite{ParticleDataGroup:2022pth}. Thus the relevant basis states are categorised as:
\begin{itemize}
	\item Single-particle bare basis states: 
	\subitem $ \ket{\Delta_1} $.
	\item Two-particle basis states: 
	\subitem $ \ket{\pi N (\boldsymbol{k})},\, \ket{\pi \Delta (\boldsymbol{k})}.  $
\end{itemize}

\begin{figure*}
		\begin{center}
			\includegraphics[width=0.85\linewidth]{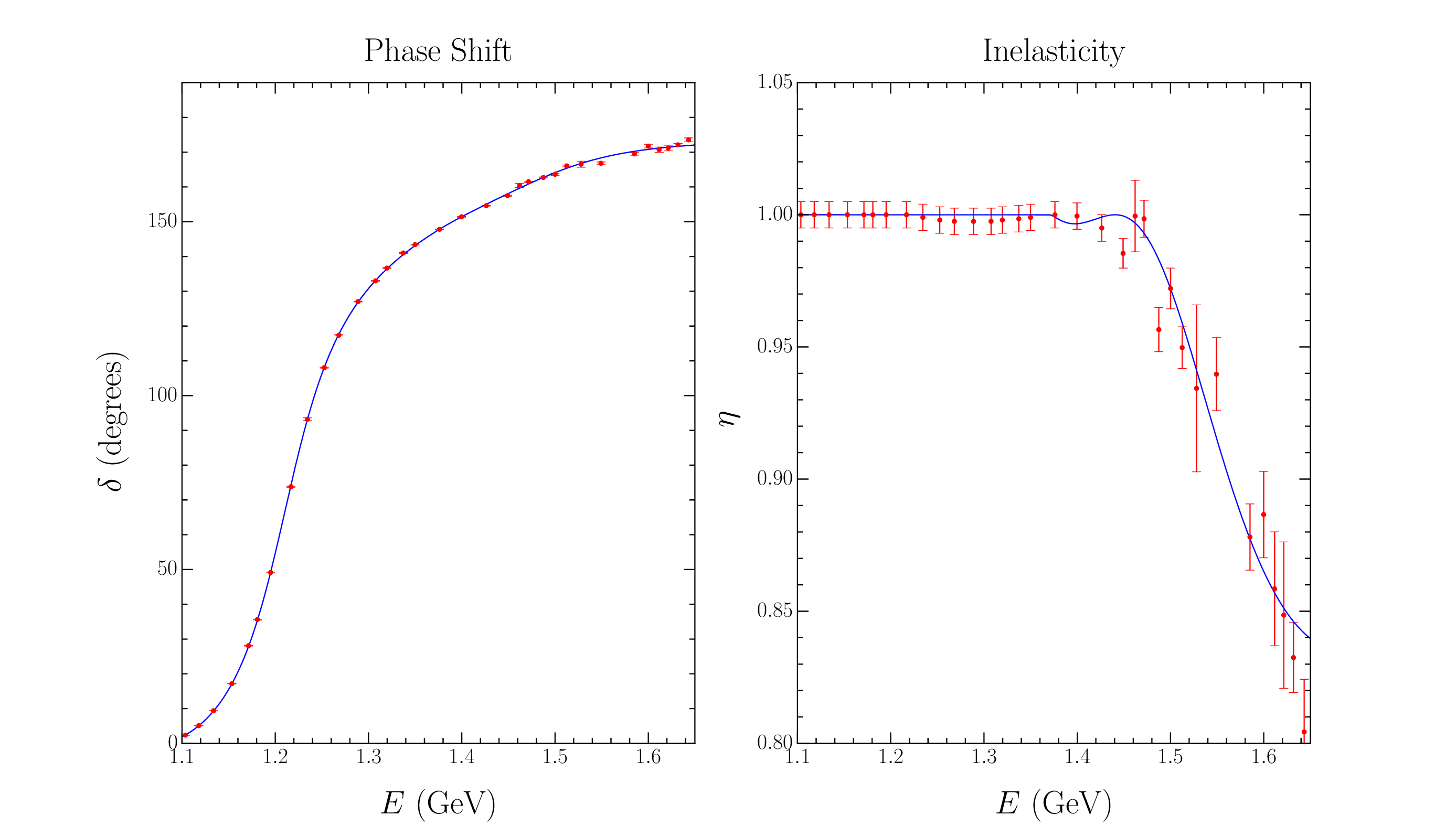}
			\caption{Plot of fits to experimental data for phase shifts $ \delta $ (Left) and inelasticities $ \eta $ (Right) in $ \pi N \to \pi N $ scattering, up to a region close to the $ \Delta(1600) $. The data we perform our fits to is shown by the red points with error bars, and the resulting fits are shown by blue solid lines. The phase shift passes through $ 90^\circ $ at the position of the $ \Delta(1232) $.}
			\label{fig:said_2ch1b}
		\end{center}
	\end{figure*}

Following the procedure outlined in Section~\ref{sec:HEFT}, the $ S $-matrix can be computed for a number of parameters, and the resulting $ \pi N \to \pi N $ scattering phase shifts $ \delta $ and inelasticities $ \eta $ can be extracted. These have then been fit to scattering data from Ref.~\cite{GWU:2023ex} as shown in Fig.~\ref{fig:said_2ch1b}, using a $ \chi^2 $ per degree of freedom ($ \chi^2/\text{dof} $) to control the optimisation. We observe excellent agreement in both scattering observables and the results of the fits are given in Table~\ref{tab:heft_params}.

\begin{table}
	\begin{center}
		\setlength{\tabcolsep}{20pt}
		\renewcommand{\arraystretch}{1.5}
		\caption{Table of HEFT parameters from the fit to SAID partial wave analysis data. The measures of how well we fit the data are shown as the $ \chi^2 $ and $ \chi^2/{\text{dof}} $, where dof is the degrees of freedom.}
		\begin{tabular}{lr}
			\hline\hline
			Parameter & Value      \\
			\hline
			$ m_{\Delta_1}/\text{GeV} $                         & $  1.3952 $ \\
			$ g^{\Delta_1}_{\pi N} $                 			& $  0.5320 $ \\
			$ g^{\Delta_1}_{\pi \Delta} $            			& $  0.6826 $ \\
			$ v_{\pi N, \pi N} $                     			& $  0.0324 $ \\
			$ v_{\pi N, \pi \Delta} $                			& $ -1.5425 $ \\
			$ v_{\pi \Delta, \pi \Delta} $           			& $ -1.6157 $ \\
			$ \Lambda^{\Delta_1}_{\pi N}/\text{GeV} $           & $  0.6377 $ \\
			$ \Lambda^{\Delta_1}_{\pi \Delta}/\text{GeV} $      & $  0.8011 $ \\
			$ \Lambda^v_{\pi N}/\text{GeV} $                    & $  0.6764 $ \\
			$ \Lambda^v_{\pi \Delta}/\text{GeV} $               & $  0.7792 $ \\
			\hline
			\text{dof} & $ 56 $ \\
			$ \chi^2 $ & $ 212.61 $ \\
			$ \chi^2/{\text{dof}} $ & $ 3.80 $ \\
			\hline\hline
		\end{tabular}
		\label{tab:heft_params}
	\end{center}
\end{table}

As was discussed in Ref.~\cite{Abell:2021awi}, the data we perform our fits to has neglected systematic errors, making it difficult to interpret the $ \chi^2/\text{dof} $ value. One can address this as in Ref.~\cite{Meissner:1999vr} by considering a small additional contribution to the errors to account for neglected systematics \cite{Abell:2021awi,Meissner:1999vr}. Taking this approach, we find that our $ \chi^2/\text{dof} $ reduces to $ 0.57 $ after assigning a $ 3\% $ uncertainty to the phase shift data. In addition to this and the good visual agreement with the W108 solution from Ref. \cite{GWU:2023ex, Workman:2012hx}, we can also consider the T-matrix poles of our model, of which we can extract two. These may be compared with those provided by the Particle Data Group (PDG) for the $ \Delta(1232) $ and $ \Delta(1600) $, given in Table~\ref{tab:heft_poles}. We find exact agreement with the lower-lying resonance pole position, and reasonable agreement with the next highest resonance.

\begin{table}
	\begin{center}
		\renewcommand{\arraystretch}{1.5}
		\caption{Table of pole positions in GeV extracted from the T-matrix corresponding to our Hamiltonian model. Particle Data Group results \cite{ParticleDataGroup:2022pth} for the first two poles of the $ \Delta(3/2^{+}) $ spectrum are included for comparison.}
		\begin{ruledtabular}
			\begin{tabular}{ccc}
				Pole \# & \multicolumn{1}{c}{This Work}            & \multicolumn{1}{c}{PDG}            \\
				\hline
				1 & $ 1.210 - 0.050i $ & $ (1.210\pm0.001) - (0.050\pm0.001)i $ \\
				2 & $ 1.475 - 0.141i $ & $ (1.52^{+0.07}_{-0.05}) - (0.14^{+0.02}_{-0.07})i $
			\end{tabular}
		\end{ruledtabular}
		\label{tab:heft_poles}
	\end{center}
\end{table}


Following these infinite-volume considerations, we perform the corresponding finite-volume calculations at unphysical pion masses, using the approach outlined in Sections~\ref{subsec:FinVolHam}, \ref{subsec:HamMat} and \ref{subsec:PionMassDep}. The various Sommer-scale lattice spacings and sizes from the PACS-CS ensembles are given in Table~\ref{tab:PACS-CS_params} and we take $ L $ to vary linearly between these values as we interpolate to build up the spectrum. 

\begin{table}
	\begin{center}
		\renewcommand{\arraystretch}{1.5}
		\caption{Table of parameters for the PACS-CS lattice ensembles used by the CSSM, characterised by the Hopping parameter $ \kappa $. The configurations are calculated on lattices with spacings $ a $ and spatial lengths $ L $, and pion masses $ m_\pi $, as determined using the Sommer parameter to set the scale on an ensemble-by-ensemble basis \cite{Hockley:2023yzn}.}
		\begin{tabular}{cccc}
			\hline
			\hline
			$ \kappa $ & $ m_\pi $ (MeV) & $ a $ (fm) & $ L $ (fm) \\
			\hline
			0.13781 & 169 & 0.0933(13) & 2.9856(416) \\
			0.13770 & 280 & 0.0951(13) & 3.0432(416) \\
			0.13754 & 391 & 0.0961(13) & 3.0752(416) \\
			0.13727 & 515 & 0.1009(15) & 3.2288(480) \\
			0.13700 & 623 & 0.1023(15) & 3.2736(480) \\
			\hline
		\end{tabular}
		\label{tab:PACS-CS_params}
	\end{center}
\end{table}

The values for the finite-volume mass and slope parameters are given in Table~\ref{tab:Bar2ch1b} where we note that the values for $ B = N,\, \Delta $ are obtained by fitting to results from the PACS-CS collaboration.

\begin{table}
	\begin{center}
		\renewcommand{\arraystretch}{1.5}
		\setlength{\tabcolsep}{12pt}
		\caption{Table of finite-volume baryon masses and slope parameters for the single bare state, $ n_b = 1 $, and two channel, $ n_c = 2 $, case. The Basis State column describes how the mass and slope parameters are used.}
		\begin{tabular}{lccc}
			\hline
			\hline
			$ B $ 			& Basis State  		& $ m_B^* $ (GeV) 	& $ \alpha_B $ (GeV$ ^{-1} $) \\
			\hline
			$ N $        	& two-particle 		& 1.002 			& 1.097 \\
			$ \Delta $   	& two-particle 		& 1.263 			& 0.972 \\
			$ \Delta_1 $ 	& single-particle 	& 1.395 			& 0.896 \\
			\hline
		\end{tabular}
		\label{tab:Bar2ch1b}
	\end{center}
\end{table}

\begin{figure}
	\includegraphics[width=\linewidth]{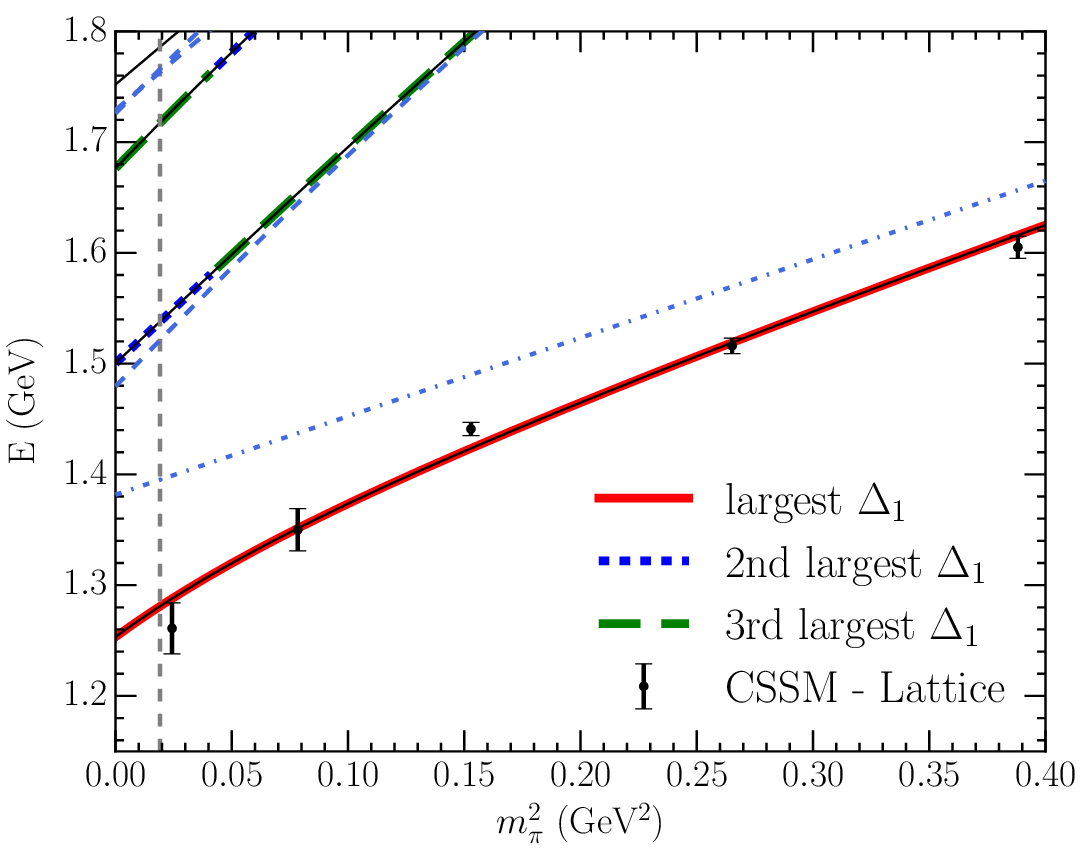}
	\caption{Plot of the $ L\sim3 $ fm finite volume energy spectrum. Solid lines are fully interacting energy eigenvalues while dashed blue lines are non-interacting energies. The blue dot-dashed line is the non-interacting energy associated with the bare basis state. Coloured dressings on interacting energy level lines show states with the largest, second largest and third largest bare basis state components, as described in the text. The vertical grey dashed line is at the physical quark mass point.} \label{fig:finVol_2ch1b}
\end{figure}

With the two-particle baryon mass parameters in hand, the full energy spectrum is calculated across a broad range of pion masses, at lattice sizes $ \sim3 $ fm. Identifying the Hamiltonian eigenstate dominated by the bare basis state, the bare mass, $ m^*_{\Delta_1} $, and slope parameter, $ \alpha_{\Delta_1} $, are constrained by recent lattice QCD results from the CSSM \cite{Hockley:2023yzn}. These results are displayed in Fig.~\ref{fig:finVol_2ch1b}. The energy levels at each value of $ m_\pi^2 $ are shown by the solid black lines. The states in the spectrum with largest overlap with the bare basis state are shown by the solid red dressings, while blue and green dressed lines represent states with the 2nd and 3rd largest $ \ket{\Delta_1} $ components respectively. Importantly, the state with the largest overlap with the bare basis state is constrained to match the lattice QCD state excited by 3-quark operators. The excellent agreement between the red solid dressed line and the lattice results follows from this constraint. 

In summary then, we have produced a Hamiltonian model which simultaneously describes experiment and the lattice results, capturing the resonance physics in both infinite and finite volumes. With this model however, we are unable to comment on the structure of the $ 2s $ excitation observed in lattice QCD \cite{Hockley:2023yzn} at $ \sim 2.15 $ GeV. 

With this proof of concept established, we turn our attention to a more sophisticated analysis which incorporates a second bare state and an additional channel, to explore the higher energy region.

\section{Results: $ L = 3\, \text{fm}$ }\label{sec:FullModel}
In order to investigate the $ 2s $ state excited on the lattice, we extend our Hamiltonian model to incorporate a second bare basis state. To capture the physics involved at the higher energy scale near the $ 2s $ excitation, we include a third meson-baryon channel in the form of $ \pi \Delta $ coupled in $ f $-wave. This has strong coupling to the $ \Delta(1920) $ and should be significant in the $ \sim 2 $ GeV region. We distinguish between our $ p $- and $ f $-wave $ \pi \Delta $ channels with a subscript $ p $ or $ f $. Our Hamiltonian basis states for this extended model are then
\begin{itemize}
	\item Single-particle bare basis states: 
	\subitem $ \ket{\Delta_1},\, \ket{\Delta_2} $.
	\item Two-particle basis states: 
	\subitem $ \ket{\pi N (\boldsymbol{k})},\, \ket{\pi \Delta_p (\boldsymbol{k})},\, \ket{\pi \Delta_f (\boldsymbol{k})}.  $
\end{itemize}

We note that while our Hamiltonian formalism doesn't allow us to include a direct 3-body $ \pi \pi N $ threshold, the resonant contributions for this process are incorporated in the 2-body channels chosen above. Thus we proceed with the assumption that our 2-body channels are suitable for capturing the relevant physics.

\begin{figure*}
	\includegraphics[width=0.85\linewidth]{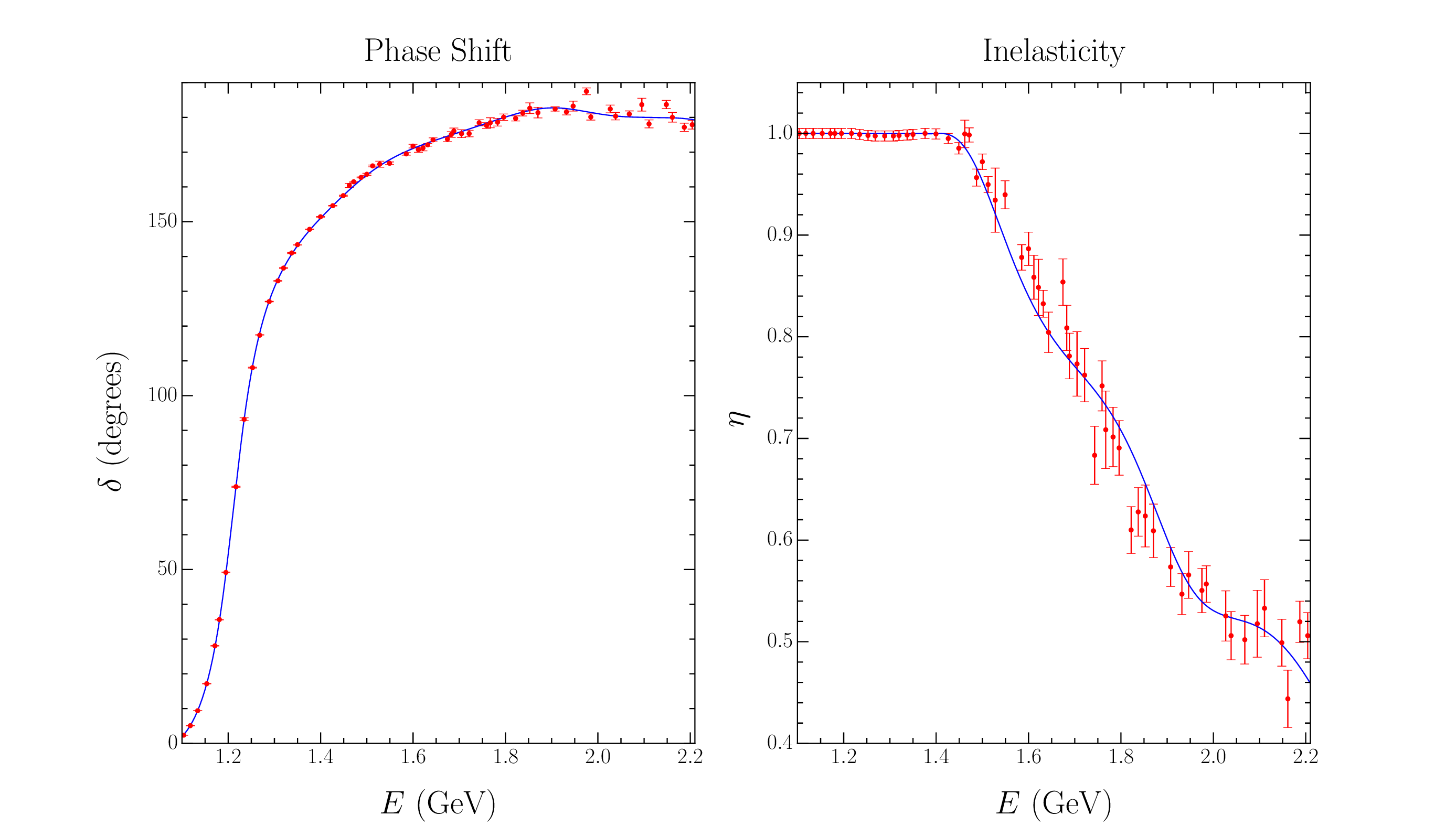}
	\caption{Plot of fits to experimental data for phase shifts $ \delta $ (Left) and inelasticities $ \eta $ (Right) in $ \pi N \to \pi N $ scattering, for centre-of-mass energies up to $ \sim 2.2 $ GeV. The data we perform our fits to is shown by the red points with error bars, and the best fit of our HEFT analysis is shown by blue solid lines.}
	\label{fig:said_3ch2b}
\end{figure*}

The Hamiltonian is then constructed from the various couplings between the basis states given by Eqs.~\eqref{eq:full_g} and \eqref{eq:full_v}, and the masses of the two bare basis states. We then constrain the Hamiltonian using a fit to available scattering data up to $ 2.3 $~GeV, ensuring that we produce the relevant resonance at $ 1.920 $ GeV. The fit to the scattering data is displayed in Fig.~\ref{fig:said_3ch2b} with the Hamiltonian parameters given in Table~\ref{tab:heft_params2}.

\begin{table}[]
	\begin{center}
		\setlength{\tabcolsep}{10pt}
		\renewcommand{\arraystretch}{1.2}
		\caption{Table of HEFT parameters constrained by scattering phase shifts and inelasticities at 61 SAID partial wave analysis data points.}
		\begin{tabular}{lr|lr}
			\hline
			\hline
			Parameter                                & Value &      
			Parameter                                & Value \\
			\hline
			$ m_{\Delta_1}/ \text{GeV} $                         & $  1.3894 $ &
			$ m_{\Delta_2}/ \text{GeV} $                         & $  2.3177 $ \\
			$ g^{\Delta_1}_{\pi N} $                 & $  0.4974 $ &
			$ g^{\Delta_2}_{\pi N} $                 & $  0.2914 $ \\
			$ g^{\Delta_1}_{\pi \Delta_p} $          & $  0.5300 $ &
			$ g^{\Delta_2}_{\pi \Delta_p} $          & $  0.2289 $ \\
			$ g^{\Delta_1}_{\pi \Delta_f} $          & $  0.0004 $ &
			$ g^{\Delta_2}_{\pi \Delta_f} $          & $  0.0075 $ \\
			$ \Lambda^{\Delta_1}_{\pi N}/ \text{GeV} $           & $  0.8246 $ &
			$ \Lambda^{\Delta_2}_{\pi N}/ \text{GeV} $           & $  1.3384 $ \\
			$ \Lambda^{\Delta_1}_{\pi \Delta_p}/ \text{GeV} $    & $  0.8376 $ &
			$ \Lambda^{\Delta_2}_{\pi \Delta_p}/ \text{GeV} $    & $  0.5428 $ \\
			$ \Lambda^{\Delta_1}_{\pi \Delta_f}/ \text{GeV} $    & $  0.5776 $ &
			$ \Lambda^{\Delta_2}_{\pi \Delta_f}/ \text{GeV} $    & $  1.0549 $ \\
			$ v_{\pi N, \pi N} $                     & $  0.0454 $ &
			$ v_{\pi N, \pi \Delta_f} $              & $ -0.0030 $ \\
			$ v_{\pi N, \pi \Delta_p} $              & $ -1.5545 $ &
			$ v_{\pi \Delta_p, \pi \Delta_f} $       & $ -0.0053 $ \\
			$ v_{\pi \Delta_p, \pi \Delta_p} $       & $ -0.9694 $ & 			
			$ v_{\pi \Delta_f, \pi \Delta_f} $       & $ -0.0001 $ \\
			$ \Lambda^v_{\pi N}/ \text{GeV} $                    & $  0.6032 $ &
			$ \Lambda^v_{\pi \Delta_f}/ \text{GeV} $             & $  1.3289 $  \\
			$ \Lambda^v_{\pi \Delta_p}/ \text{GeV} $             & $  0.8058 $ && \\
			\hline
			dof     & $ 99 $ && \\
			$ \chi^2 $ & $ 581.81 $ && \\
			$ \chi^2/{\text{dof}} $ & $ 5.88 $ && \\
			\hline
			\hline
		\end{tabular}
		\label{tab:heft_params2}
	\end{center}
\end{table}

Again, there is excellent visual agreement between our fit result and the reported partial wave analysis data. While the $ \chi^2/\text{dof} $ is once again quite large, we treat the uncertainties in the same way as in Section~\ref{sec:LowEnergyModel} and achieve a reduction to $ 1.07 $. The poles achieved with this fit are given in Table~\ref{tab:heft_poles2} and once again agree remarkably well with the PDG, though both the real and imaginary parts of the second pole sit just beyond the bounds of the corresponding PDG pole.

\begin{table}
	\begin{center}
		\renewcommand{\arraystretch}{1.5}
		\caption{Table of pole positions in GeV extracted from the T-matrix corresponding to our full Hamiltonian model. Particle Data Group results \cite{ParticleDataGroup:2022pth} for the first three poles of the $ \Delta(3/2^{+}) $ spectrum are included for comparison.}
		\begin{tabular}{ccc}
			\hline
			\hline
			Pole \#   & Our Value     &  PDG Value \\
			\hline
			1  & $ 1.211 - 0.049i $ & $ (1.210\pm0.001) - (0.050\pm0.001)i $ \\
			2  & $ 1.46 - 0.17i $ & $ (1.52^{+0.07}_{-0.05}) - (0.14^{+0.02}_{-0.07})i $ \\
			3  & $ 1.91 - 0.19i $ & $ (1.90\pm0.05) - (0.15\pm0.10)i $ \\
			\hline
		\end{tabular}
		\label{tab:heft_poles2}
	\end{center}
\end{table}

\begin{figure*}
		\includegraphics[width=0.8\linewidth]{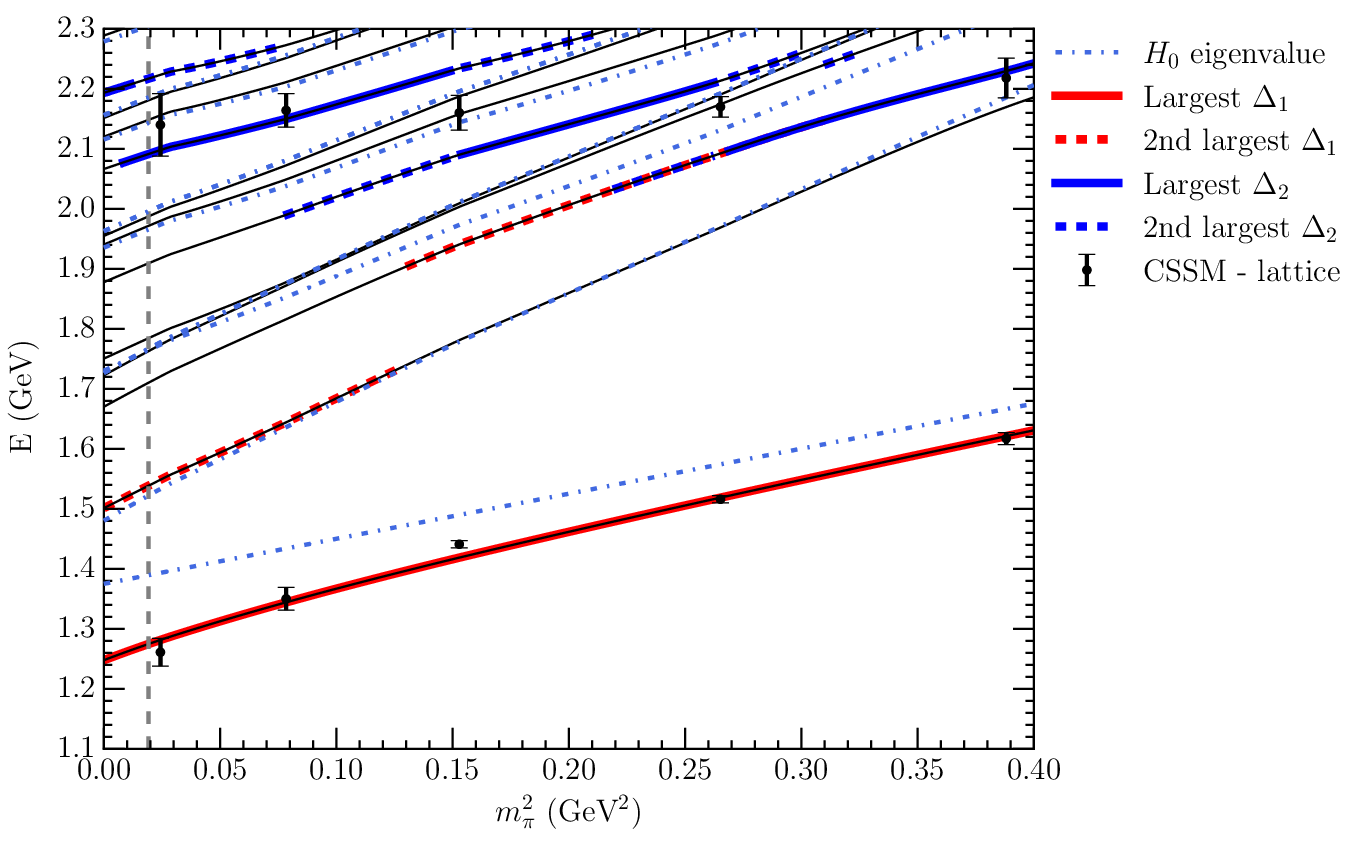}
		\caption{Plot of the finite-volume spectrum for $ L \sim 3 $ fm using 3 channels and 2 bare states. Solid colours denote states with the largest overlaps with the first and second bare basis states (red for the lower lying and blue for the higher energy bare state). Dashed dressings show the states with second largest overlap. Dot-dashed blue lines show the non-interacting states in the spectrum. The vertical grey dashed line is at the physical quark mass point.}
		\label{fig:finVol_3fm}
	\end{figure*}

The finite-volume spectrum is generated in the same manner as discussed in Section~\ref{sec:LowEnergyModel} using baryon masses given in Table~\ref{tab:Bar3ch2b} and is shown in Fig.~\ref{fig:finVol_3fm}. Considering the ground state first, we see that we have retained an excellent quark model dominated description of the $ \Delta(1232) $. 

However this is not to say that the ground state has negligible mixing with the other basis states. To see this, one can consider plots of the eigenvector components for each state in the spectrum, such as those shown in Fig.~\ref{fig:eigvec_comps1}. Note that  the ``Overlap" in these eigenvector plots is a measure of the modulus-square of the coefficient for each basis state, as in Eq.~\eqref{eq:overlap} for the bare basis states. The coefficients for a given energy eigenstate are normalised such that the sum of the overlap probabilities over all basis states is unity. 

At the physical pion mass, the first eigenvector, corresponding to the ground state of the spectrum, has $ \sim 75 \%  $ component coming from the first bare basis state $ \Delta_1 $. The remaining $ \sim 25 \% $ is composed of $ \pi N $ and $ \pi \Delta $ in $ p $-wave. Additionally, we note that further separating the eigenvectors into their individual contributions from each basis state (essentially separating each of the two-particle species into their different back-to-back momentum components) as in the right-hand column of Fig.~\ref{fig:eigvec_comps1} yields a further breakdown of the structural information. 

\begin{table}
	\begin{center}
		\renewcommand{\arraystretch}{1.5}
		\setlength{\tabcolsep}{12pt}
		\caption{Table of finite-volume baryon mass and slope parameters for the two bare state, $ n_b = 2 $, and three channel, $ n_c = 3 $ case. The Basis State column describes how the mass and slope parameters are used.}
		\begin{tabular}{cccc}
			\hline
			\hline
			$ B $ 			& Basis State 		& $ m_B^* $ (GeV) 	& $ \alpha_B $ (GeV$ ^{-1} $) \\
			\hline
			$ N $ 			& two-particle 		& 1.002 			& 1.097 \\
			$ \Delta $ 		& two-particle 		& 1.263 			& 0.972 \\
			$ \Delta_1 $ 	& single-particle 	& 1.389 			& 0.751 \\
			$ \Delta_2 $ 	& single-particle 	& 2.318 			& 0.203 \\
			\hline
		\end{tabular}
		\label{tab:Bar3ch2b}
	\end{center}
\end{table}

Considering the lattice QCD results at $ \sim 2.15 $ GeV corresponding to the $ 2s $ state, we see that there is good overall agreement between these and the blue-dressed line indicating the HEFT energy eigenstate with largest $ \ket{\Delta_2} $ basis state component. Thus this dressing indicates where the second quark model-like state in the spectrum is to be found on the lattice. 

However, not all of the $ 2s $ excitation lattice points lie on a blue solid or dashed line. The two points at $m_\pi^2 \sim 0.15 $ and $ 0.27$ GeV${}^2$ sit between the blue lines instead of selecting one of the two. This suggests that the lattice QCD analysis has reported an energy associated with a superposition of two bare-basis-state dominated energy eigenstates. 

Remarkably, our HEFT analysis actually predicts this superposition of energy eigenstates. In both cases, the lattice QCD results sit at the precise point where the bare-basis-state dominated energy eigenstate jumps from one eigenstate energy level to another. Thus the lattice QCD points are at the location where the 3-quark interpolating field has an approximately equal overlap with two nearby energy eigenstates. Thus a superposition of nearby states is excited and an average mass is reported in the lattice analysis. 

Note how this situation contrasts the other three lattice QCD points. These points are located away from the bare-basis-state cross-overs and sit well on the bare-basis-state dominated HEFT eigenstate. Of the three points, the lightest point is closest to a cross over and may have a small contamination from the higher-energy second bare-basis-state dominated energy eigenstate.

In this light, we have succeeded in describing the finite-volume results from lattice QCD, using our infinite-volume inspired Hamiltonian.

\subsection{Eigenvectors 1, 2, 3}
At this point we can begin identifying key structural features of the energy eigenstates using the superposition of basis states provided in the Hamiltonian eigenvectors. Fig.~\ref{fig:eigvec_comps1} presents results for the composition of the three lowest energy eigenstates. The left-hand column shows a sum over all momentum basis states for a given channel, while the right-hand column reports individual momentum components. 

The dominant component for the lowest state (described by eigenvector 1) is $ \Delta_1 $, with the overall $ \pi N $ and $ \pi \Delta_p $ contributions coming from $ n = 1 $ and $ n = 2 $ momenta. So we can conclude that the state in the HEFT spectrum most closely associated with the $ \Delta(1232) $ is mostly composed of a quark model-like bare basis state, with a relatively small admixture of low-lying two-particle states. This explains why local 3-quark operators were sufficient to excite the ground state on the lattice.

Notice also that there is a significant mass gap between the first non-interacting energy level and the ground state in Fig.~\ref{fig:finVol_3fm}. We can now understand this based on the eigenvector components. The significant $ \Delta_1 $ component and the large coupling strengths $ g^{\Delta_1}_{\pi N} $ and $ g^{\Delta_1}_{\pi \Delta_p} $ lead to significant dressing of the bare basis state mass of $ m_{\Delta_1} = 1.3894 $ GeV at the physical point, shifting it down to the interacting energy level observed. 

Interestingly, these mass gaps are not present for all states in the spectrum. Indeed, the very next state (2nd solid line in Fig.~\ref{fig:finVol_3fm}) lies almost exactly on the corresponding non-interacting line. This is reflected in the eigenvector components in the second row of Fig.~\ref{fig:eigvec_comps1}. Here, there is clear dominance of the $ \pi N $ basis states (predominantly composed of the $ n=1 $ momentum state) with only a small contribution from the first bare state. This leads to a much smaller mass gap, as the smaller $ \pi N $ and $ \pi \Delta $ channel coupling strengths are insufficient to drive the mass away from the non-interacting energy level.

Climbing to the next state in the spectrum, described by eigenvector 3, the first bare state contribution is significantly diminished and we see the second bare state starting to become more involved. The larger couplings between these bare states and the meson-baryon channels again drives the mass away from the non-interacting level. Indeed, the composition of this state at $ \sim 1.7 $ GeV is highly mixed, with the dominant contributions coming from all three two-particle channels.

\subsection{Eigenvectors 4, 5, 6}

The 4th and 5th eigenvectors have no significant bare state components and we see that the interacting energy levels lie close to the non-interacting levels. Interestingly, the 4th eigenvector is initially coincident with a non-interacting energy level at the physical point and gradually diverges as we increase in pion mass. This is also reflected in the eigenvector component plots (top panels of Fig.~\ref{fig:eigvec_comps2}) where less trivial two-particle mixing and also mixing with the second bare state enter at larger pion masses. The evolution of eigenvector 5 in $ m_\pi^2 $ is almost an inverse of this with the initial mixture of $ \pi N $ and $ \pi \Delta $ basis states becoming dominated by $ \pi \Delta $ beyond $ m_\pi^2 \sim 0.15 $ GeV$ ^2 $. This explains why the line corresponding to eigenvector 5 begins to overlap with the nearby non-interacting state in Fig.~\ref{fig:finVol_3fm}. 

The 6th eigenvector is then the next most interesting where we again observe some non-trivial bare state component (now almost exclusively from the second bare state $ \Delta_2 $) along with the indicative mass gap in Fig.~\ref{fig:finVol_3fm}.

\subsection{Eigenvectors 7--12}
Similar analysis of the 7th and 8th eigenvectors in Fig.~\ref{fig:eigvec_comps3} can be used to understand their near degeneracy with the corresponding non-interacting levels and the large mass gap for the 9th eigenvector is similarly associated with the clear presence of the second bare state. The 10th, 11th and 12th eigenvectors follow this same pattern.

\subsection{Momentum Modes}
Comparing the eigenvector components between the left and right columns of Fig.~\ref{fig:eigvec_comps1}, and likewise for Fig.~\ref{fig:eigvec_comps2}, it appears that typically only a single momentum mode in each two-particle channel contributes in any significant way. The exceptions to this appear to be the 3rd and 6th eigenvectors where $ n=1,\, 2 $ and $ 3 $ momentum modes of the same two-particle species are mixed (for example the $ \pi N (n=1,2) $ basis states for the 3rd eigenvector). This dominance by the lowest lying momentum modes is partly due to the inverse proportionality of the lattice size $ L $ and the discretised momentum $ k_n $ in Eq.~\eqref{eq:DiscreteMom}. This indicates that larger lattice sizes will allow for more densely packed momentum states, and in particular, larger volumes allow more momentum basis states to be kinematically allowed participants for low-lying energy levels. This will be further explored in Section~\ref{sec:CompLat}.

\subsection{HEFT as a guide for lattice QCD}
The eigenvector components for each energy eigenstate observed in HEFT can be used to guide lattice QCD studies in their choice of interpolating fields. For example, the ground state HEFT energy aligns very well with the observed $ \Delta(1232) $ resonance and has significant overlap with the first bare state. This explains why the lattice calculation of the $ \Delta(1232) $ mass in Ref.~\cite{Hockley:2023yzn} was effective when using 3-quark interpolators.


When it comes to the HEFT eigenstates nearby to the $ \Delta(1600) $, namely eigenvectors 2, 3, 4 and 5, we observe smaller bare state components than for those eigenstates near the $ 2s $ excitation around $ \sim 2.15 $ GeV. This explains why the lattice study of Ref.~\cite{Hockley:2023yzn} was unable to yield eigenstates in the vicinity of the $ \Delta(1600) $; using 3-quark operators alone means their interpolators couple more strongly to the states in the $ \sim 2.15 $ GeV region which have largest $ \Delta_2 $ component. 

\begin{figure*}
		\includegraphics[width=0.453\linewidth]{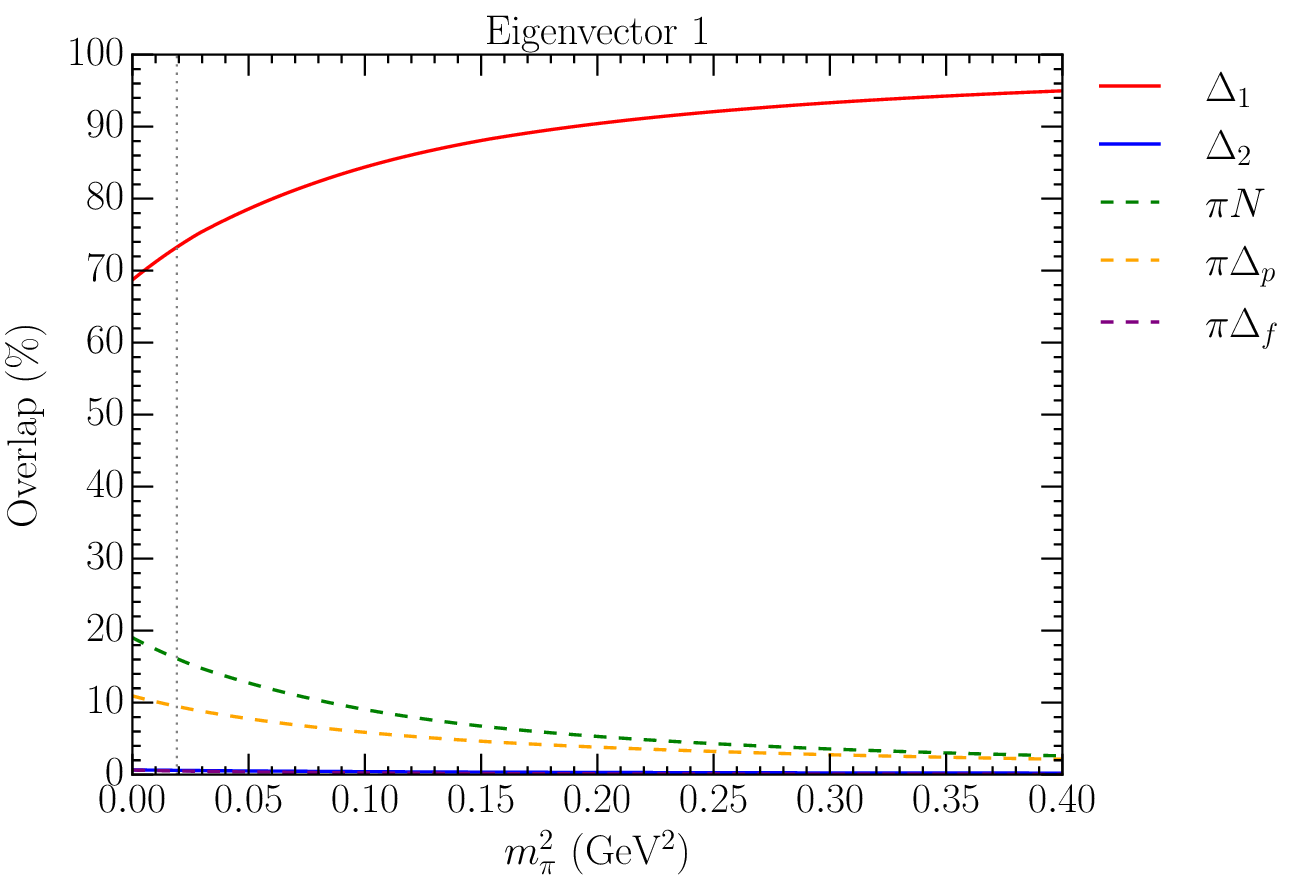}
		\includegraphics[width=0.5\linewidth]{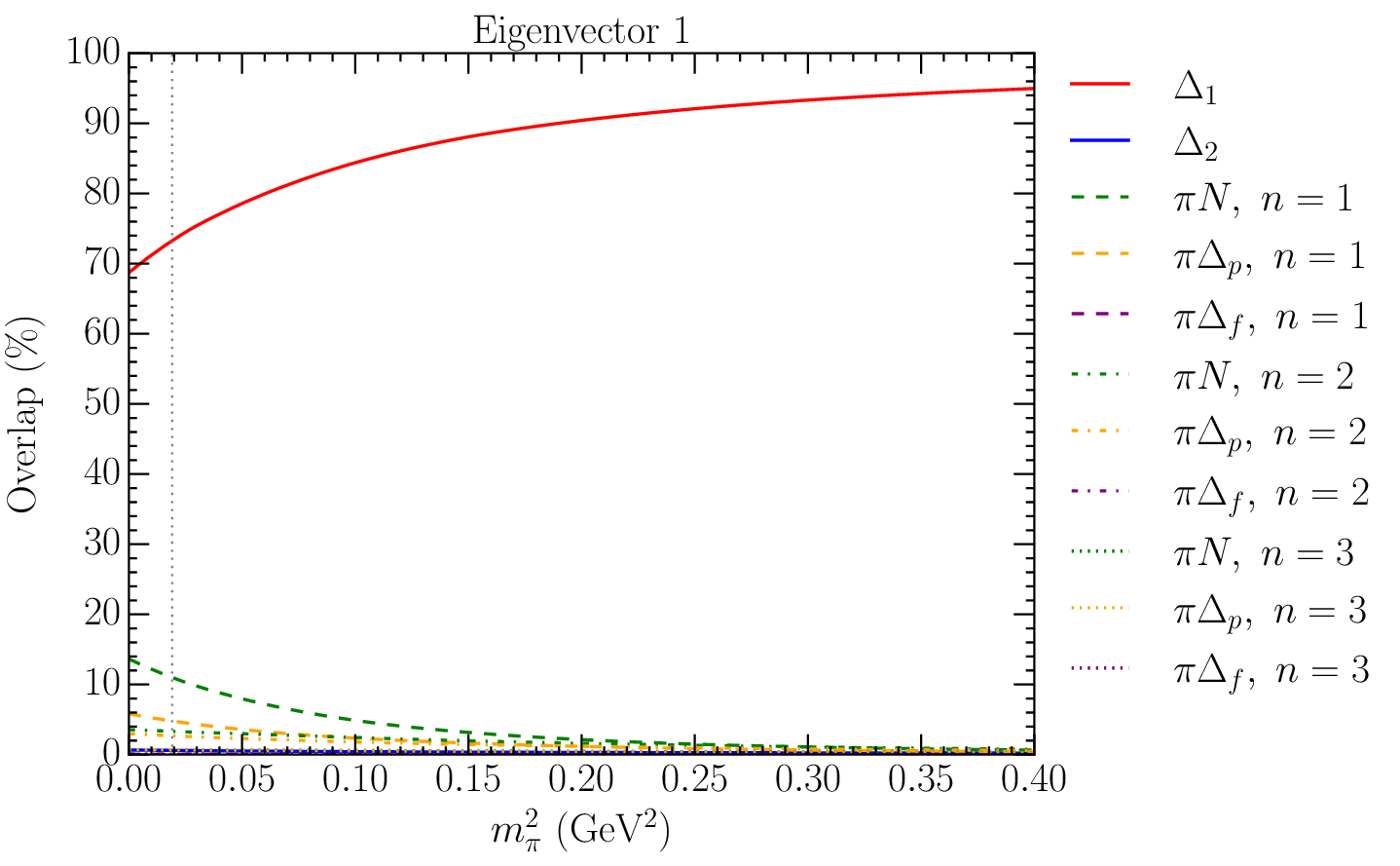}
		\includegraphics[width=0.453\linewidth]{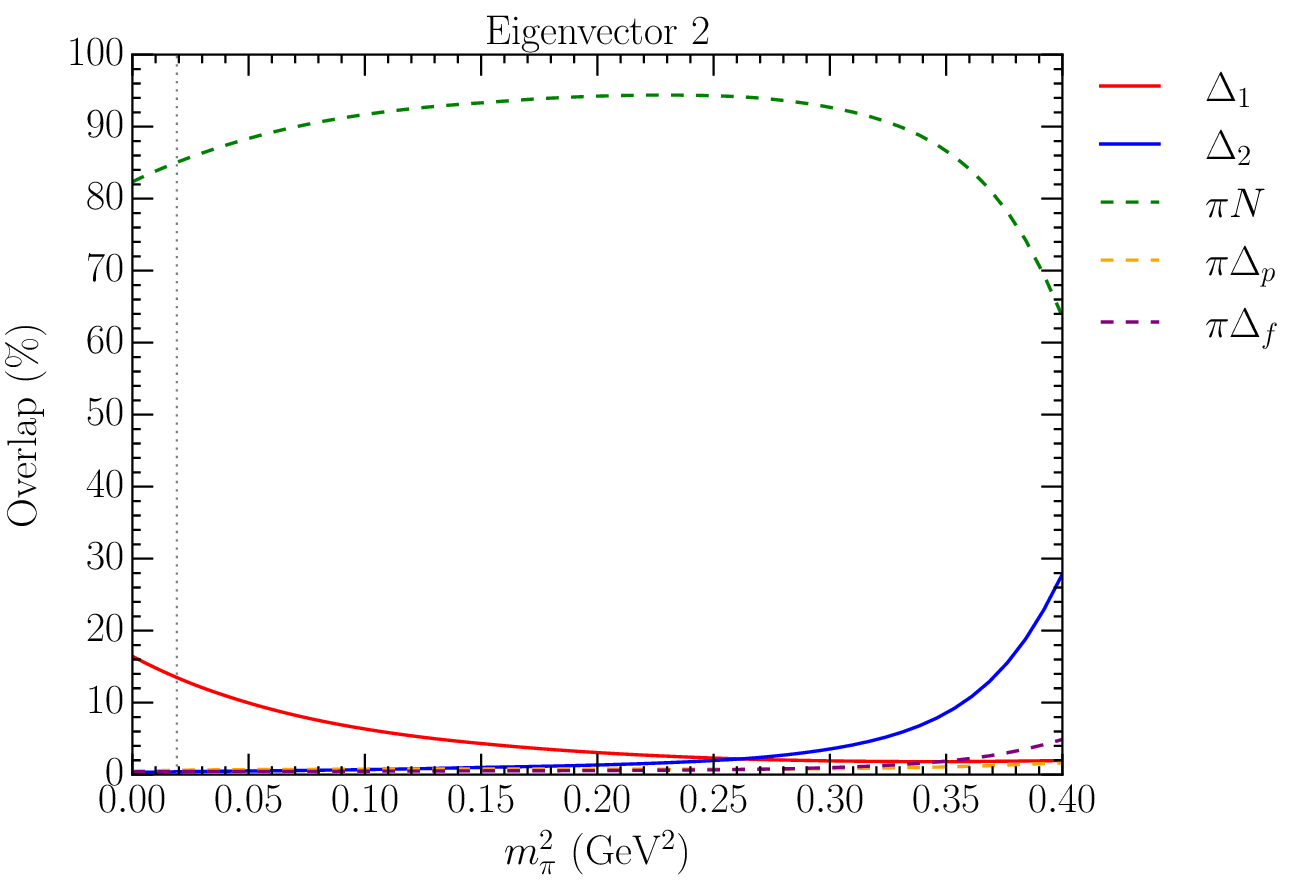}
		\includegraphics[width=0.5\linewidth]{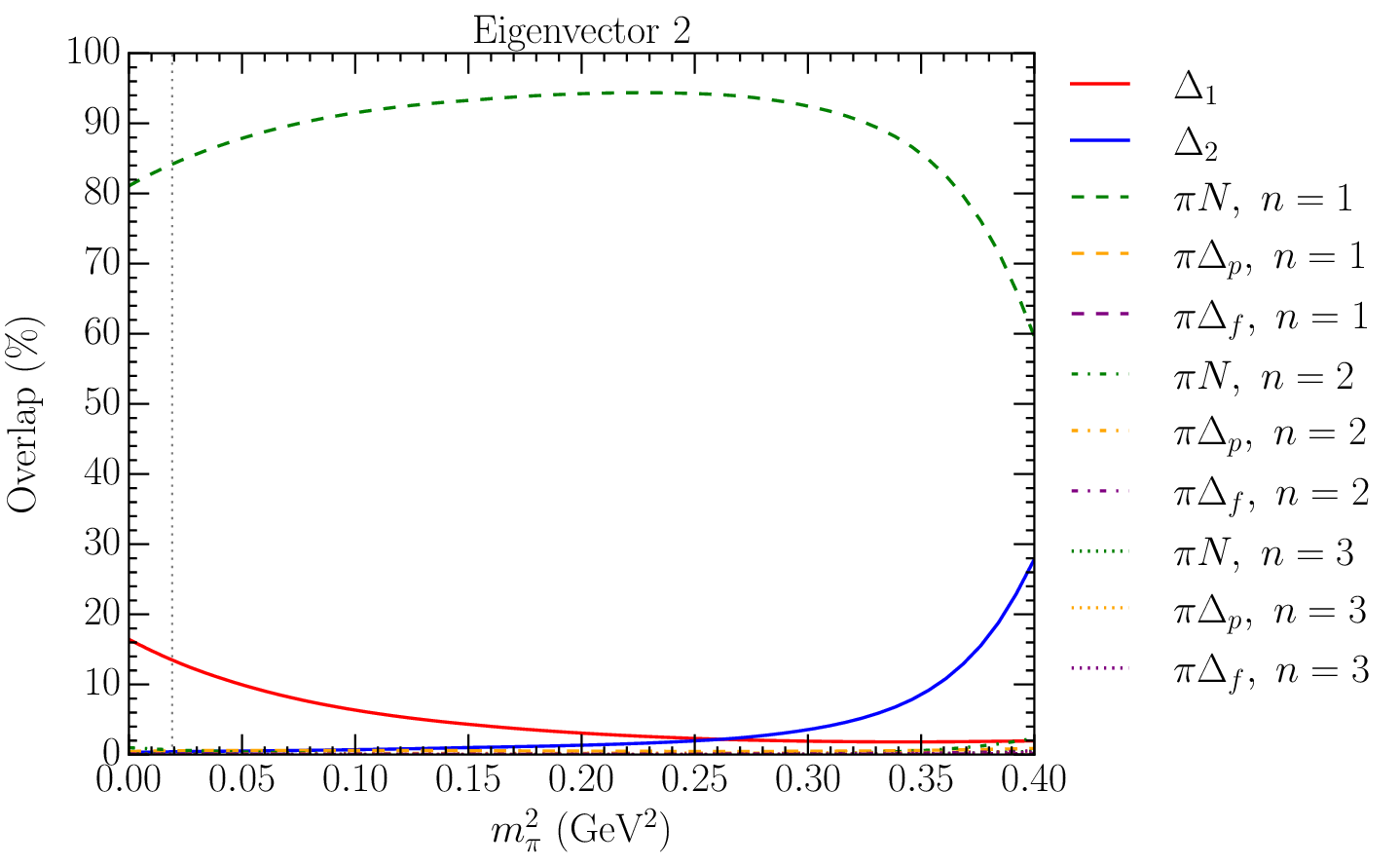}
		\includegraphics[width=0.453\linewidth]{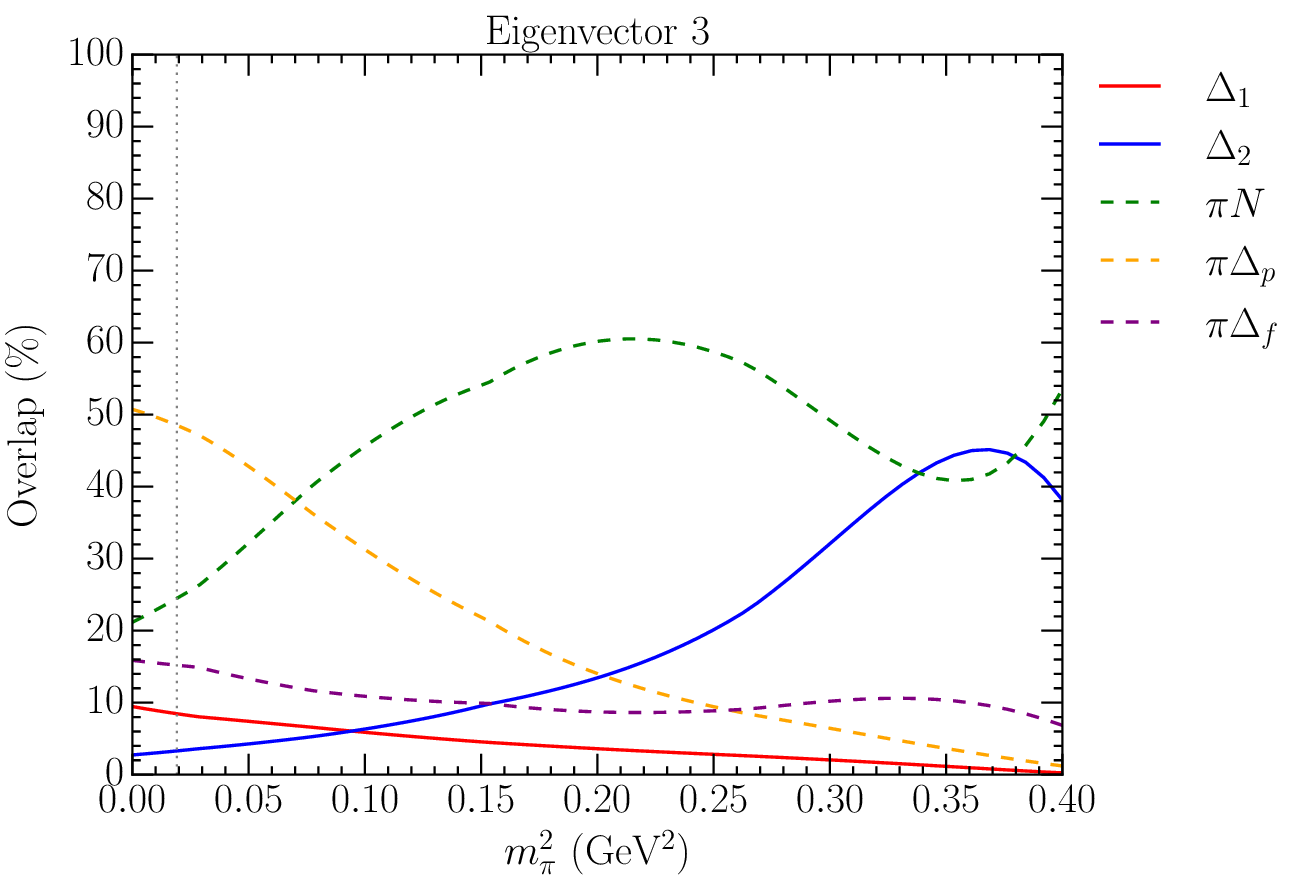}
		\includegraphics[width=0.5\linewidth]{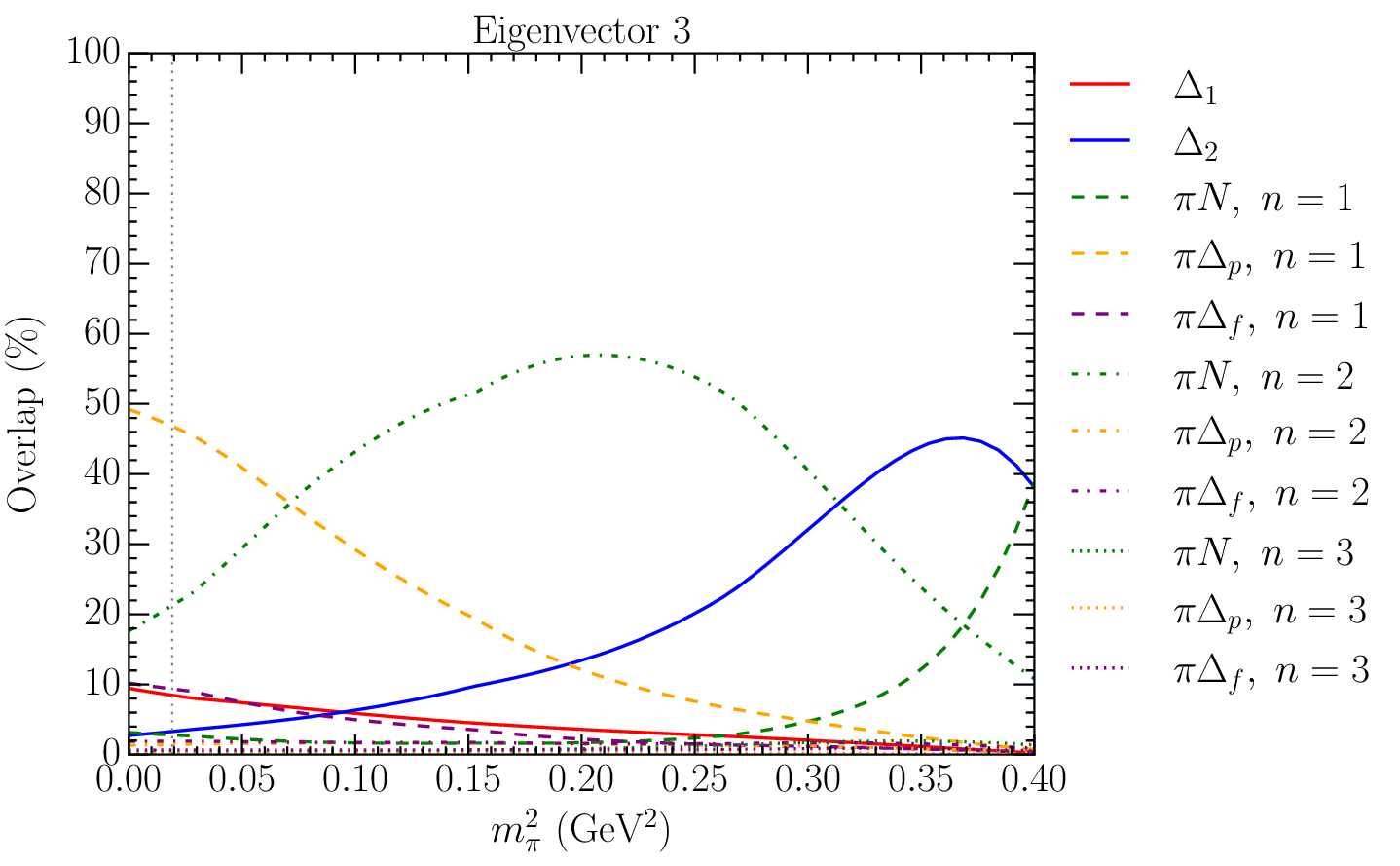}
		\caption{Plots of the first three eigenvectors describing the basis state components of the energy eigenstates. The ``Overlap" with an individual basis state is equal to the modulus-square of the coefficient of that basis state in the superposition describing the eigenstate, normalised to unity. Note that the two-particle components have their back-to-back momenta $ k_n $ summed over in the left column of plots, while the right column shows the composition in terms of separate two-particle basis state contributions with momenta $ k_n $ up to $ n=3 $. The vertical grey dotted line is at the physical quark mass point.}
		\label{fig:eigvec_comps1}
	\end{figure*}

\begin{figure*}
		\includegraphics[width=0.453\linewidth]{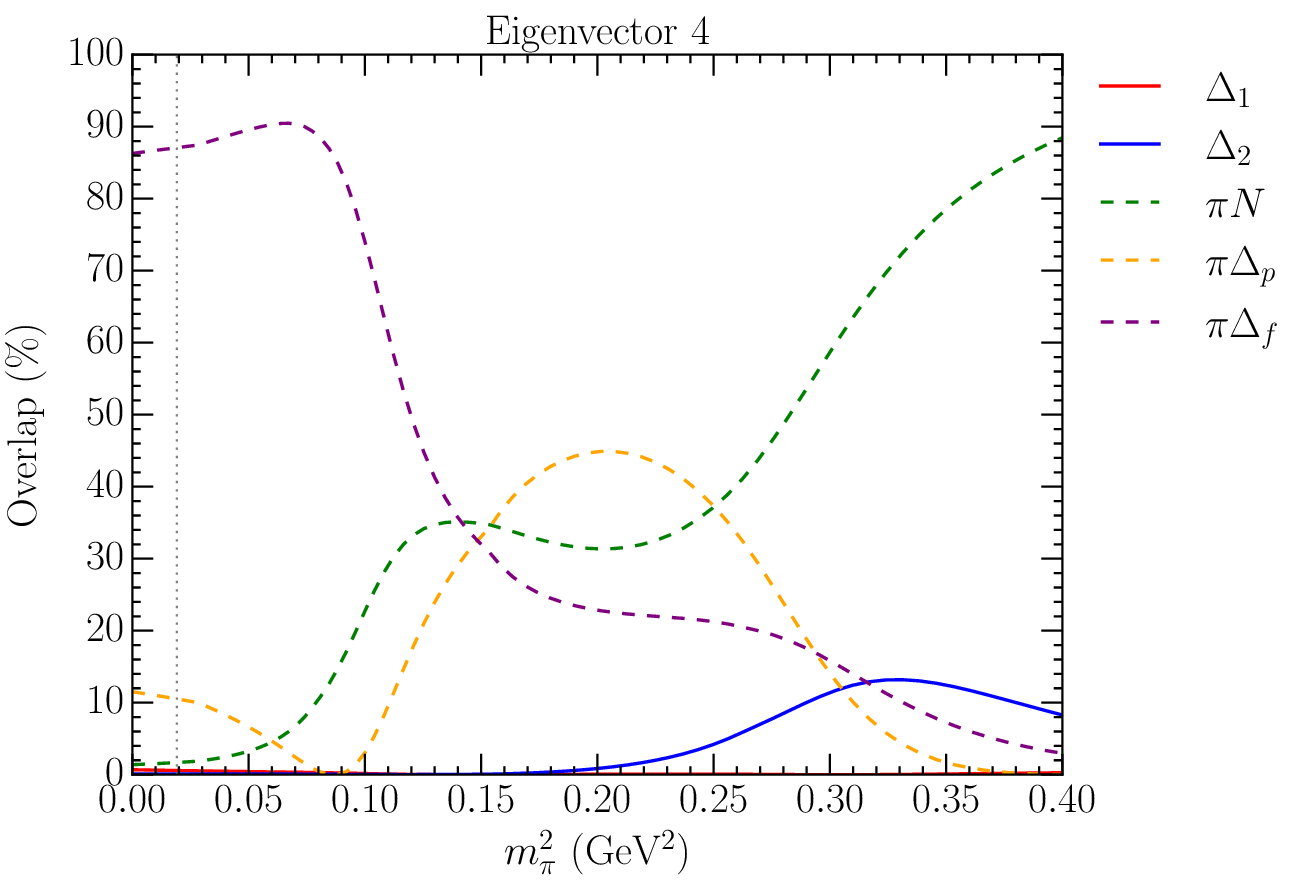}
		\includegraphics[width=0.5\linewidth]{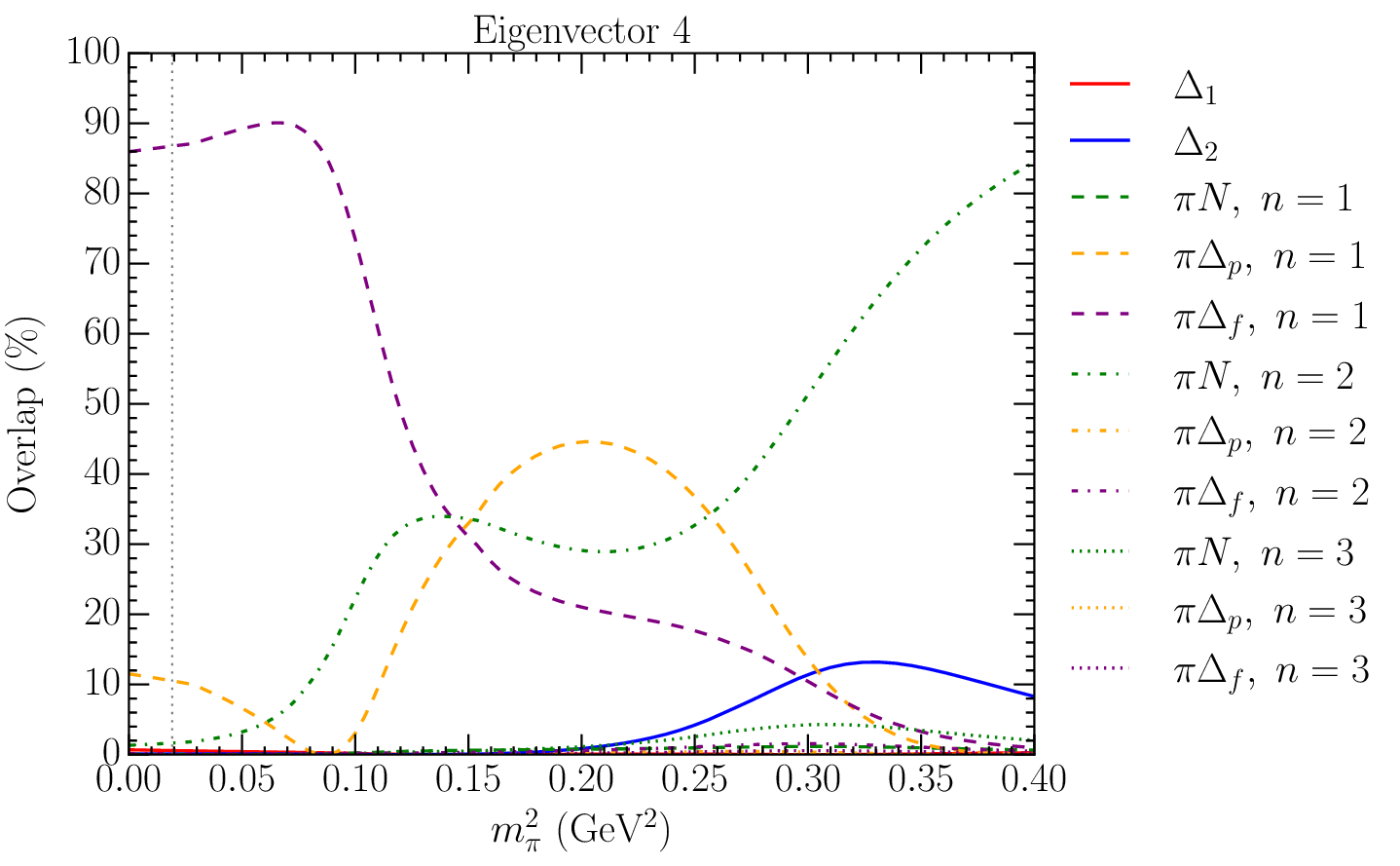}
		\includegraphics[width=0.453\linewidth]{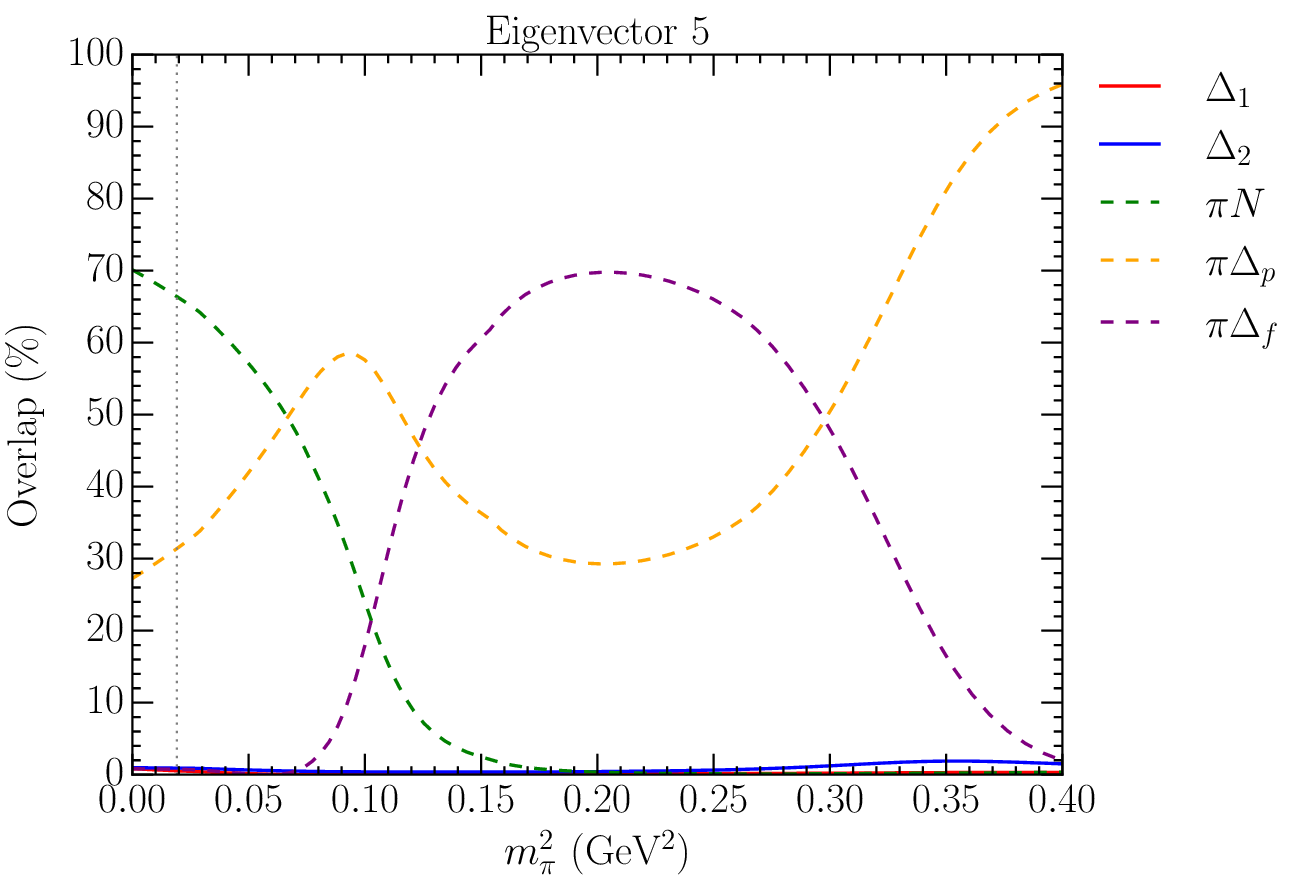}
		\includegraphics[width=0.5\linewidth]{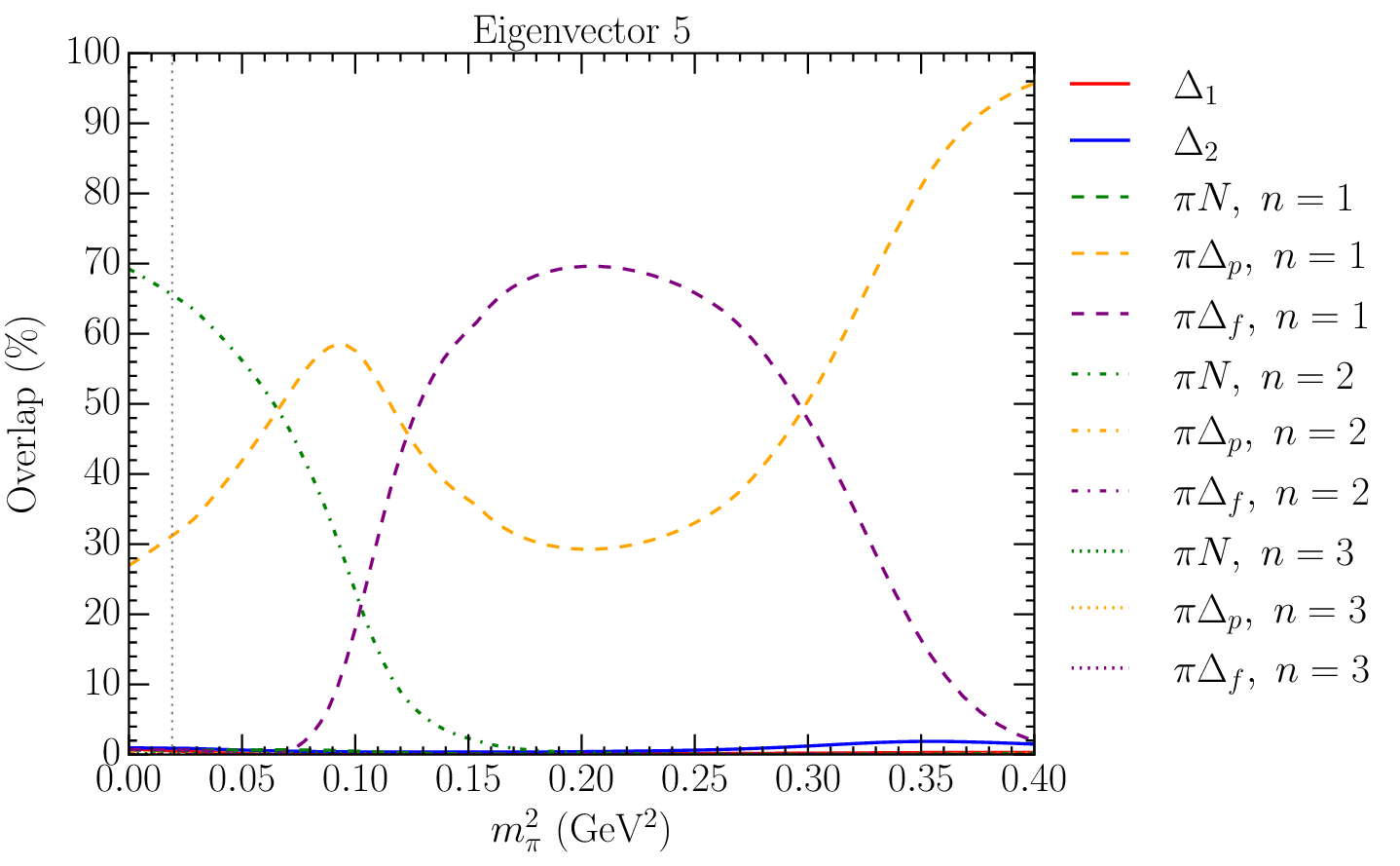}
		\includegraphics[width=0.453\linewidth]{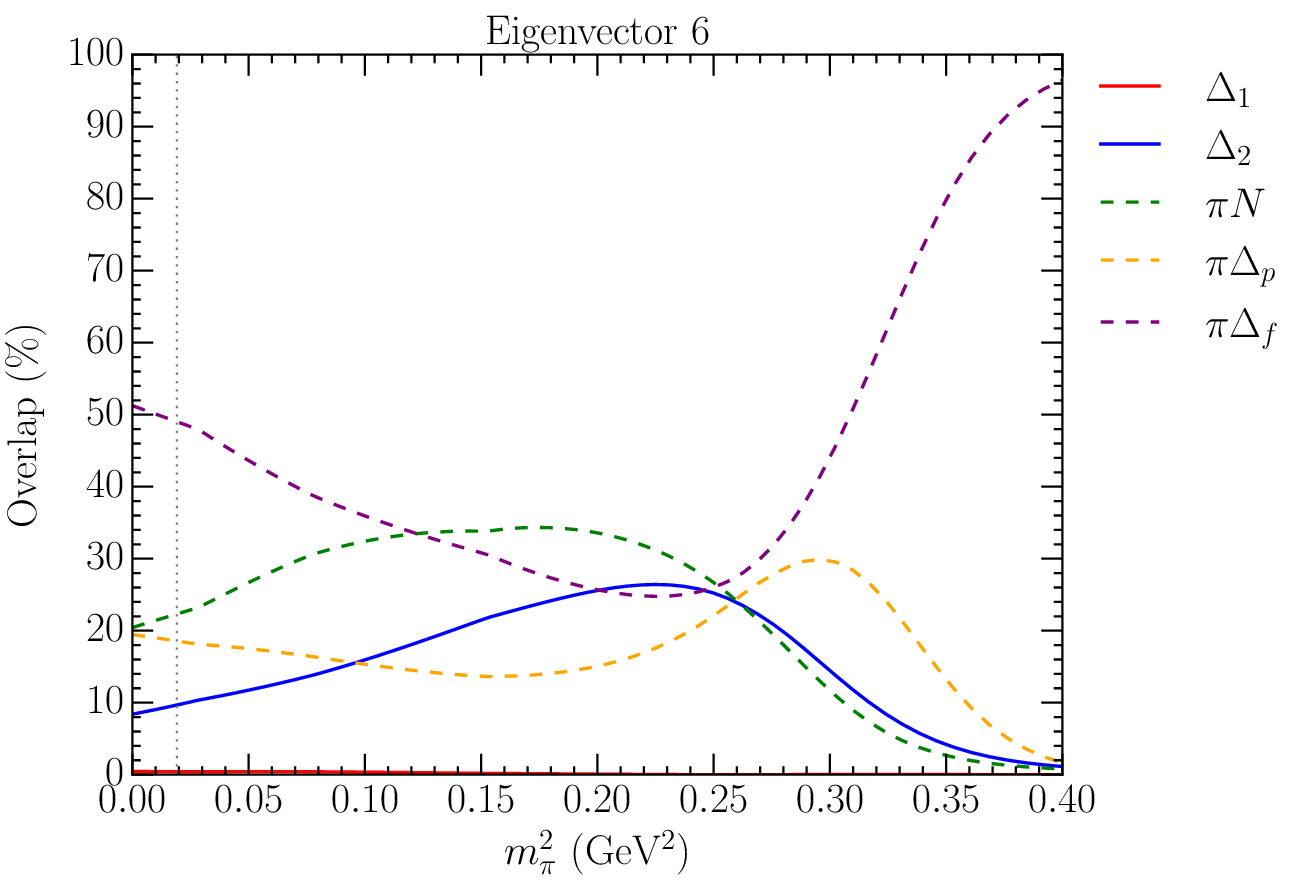}
		\includegraphics[width=0.5\linewidth]{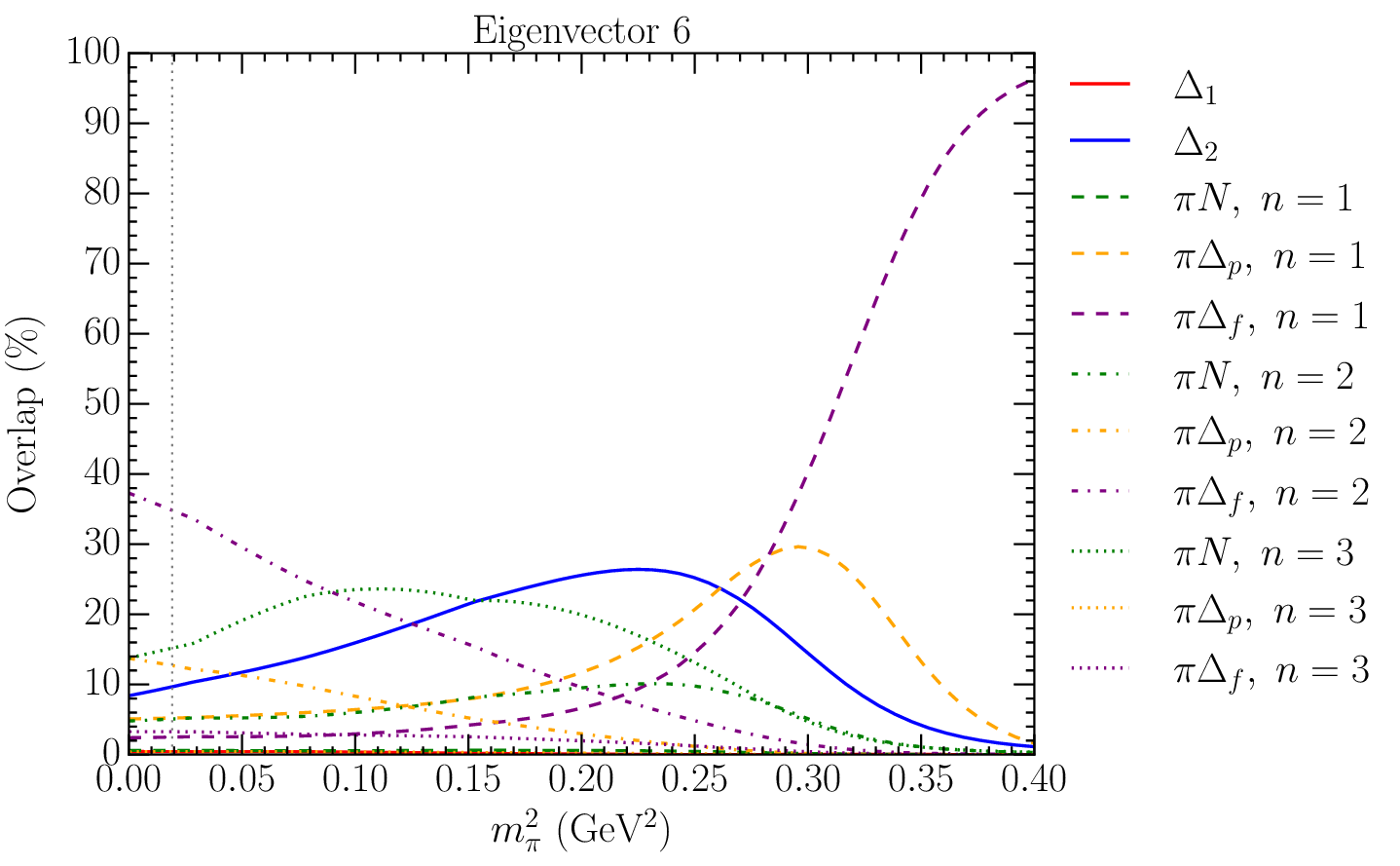}
		\caption{Same as in Fig.~\ref{fig:eigvec_comps1} but for eigenvectors 4, 5 and 6.}
		\label{fig:eigvec_comps2}
	\end{figure*}

\begin{figure*}
		\includegraphics[width=0.453\linewidth]{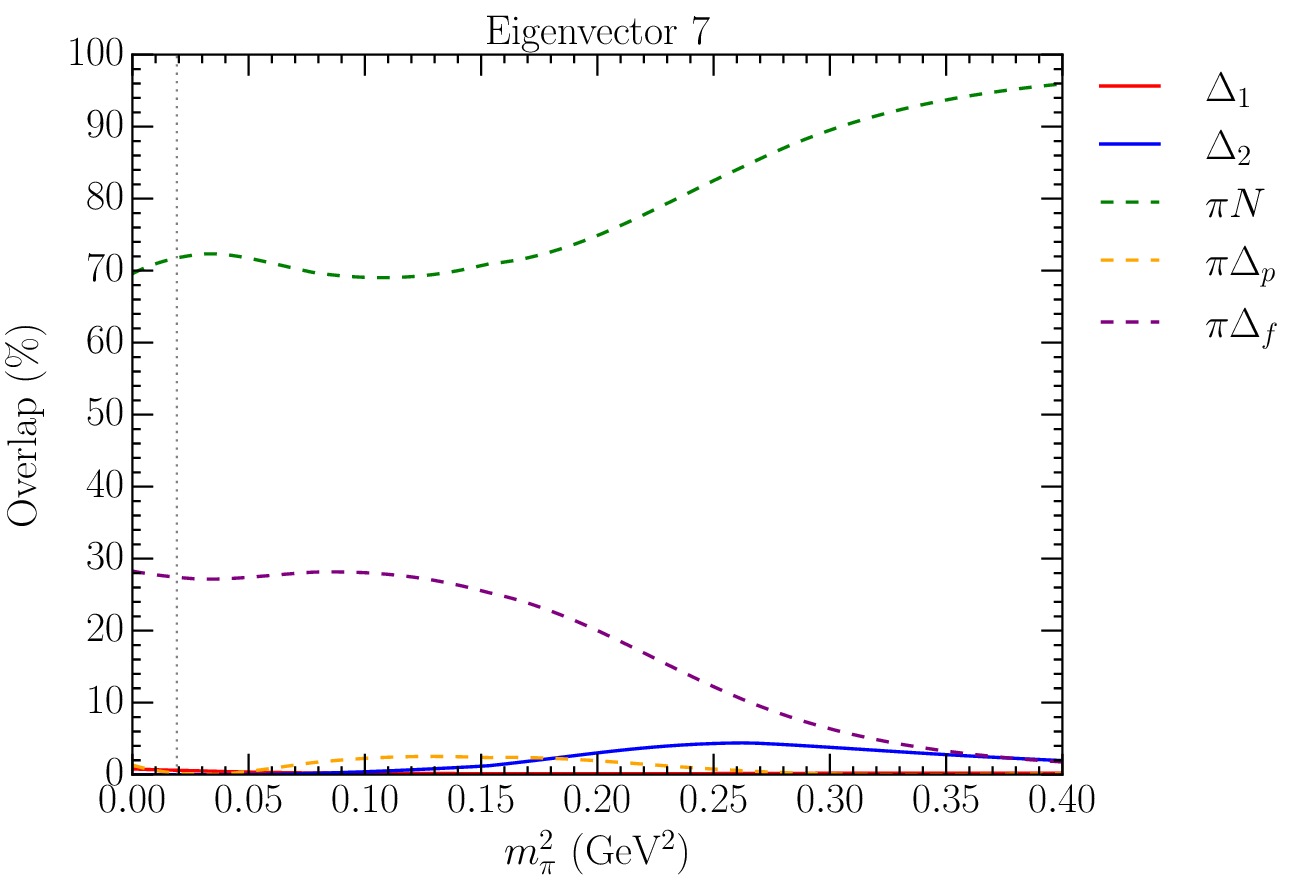}
		\includegraphics[width=0.453\linewidth]{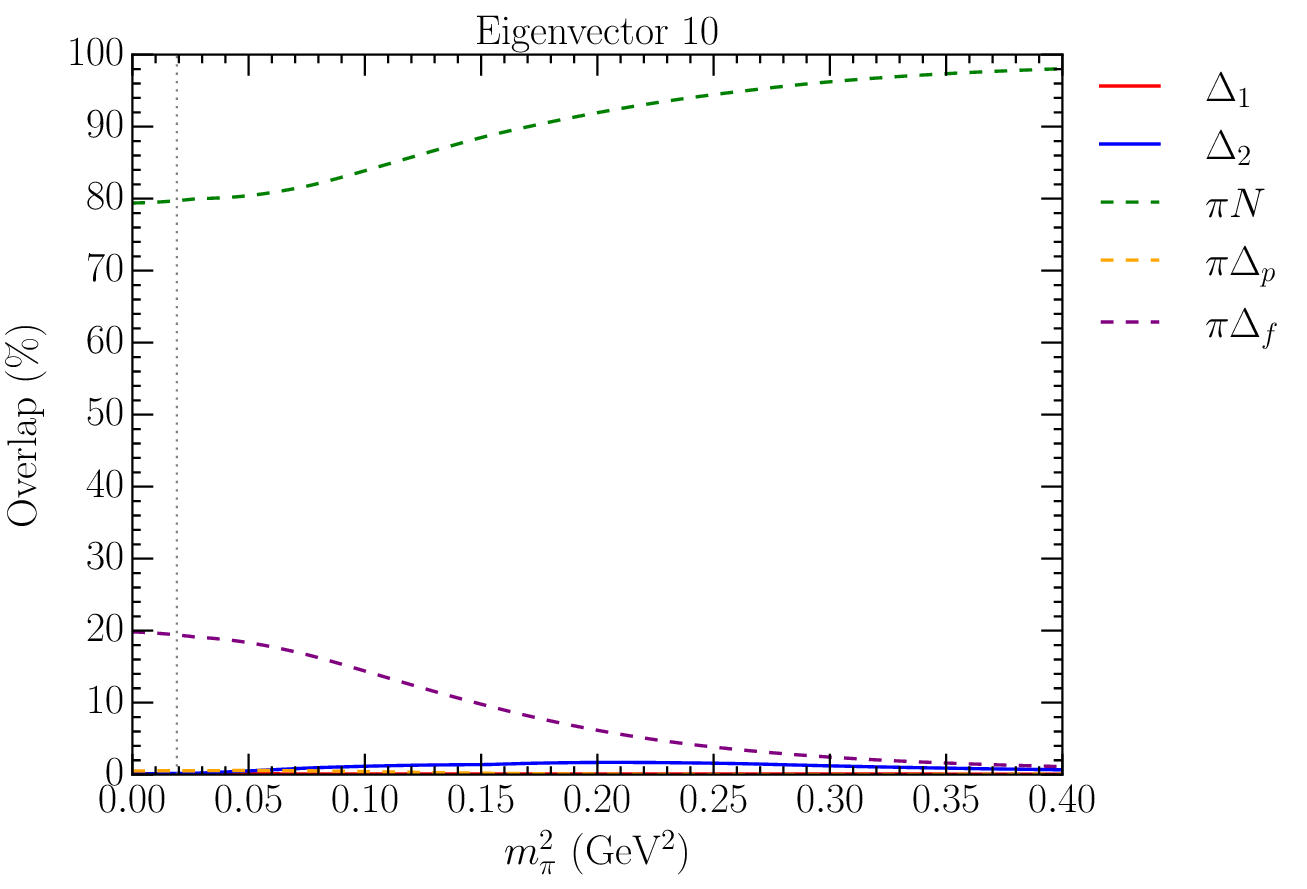}
		\includegraphics[width=0.453\linewidth]{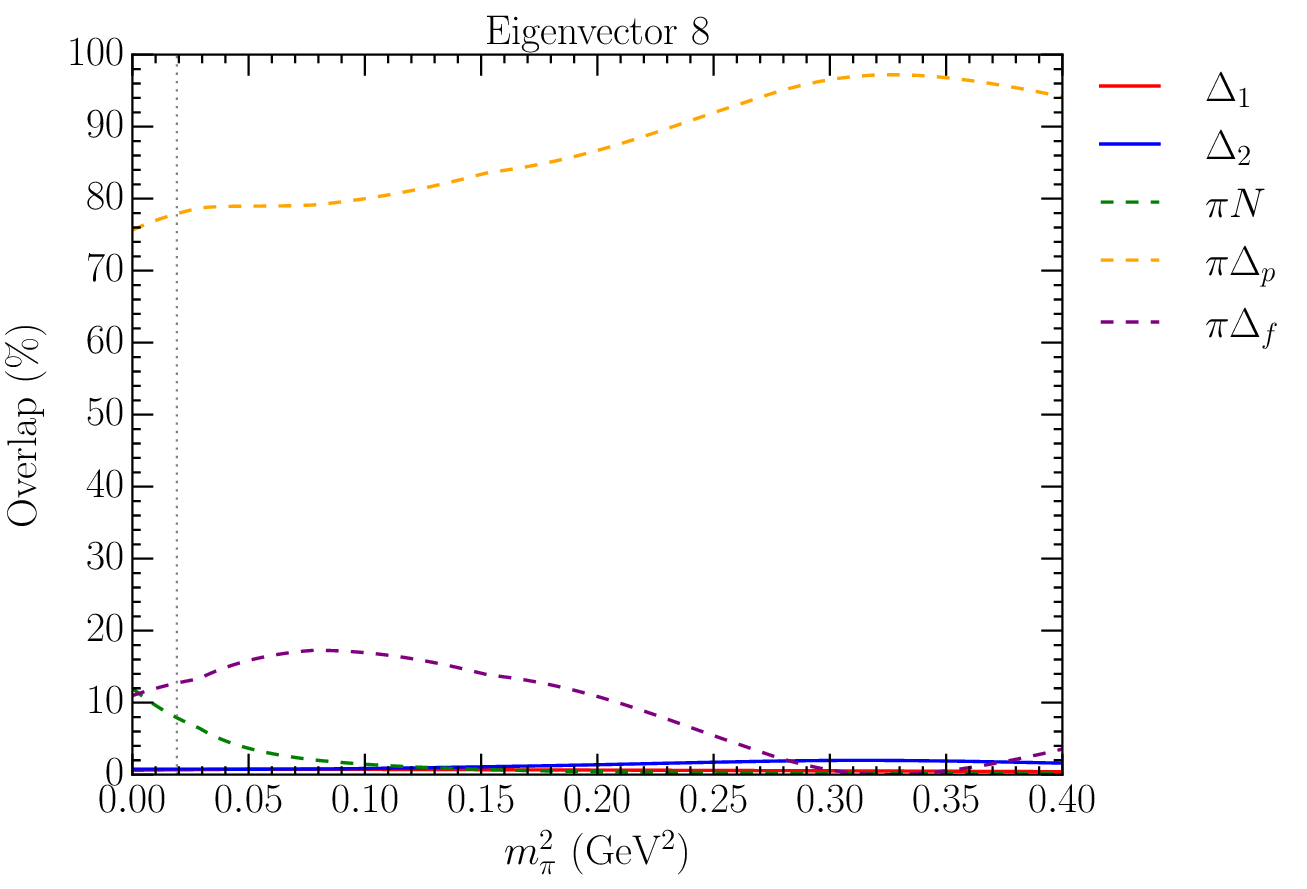}
		\includegraphics[width=0.453\linewidth]{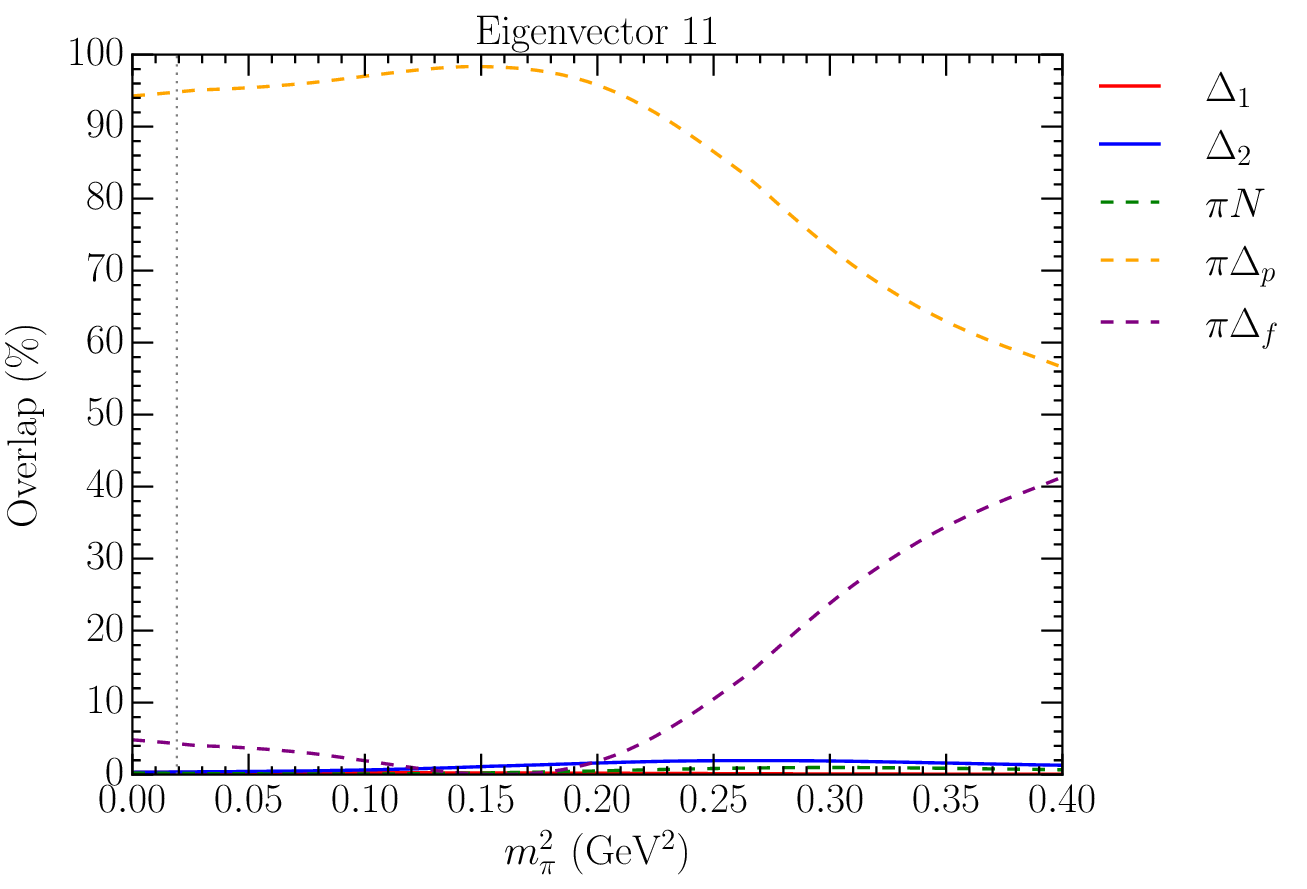}
		\includegraphics[width=0.453\linewidth]{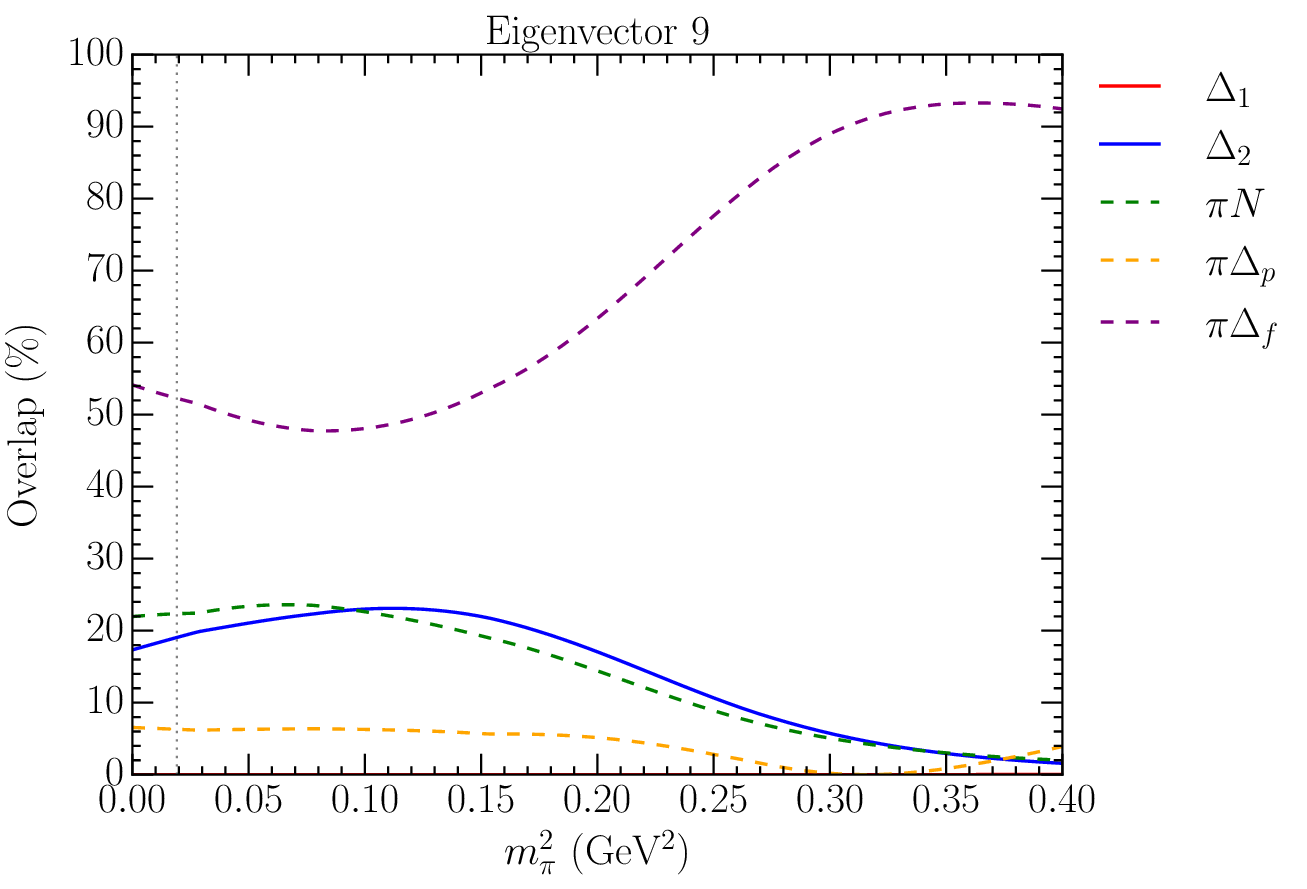}
		\includegraphics[width=0.453\linewidth]{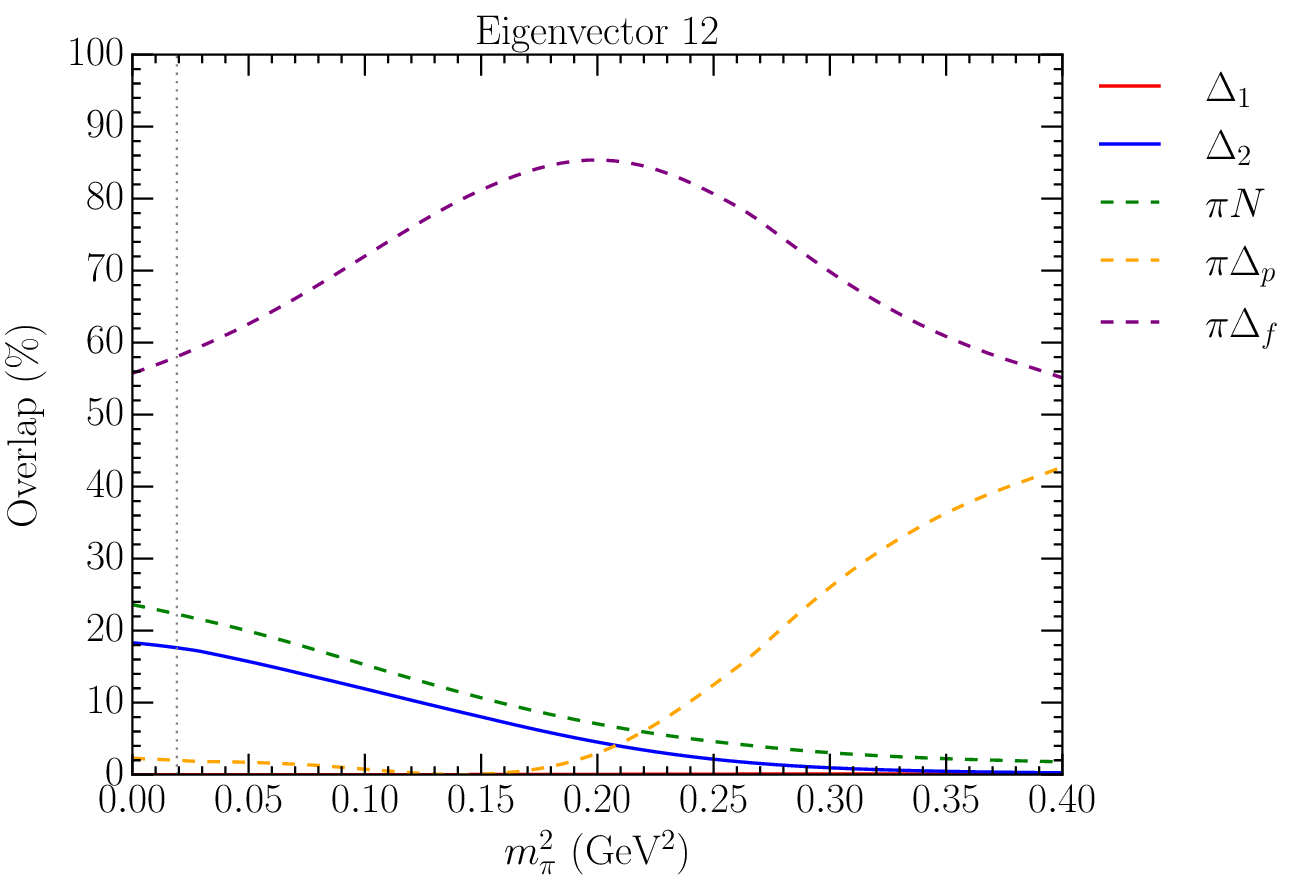}
		\caption{Plots of the eigenvector components describing the basis state composition of the energy eigenstates. Here the momenta, $ k_n $, are summed over for each two-particle channel. Eigenvectors 7--12 are arranged in a column-ordered manner as in Figs.~\ref{fig:eigvec_comps1} and \ref{fig:eigvec_comps2}.}
		\label{fig:eigvec_comps3}
	\end{figure*}

This indicates that future lattice studies looking to extract the $ \Delta(1600) $ resonance properties require momentum projected two-particle interpolators in the $ \pi N $ and $ \pi \Delta $ channels with momenta as indicated in the right-hand columns of Figs.~\ref{fig:eigvec_comps1} and \ref{fig:eigvec_comps2} for eigenstates 2 through 5. Here, the $ \pi \Delta $ channel will capture the most important resonant contributions of the three-body $ \pi\pi N $ channel. 

	\clearpage
	
\section{Comparison with other lattice QCD results} \label{sec:CompLat}
In this final section of the paper, we use the HEFT Hamiltonian, constrained with lattice QCD results from Ref.~\cite{Hockley:2023yzn}, to recalculate the finite-volume eigenvalues and eigenvectors at various lattice sizes in order to make contact with the results of Refs.~\cite{Bulava:2010yg,Khan:2020ahz,Alexandrou:2023elk,Morningstar:2021ewk,Andersen:2017una}. To ensure direct comparisons are valid, we match the masses and slope parameters of the two-particle basis state participants ($ N, \Delta $) to reproduce the corresponding non-interacting energy levels of these studies. As in Refs.~\cite{Abell:2021awi, Abell:2023nex, Abell:2023qgj} we hold the single-particle bare baryon masses and slope parameters fixed as the finite-volume effects are contained in the interactions. In doing so, we find general agreement between our HEFT predictions and the results of the studies considered herein.

\subsection{CLS Consortium Comparison}
\begin{figure*}[]
	\centering
	\includegraphics[width=0.325\linewidth]{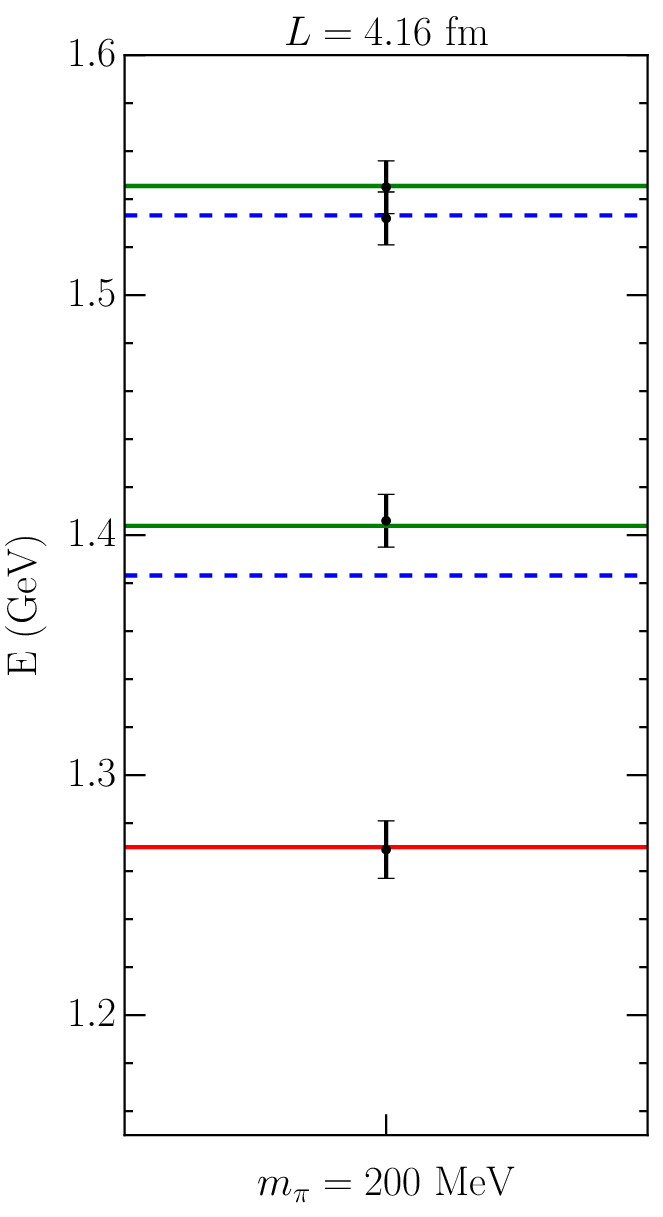}
	\hfil
	\includegraphics[width=0.441\linewidth]{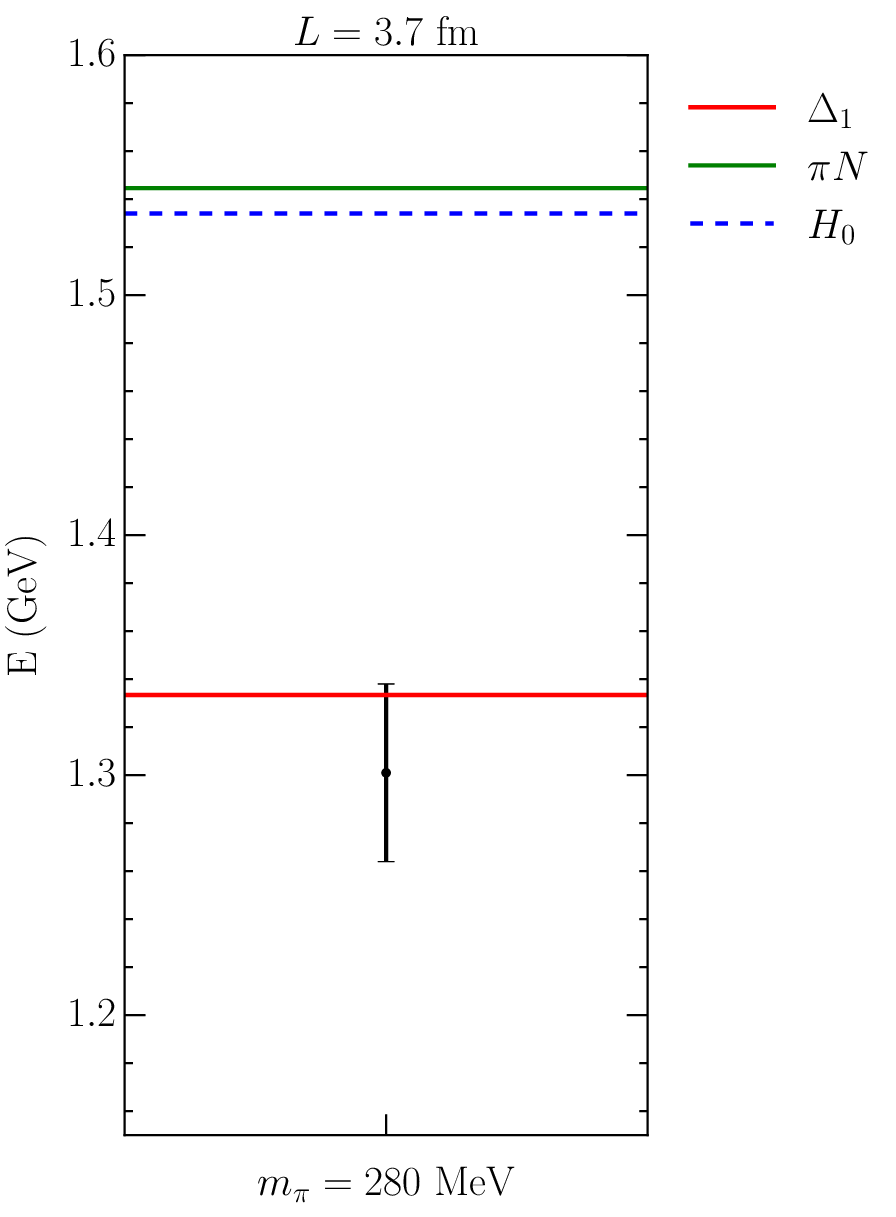}
	\caption{$ \pi N $ energy levels from Ref.~\cite{Morningstar:2021ewk} (Left) and Ref.~\cite{Andersen:2017una} (Right). Lattice results are shown by the data points. Our HEFT predictions are shown as solid lines, with red solid lines showing states dominated by the bare basis state $ \Delta_1 $, and green solid lines corresponding to $ \pi N $ dominated states. Blue dashed lines denote the non-interacting energy levels.}
	\label{fig:cls}
\end{figure*}

\begin{figure*}
	\begin{center}
		\includegraphics[width=0.9\linewidth]{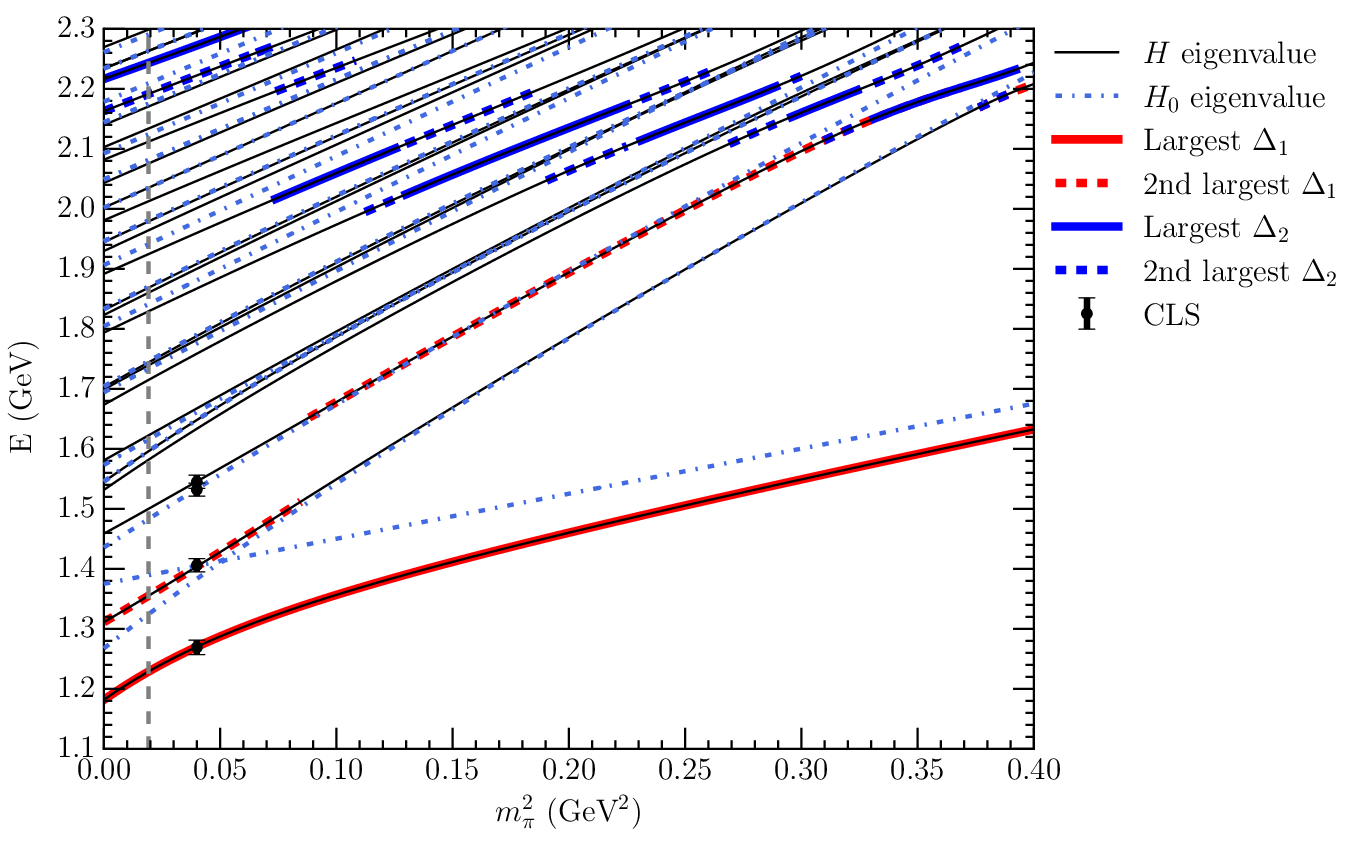}
		\caption{Finite-volume spectrum for $ L = 4.16 $ fm and with $ \pi N $ non-interacting energies matched to the non-interacting energy levels of the CLS Consortium \cite{Morningstar:2021ewk}.}
		\label{fig:finVol_cls}
	\end{center}
\end{figure*}

The studies in Refs.~\cite{Morningstar:2021ewk,Andersen:2017una} yield values for $ \pi N $ energy levels using two-particle momentum-projected operators on two gauge-field ensembles from the CLS Consortium. The first ensemble has $ m_\pi = 200 $ MeV and $ L = 4.16 $ fm and taking $ m_N $ and $ \alpha_N $ so as to match the non-interacting energy levels of Ref.~\cite{Morningstar:2021ewk}, we generate the corresponding HEFT spectrum. This is shown in the left-hand plot of Fig.~\ref{fig:cls} with the three lowest-lying energy levels of Ref.~\cite{Morningstar:2021ewk} overlaid. Likewise, the right-hand plot in Fig.~\ref{fig:cls} shows the comparison with the ensemble used in Ref.~\cite{Andersen:2017una}, with $ m_\pi = 280 $ MeV, $ L = 3.7 $ fm. By extending the baryon masses beyond the results of Ref.~\cite{Morningstar:2021ewk}, we generate the HEFT finite-volume spectrum across a range of $ m_\pi^2 $ values and this is given in Fig.~\ref{fig:finVol_cls}.

\begin{figure*}[]
	\centering
	\includegraphics[width=\linewidth]{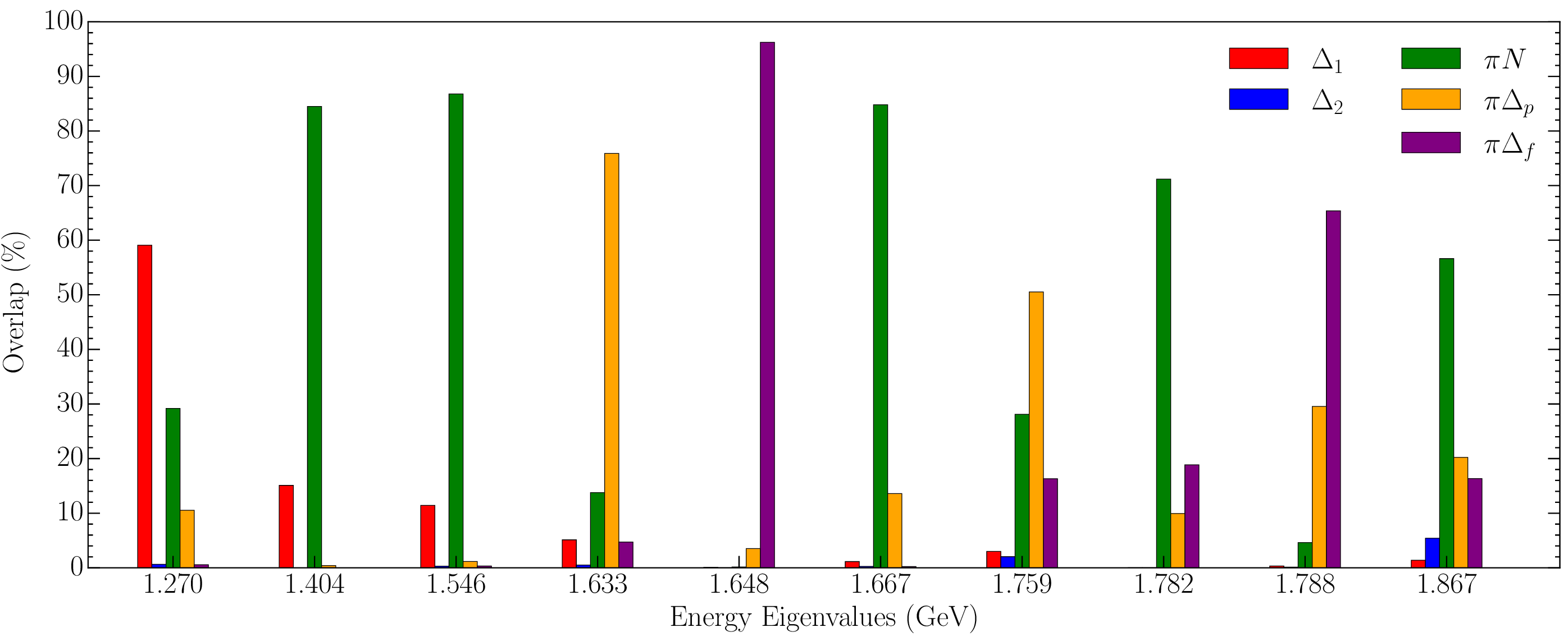}
	\caption{Eigenvectors calculated at the CLS pion mass of $ m_\pi = 200 $ MeV and decomposed into their basis state content. All two-particle channels have had their momenta summed over.}
	\label{fig:cls_components}
\end{figure*}

\begin{figure*}
	\begin{center}
		\includegraphics[width=0.453\linewidth]{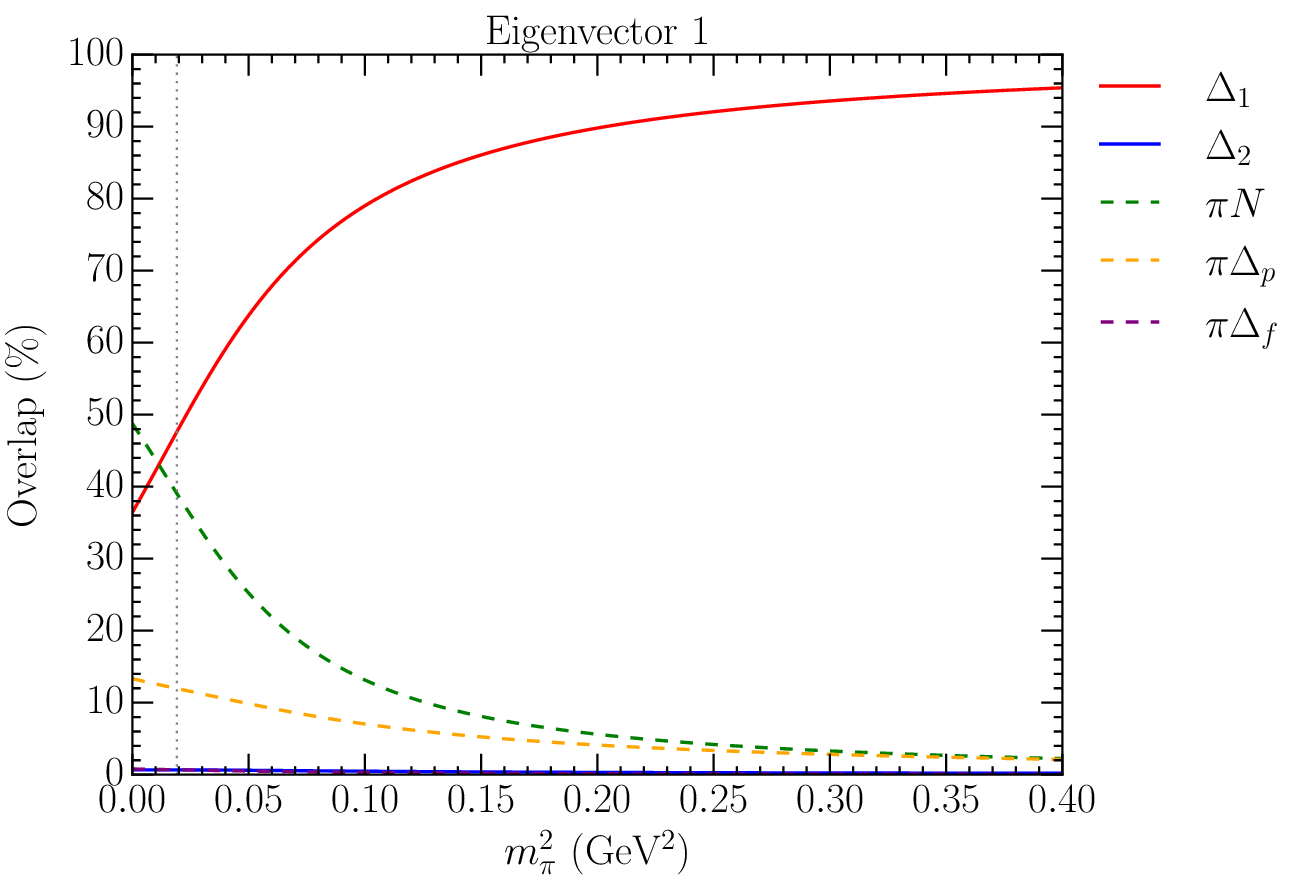} 
		\includegraphics[width=0.5\linewidth]{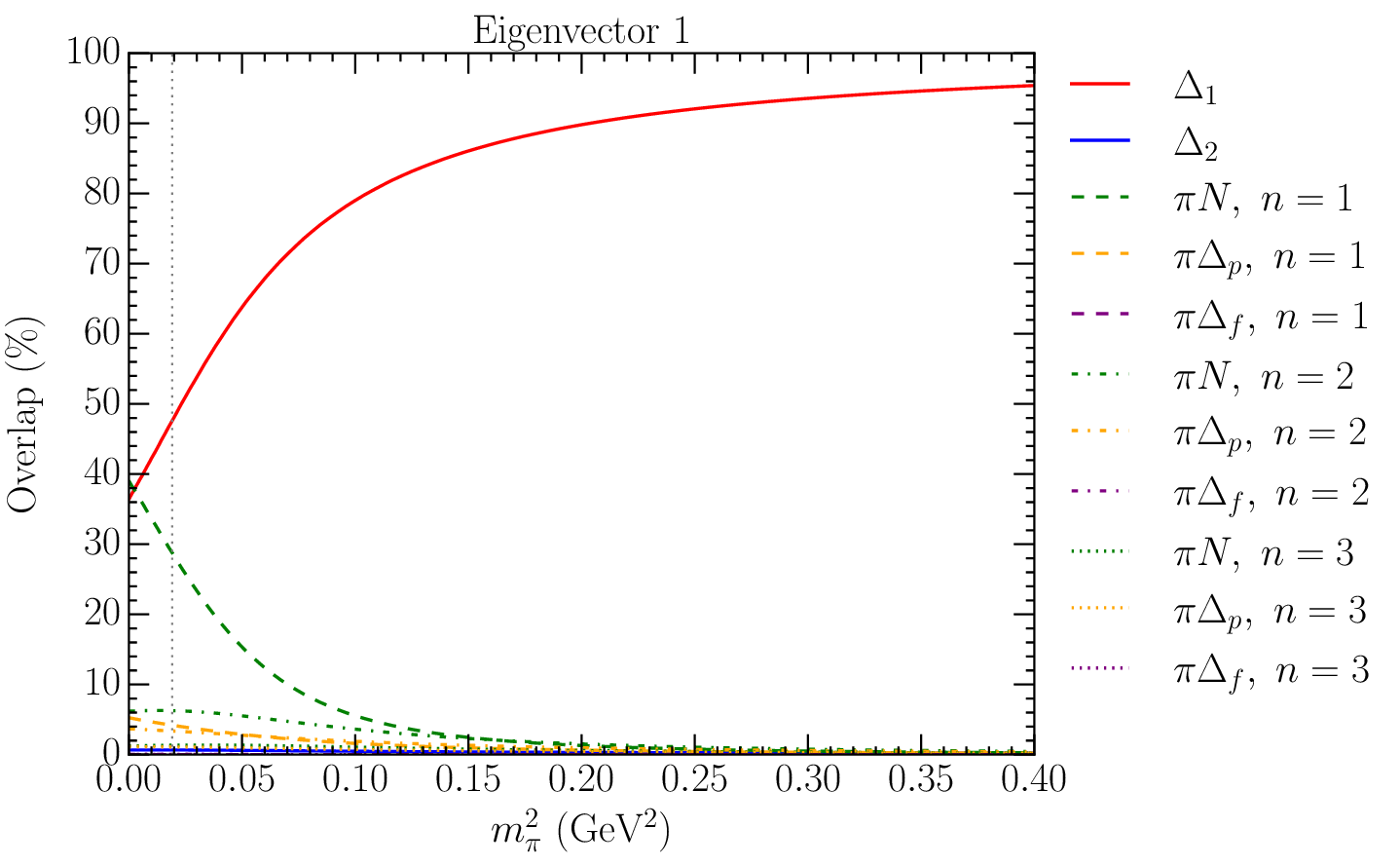}
		\includegraphics[width=0.453\linewidth]{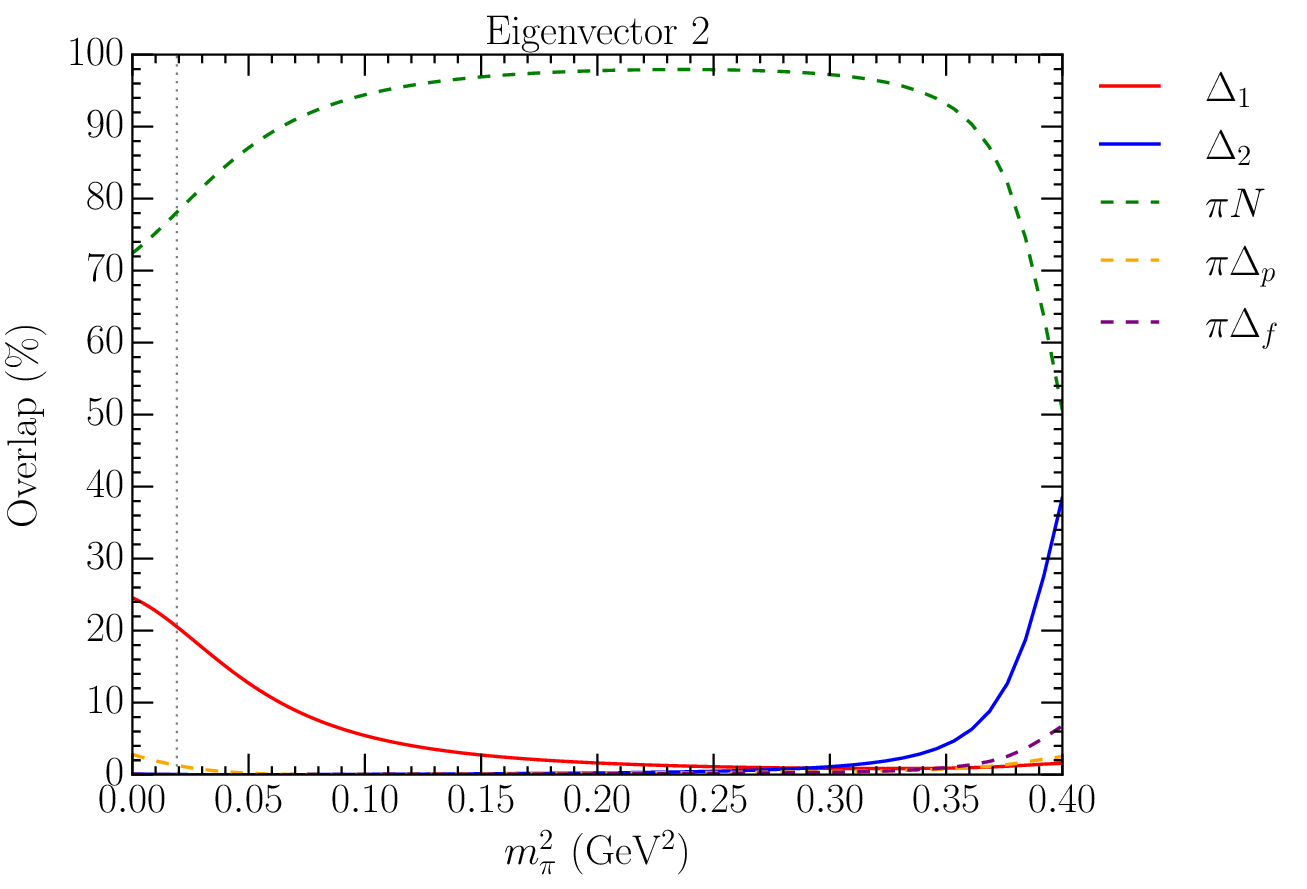}
		\includegraphics[width=0.5\linewidth]{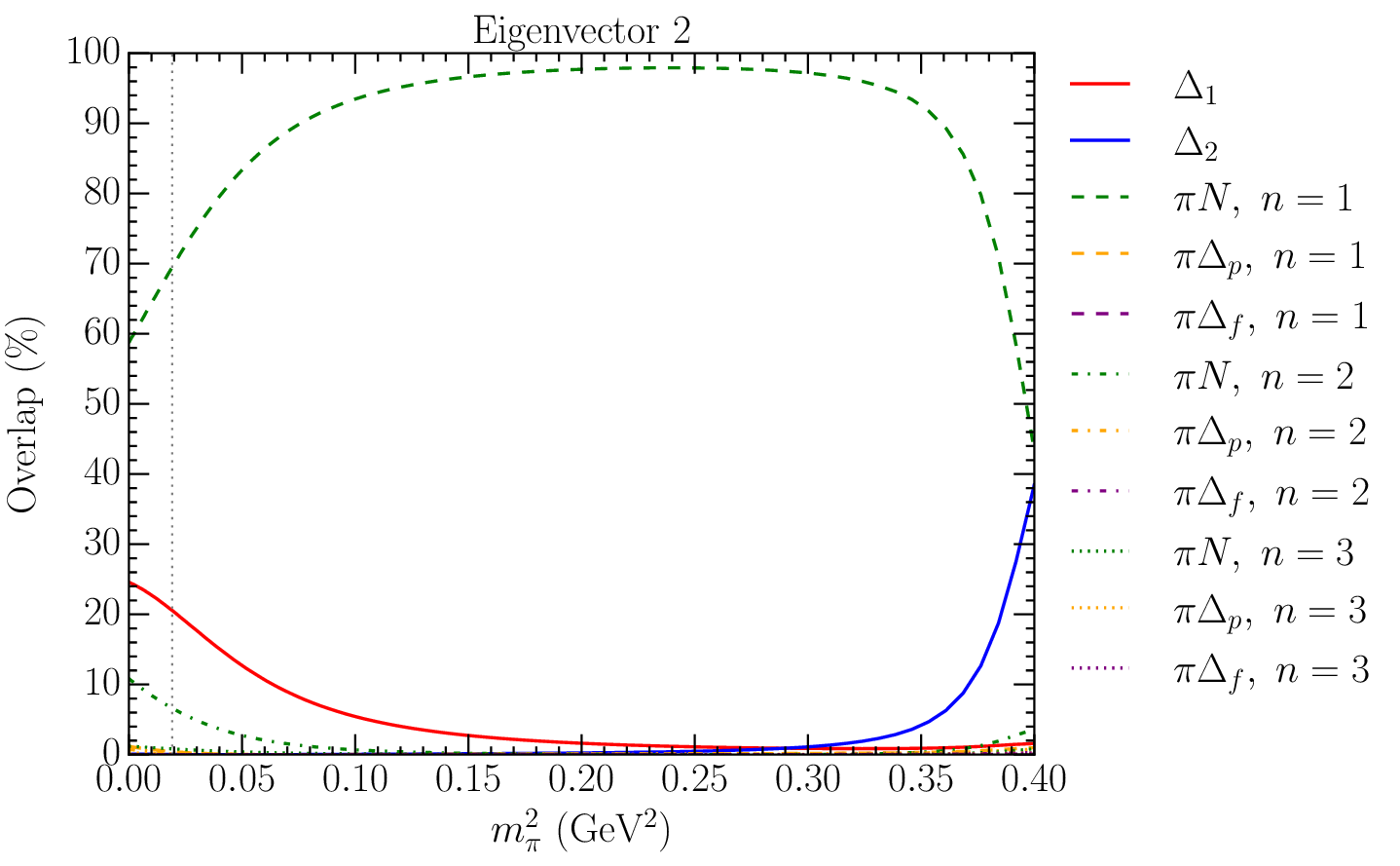}		
		\includegraphics[width=0.453\linewidth]{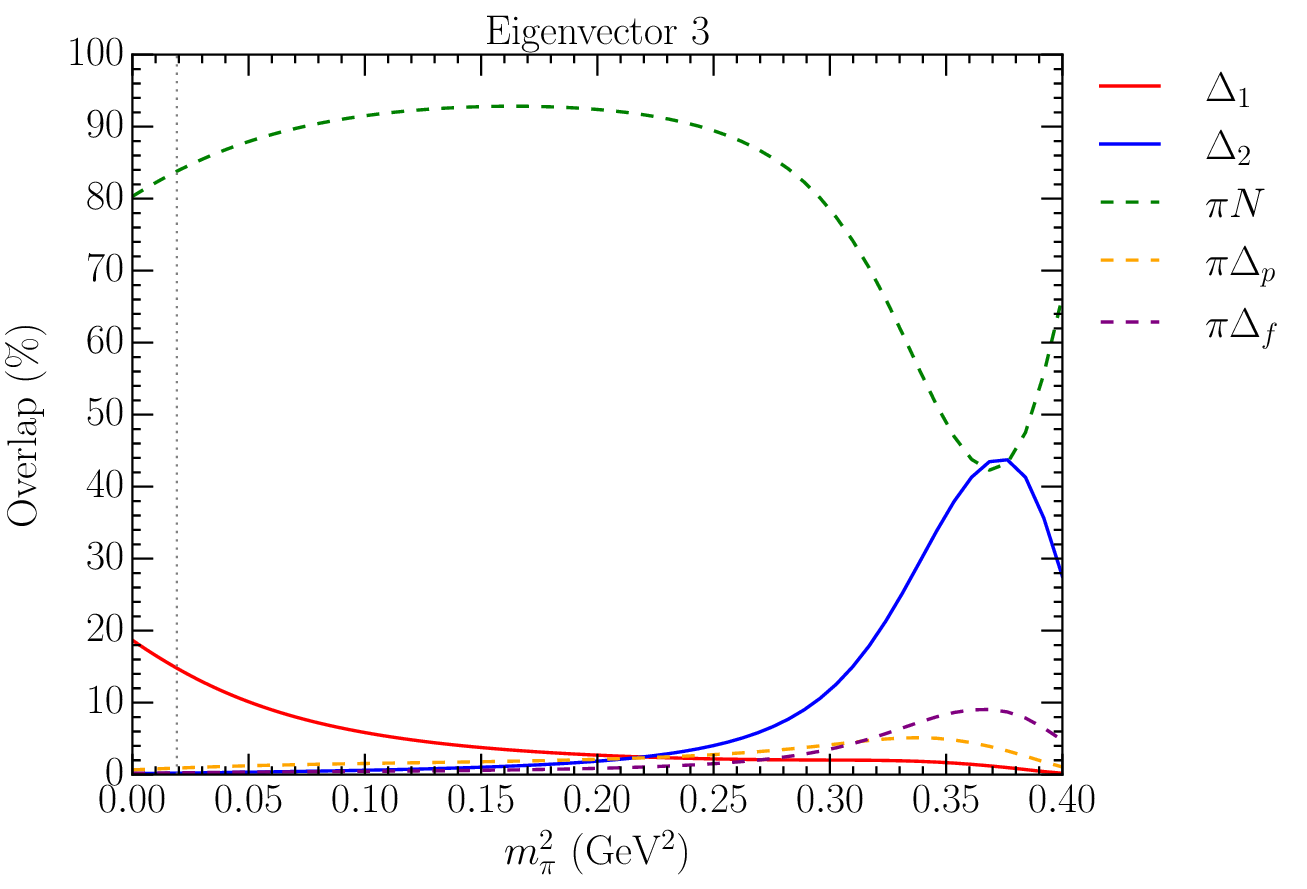}
		\includegraphics[width=0.5\linewidth]{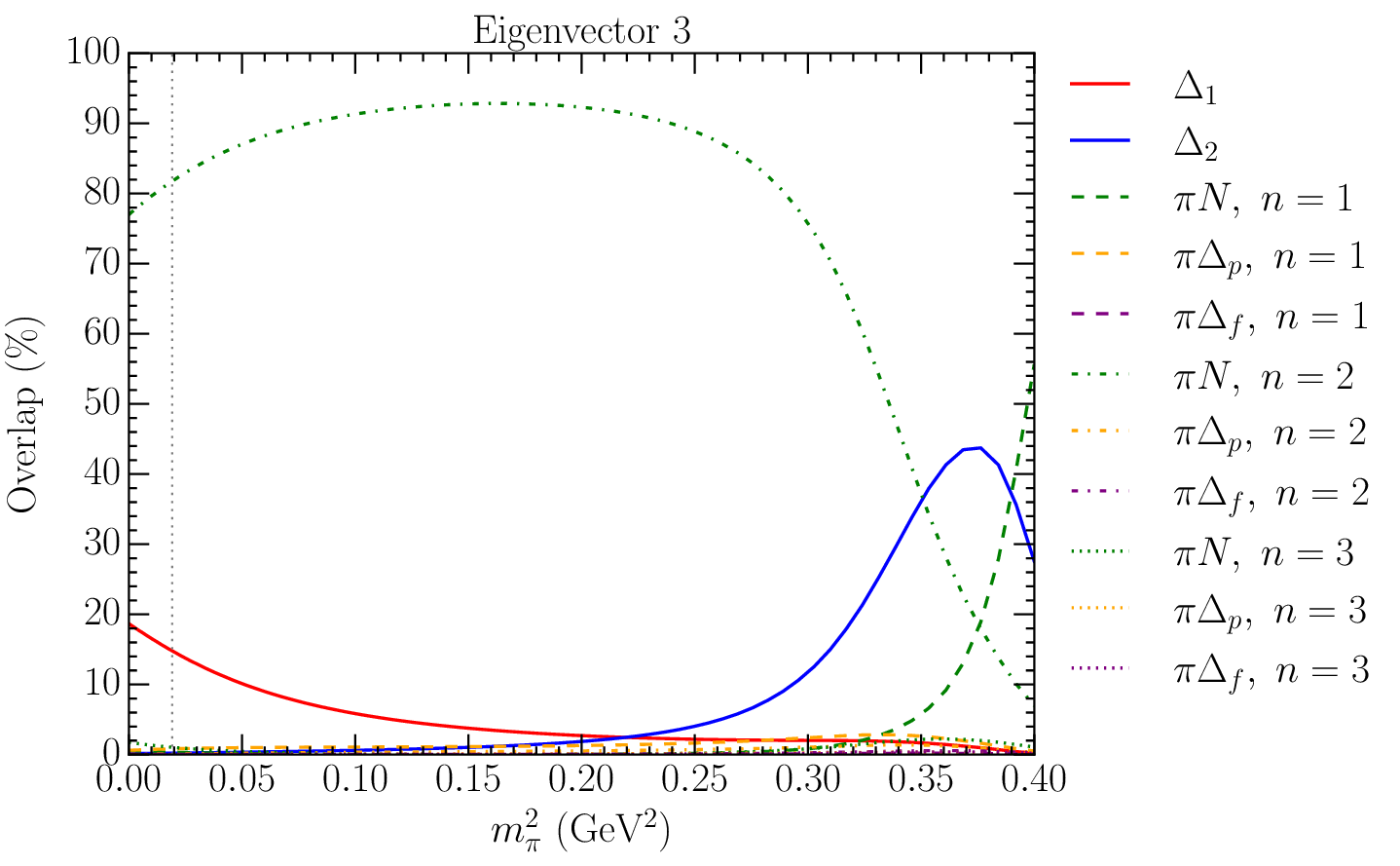}
		\caption{Eigenvector components for the first three fully interacting HEFT eigenstates in the $ L = 4.16 $ fm calculation. The left-hand column shows the basis state content with momenta summed over in each 2-particle channel, while the righ-hand column gives the full basis state decomposition without summation.}
		\label{fig:eigvec_comps_cls}
	\end{center}
\end{figure*}

\begin{figure*}
	\begin{center}
		\includegraphics[width=0.453\linewidth]{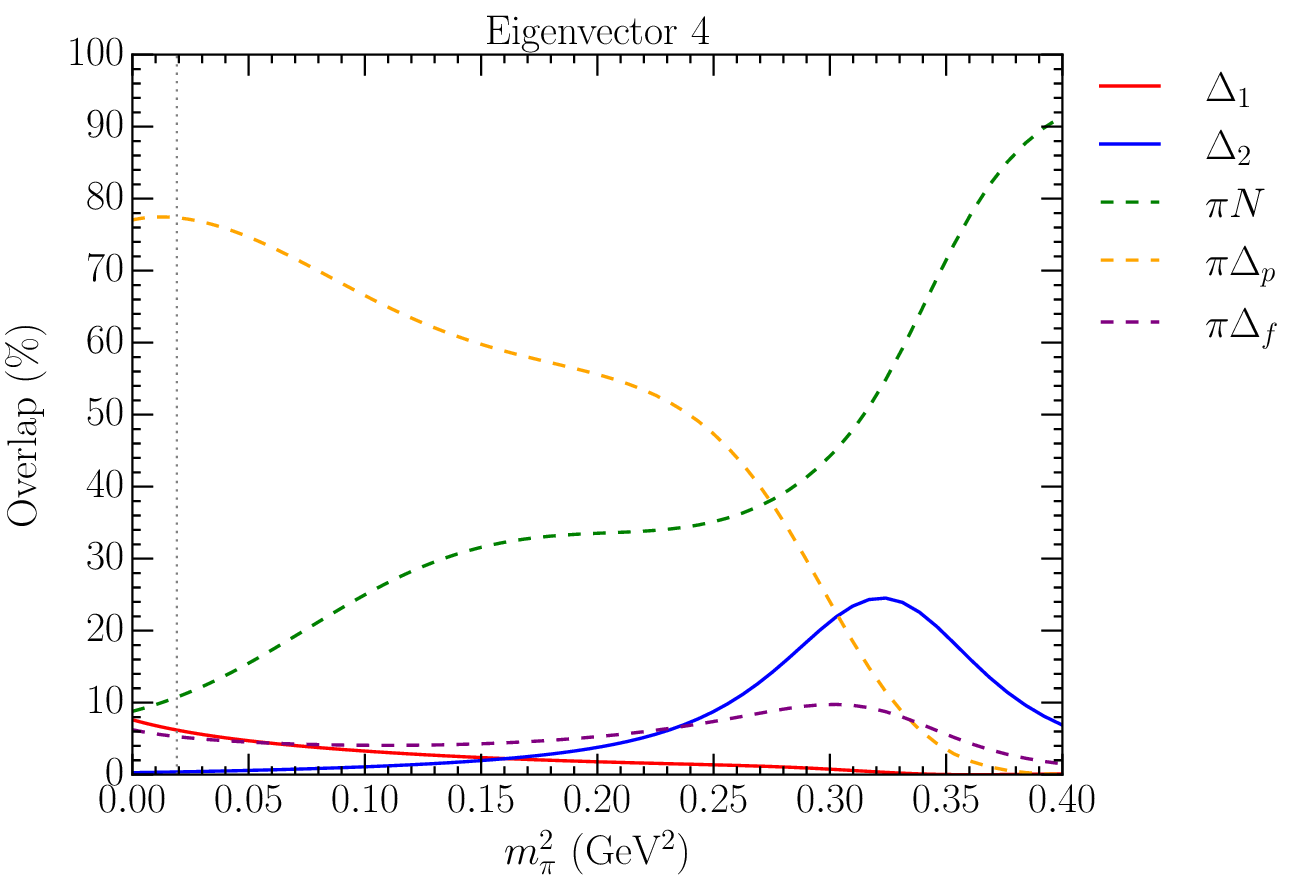} 
		\includegraphics[width=0.5\linewidth]{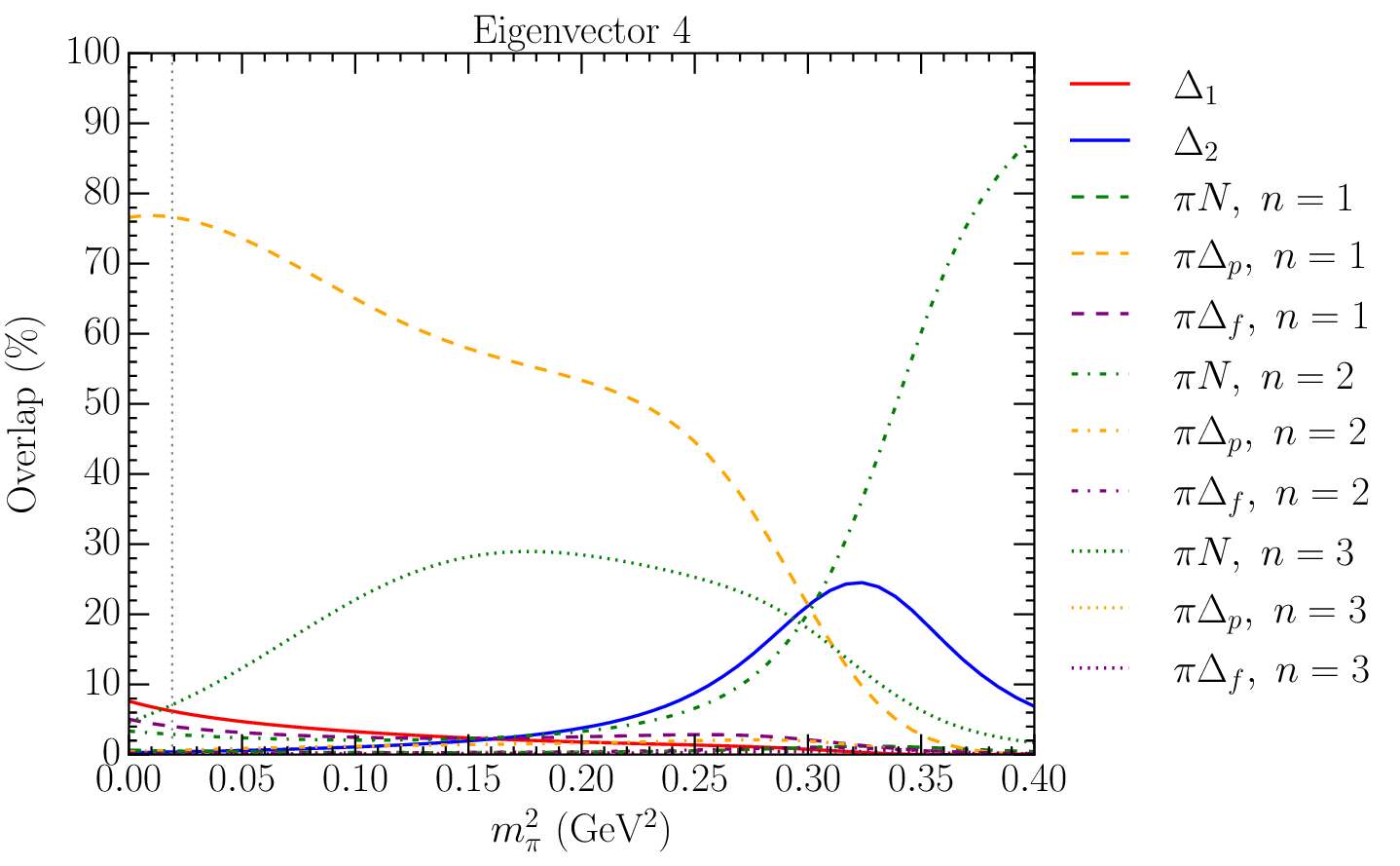}
		\includegraphics[width=0.453\linewidth]{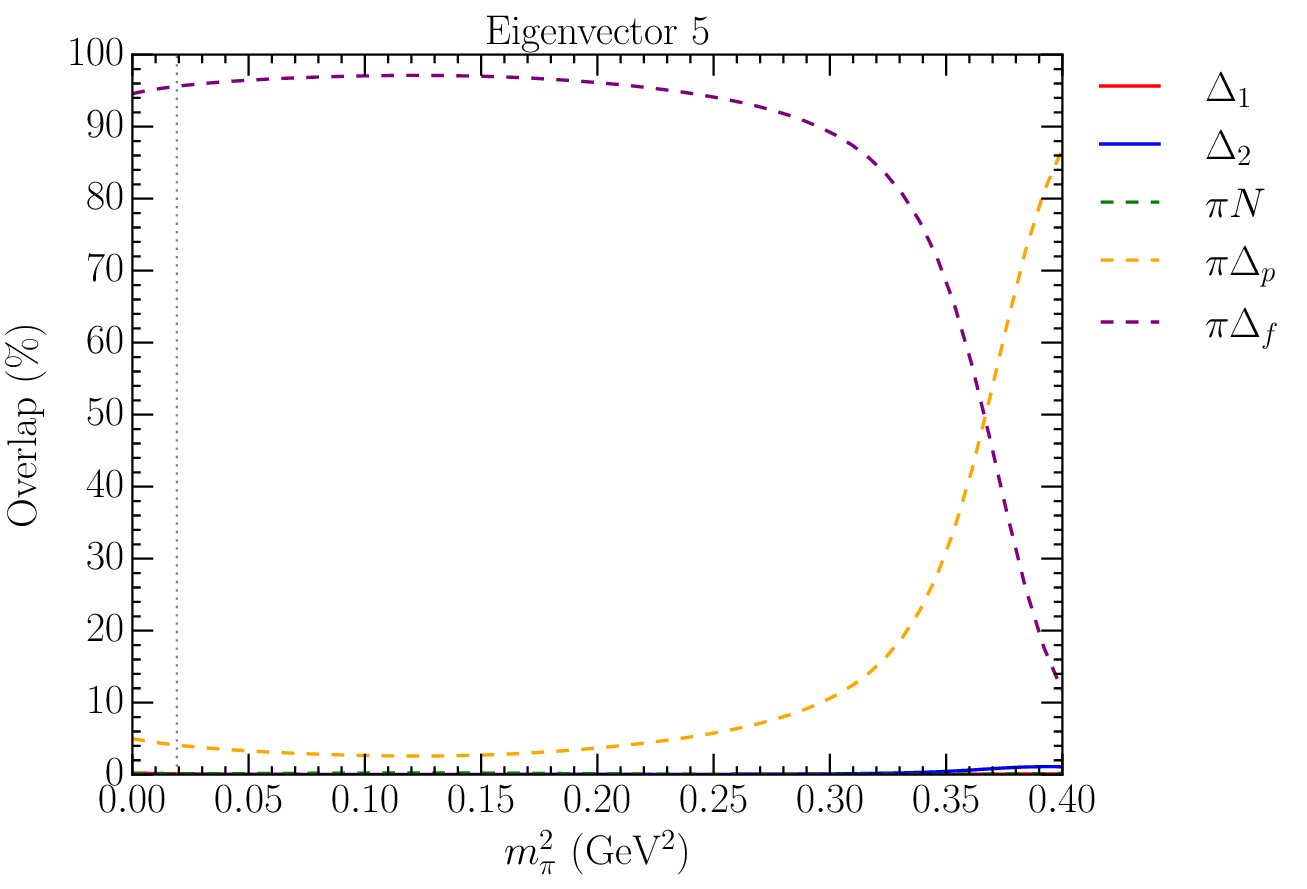}
		\includegraphics[width=0.5\linewidth]{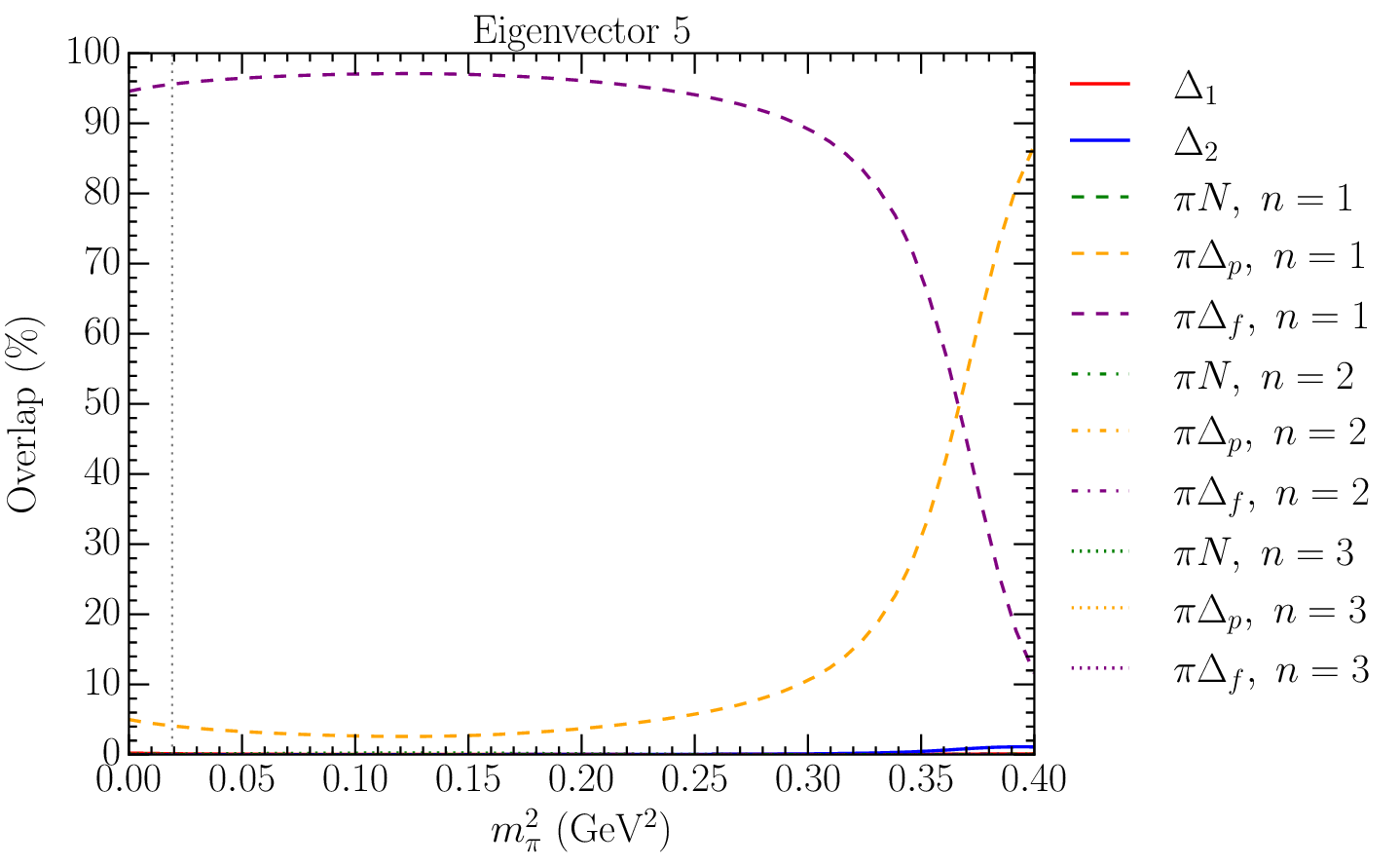}		
		\includegraphics[width=0.453\linewidth]{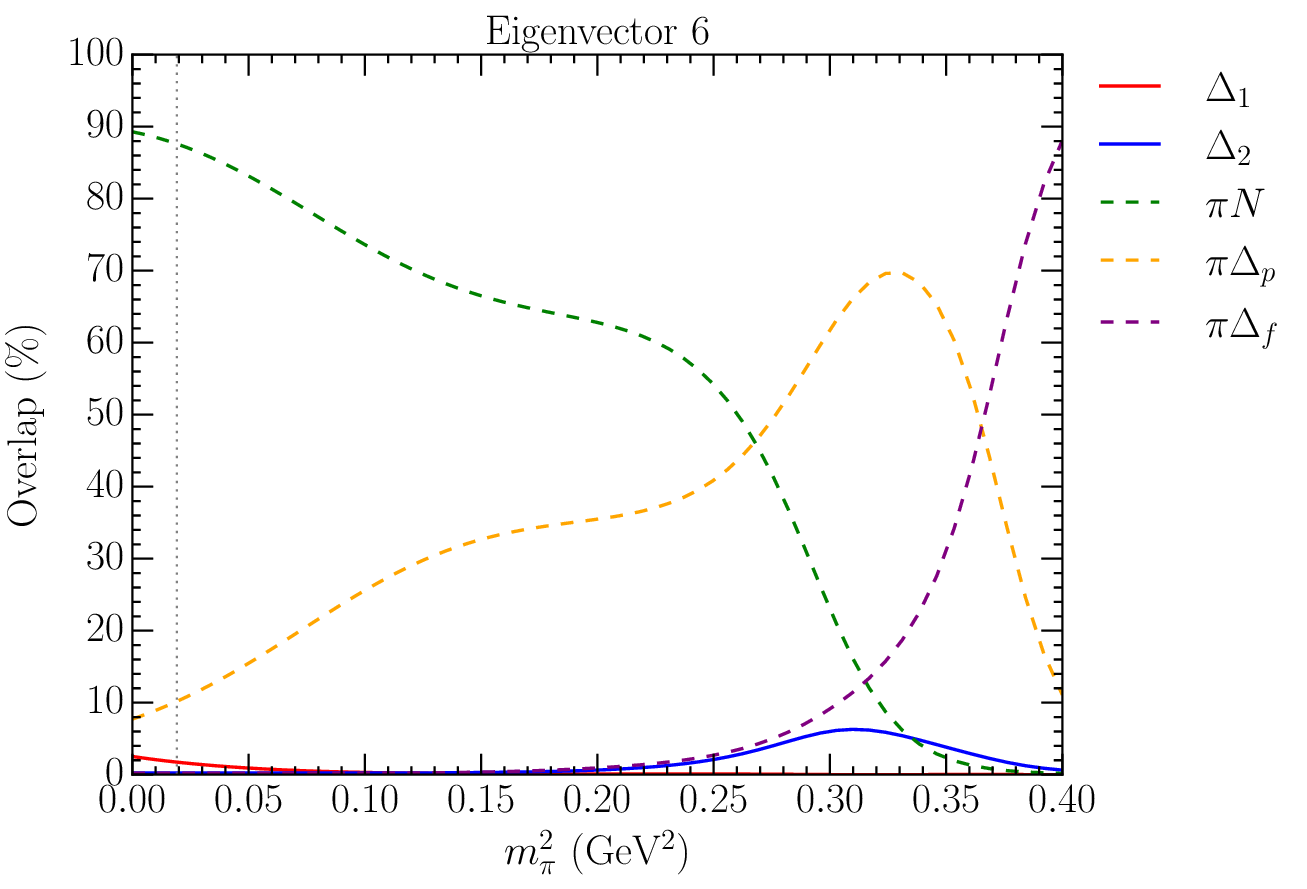}
		\includegraphics[width=0.5\linewidth]{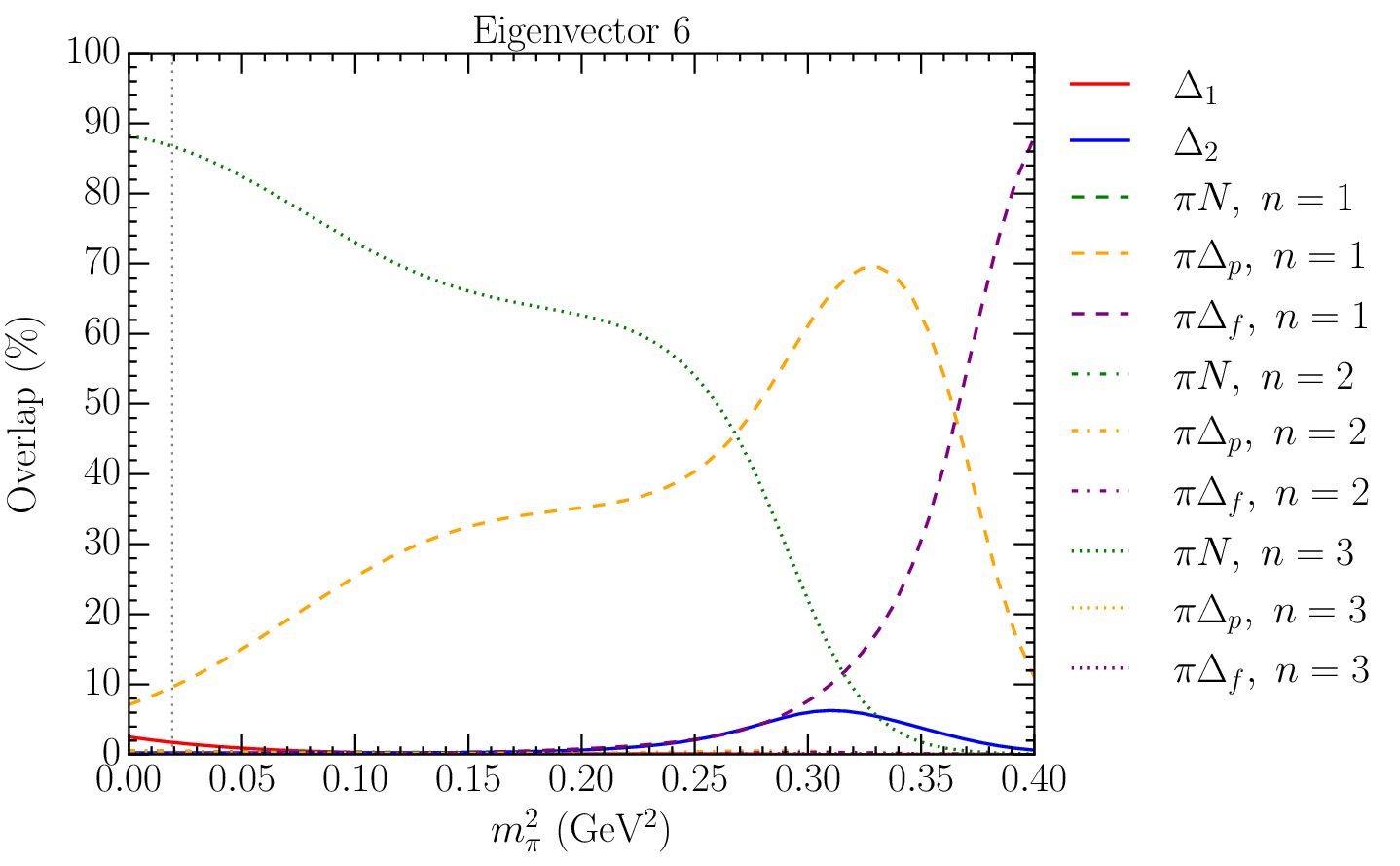}
		\includegraphics[width=0.453\linewidth]{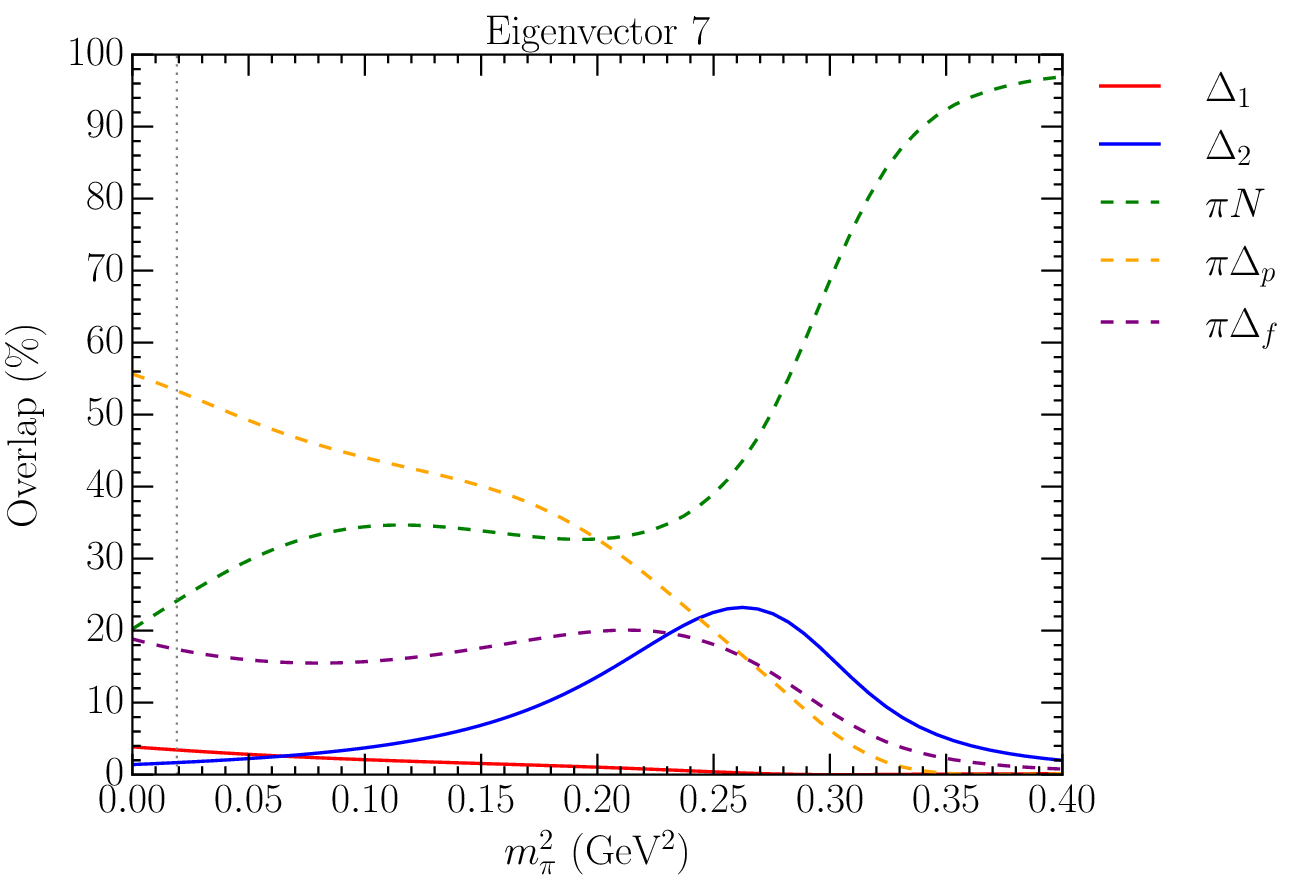}
		\includegraphics[width=0.5\linewidth]{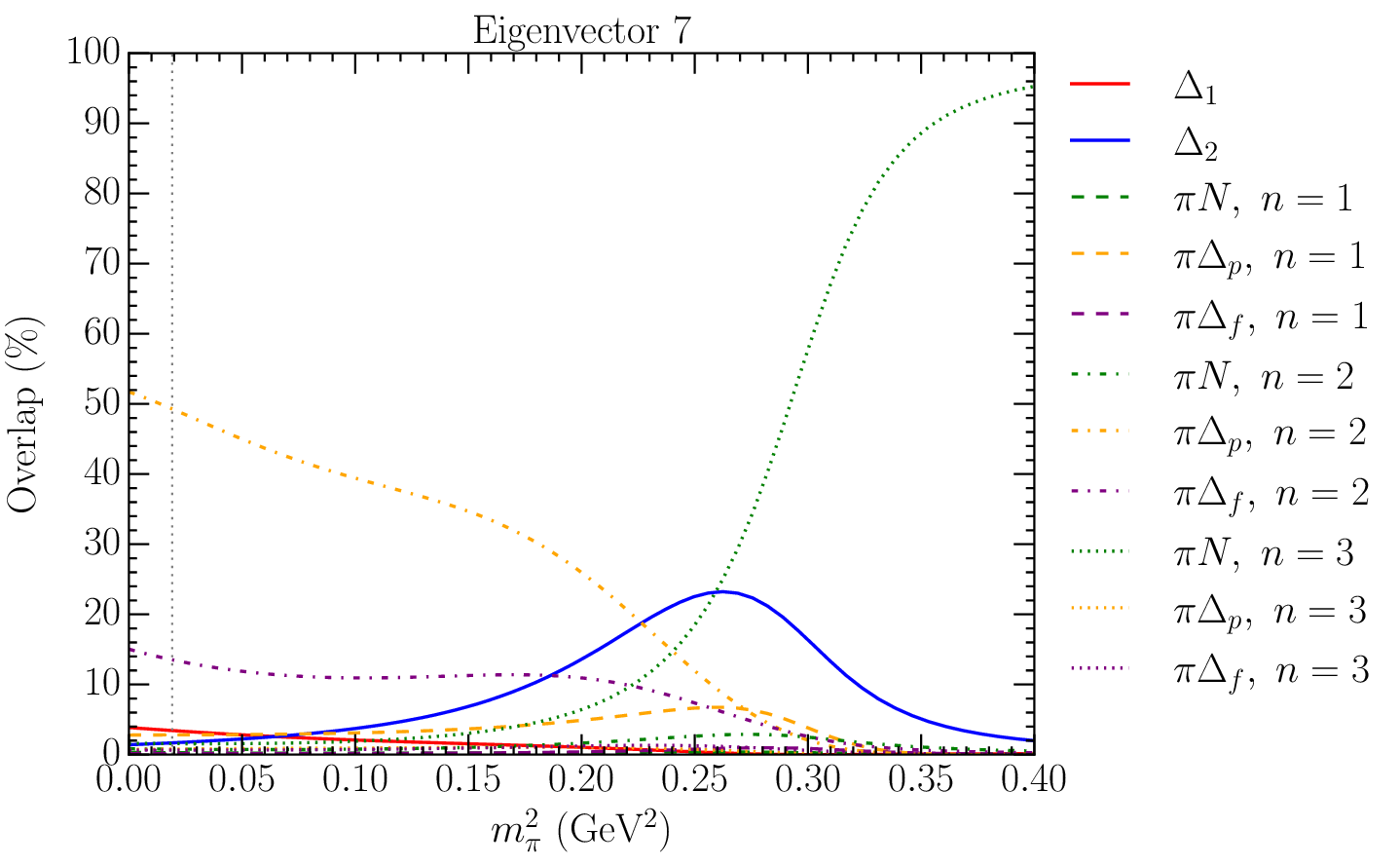}
		\caption{Same as for Fig.~\ref{fig:eigvec_comps_cls} but for eigenvectors 4-7, with energy eigenvalues in the region of the $ \Delta(1600) $ mass.}
		\label{fig:eigvec_comps_cls2}
	\end{center}
\end{figure*}

The HEFT energy levels are in excellent agreement with the lattice results, particularly for the $ m_\pi = 200 $ MeV ensemble. While the $ m_\pi = 200 $ MeV plot in Fig.~\ref{fig:cls} and the plot in Fig.~\ref{fig:finVol_cls} show four lattice results, only one of the two highest energies in the cubic group $ H_g $ has $ J = 3/2 $ \cite{Morningstar:2022}. In Fig.~\ref{fig:finVol_cls}, note that while the states dominated by $ \Delta_2 $ appear to shift higher at low pion masses, there are still significant mixtures of this basis state in the $ \sim 2.0-2.2 $ GeV region, as revealed by considering the eigenvector components of each energy eigenstate illustrated in Figs.~\ref{fig:cls_components}, \ref{fig:eigvec_comps_cls} and \ref{fig:eigvec_comps_cls2}. Fig.~\ref{fig:cls_components} gives a snapshot of the eigenvector components at the CLS pion mass of $ m_\pi = 200 $ MeV, while the plots of Figs.~\ref{fig:eigvec_comps_cls} and Figs.~\ref{fig:eigvec_comps_cls2} show the evolution of the eigenstate components in $ m_\pi^2 $.

In Fig.~\ref{fig:cls_components}, we find that the first three finite-volume eigenstates matching the CLS lattice points have significant overlap with the $ \pi N $ channel and the lower-lying bare basis state $ \Delta_1 $. With the CLS Consortium's interpolator basis including $ \pi N $ interpolators with projected momenta as per the right-hand column of Fig.~\ref{fig:eigvec_comps_cls}, this explains the excellent agreement between their results and our HEFT prediction. 

Considering those energy eigenstates with energies near the mass of the $ \Delta(1600) $ resonance, in particular the eigenstates with eigenvalues in the range $ 1.633 - 1.759 $ GeV in Fig.~\ref{fig:cls_components}, we find that they are either dominated by a single scattering channel (e.g. the fifth eigenstate) or composed of a mixture of basis states with only a small admixture of the bare basis states. This again indicates that in order to excite energies in the finite-volume spectrum corresponding to the $ \Delta(1600) $ on the lattice, one must include interpolators which capture the rich resonant properties of the $ \Delta(1600) $. Taking into account the dominant momentum basis states in the right-hand columns of Figs.~\ref{fig:eigvec_comps_cls} and \ref{fig:eigvec_comps_cls2}, we expect such excitations to respond well to a basis containing single-hadron interpolators in combination with both $ \pi N $ and $ \pi \Delta $ multi-hadron interpolators with momenta up to and including $ n = 3 $. The plots for eigenvector 7 show that $ \pi N (k_n) $ basis states beyond $ n = 3 $ become important for exciting the 7th energy eigenstate (and higher) on the lattice.

In terms of the physical $ \Delta(1600) $ resonance, this reinforces our key finding that the $ \Delta(1600) $ is not strongly associated with a 3-quark core. Rather, it relies heavily on strong rescattering in the $ \pi N $ and $ \pi \Delta $ channels.

As a final note, the interacting energy levels in Fig.~\ref{fig:finVol_cls} give a clear example of how the density of states increases with the lattice size. This is readily apparent by comparing with Fig.~\ref{fig:finVol_3fm}, where the smaller lattice volume leads to sparser energy eigenstates. This will become especially relevant when comparing with the HSC lattice QCD results on very small lattices, with $ L \sim 2 $ fm.

\begin{figure*}
	\begin{center}
		\includegraphics[width=0.453\linewidth]{./Figures/eigvec_components_cls/eigvec4_components.eps} 
		\includegraphics[width=0.453\linewidth]{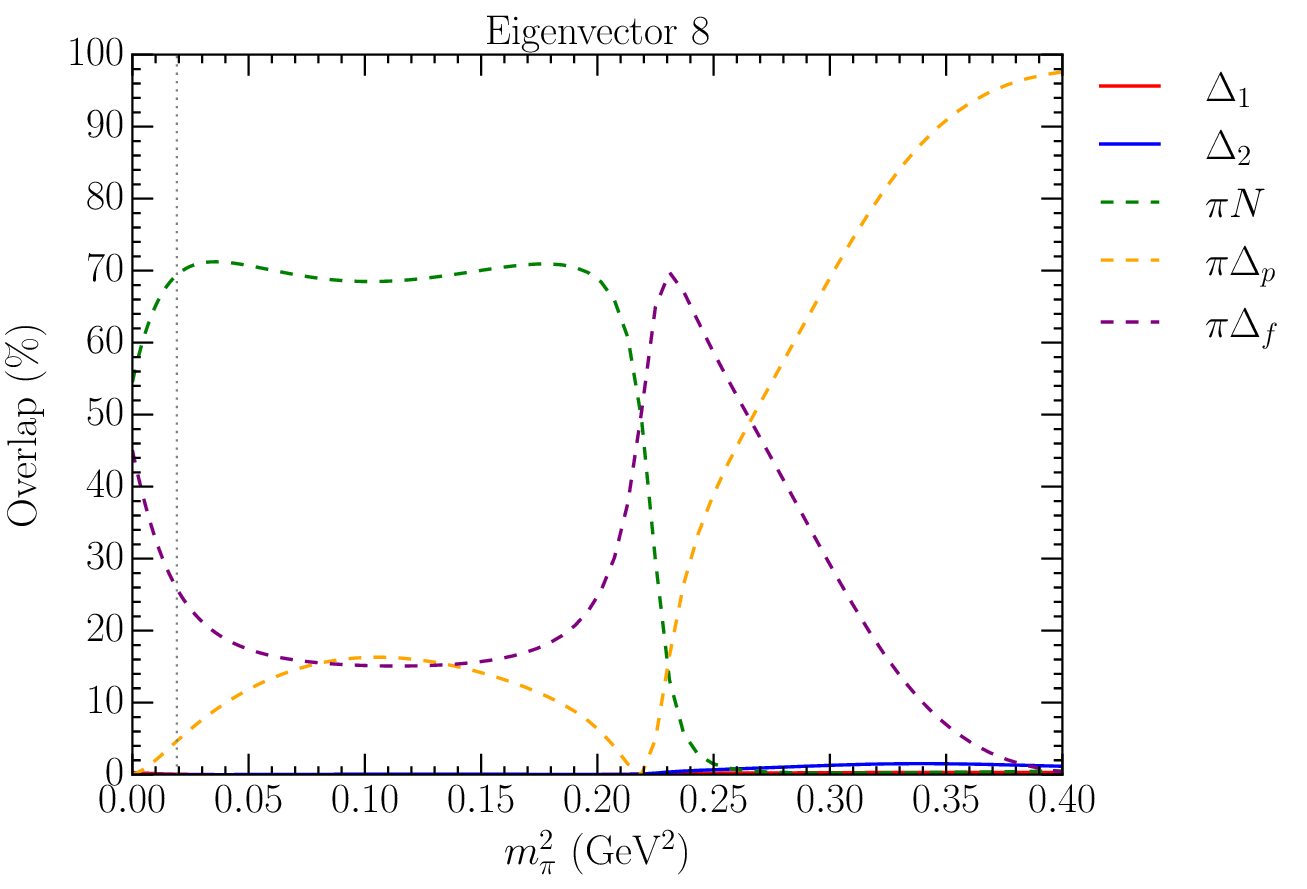}
		\includegraphics[width=0.453\linewidth]{./Figures/eigvec_components_cls/eigvec5_components.eps}
		\includegraphics[width=0.453\linewidth]{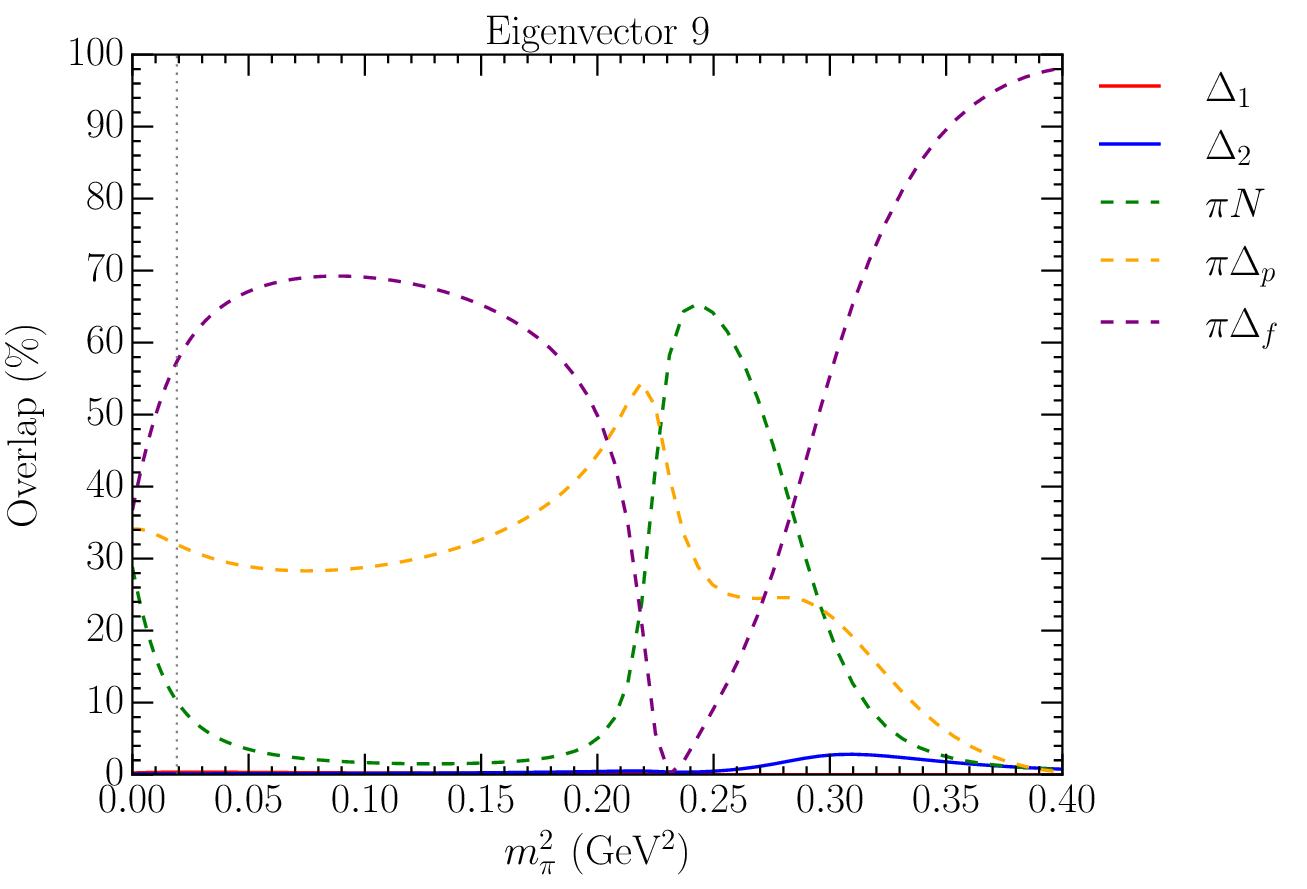}		
		\includegraphics[width=0.453\linewidth]{./Figures/eigvec_components_cls/eigvec6_components.eps}
		\includegraphics[width=0.453\linewidth]{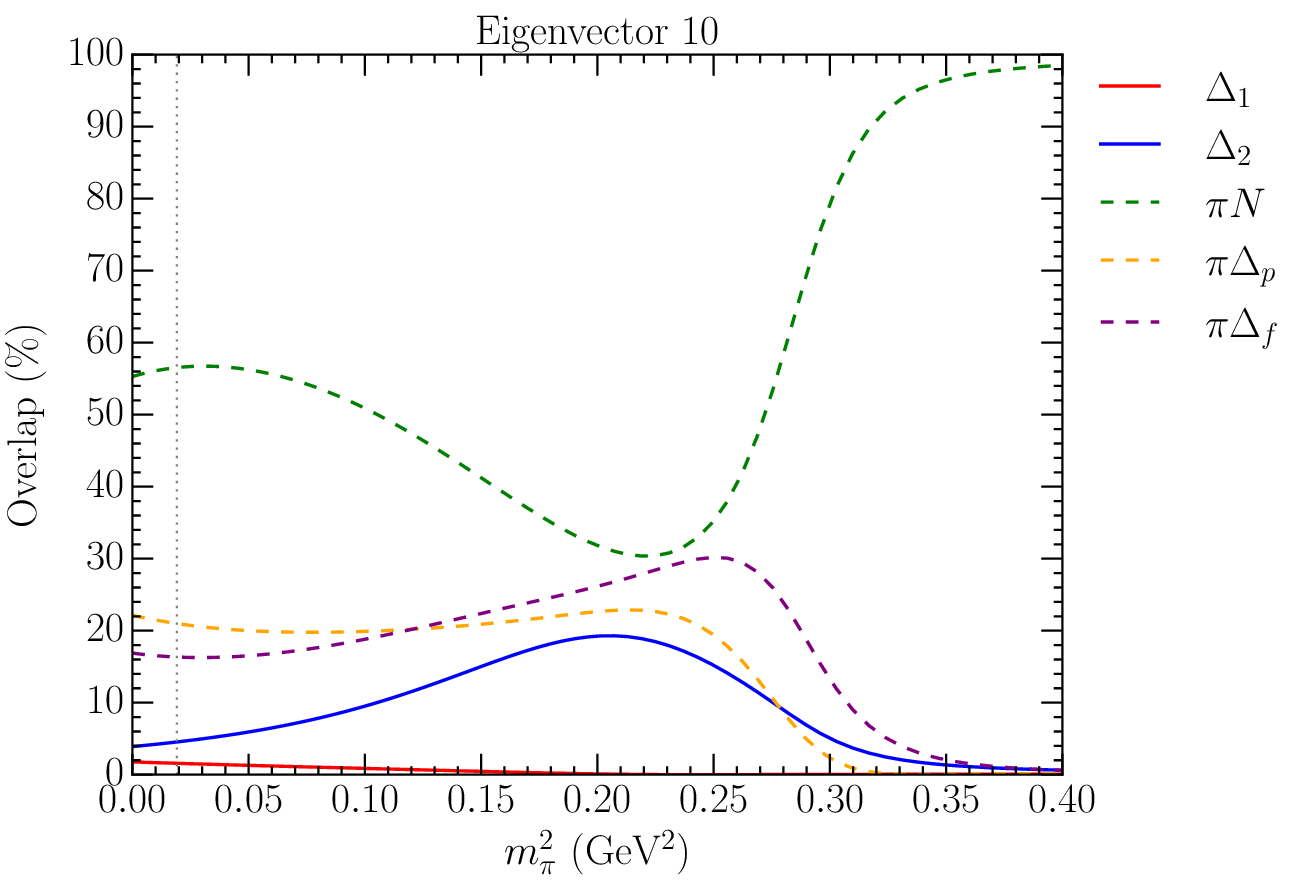}
		\includegraphics[width=0.453\linewidth]{./Figures/eigvec_components_cls/eigvec7_components.eps}
		\includegraphics[width=0.453\linewidth]{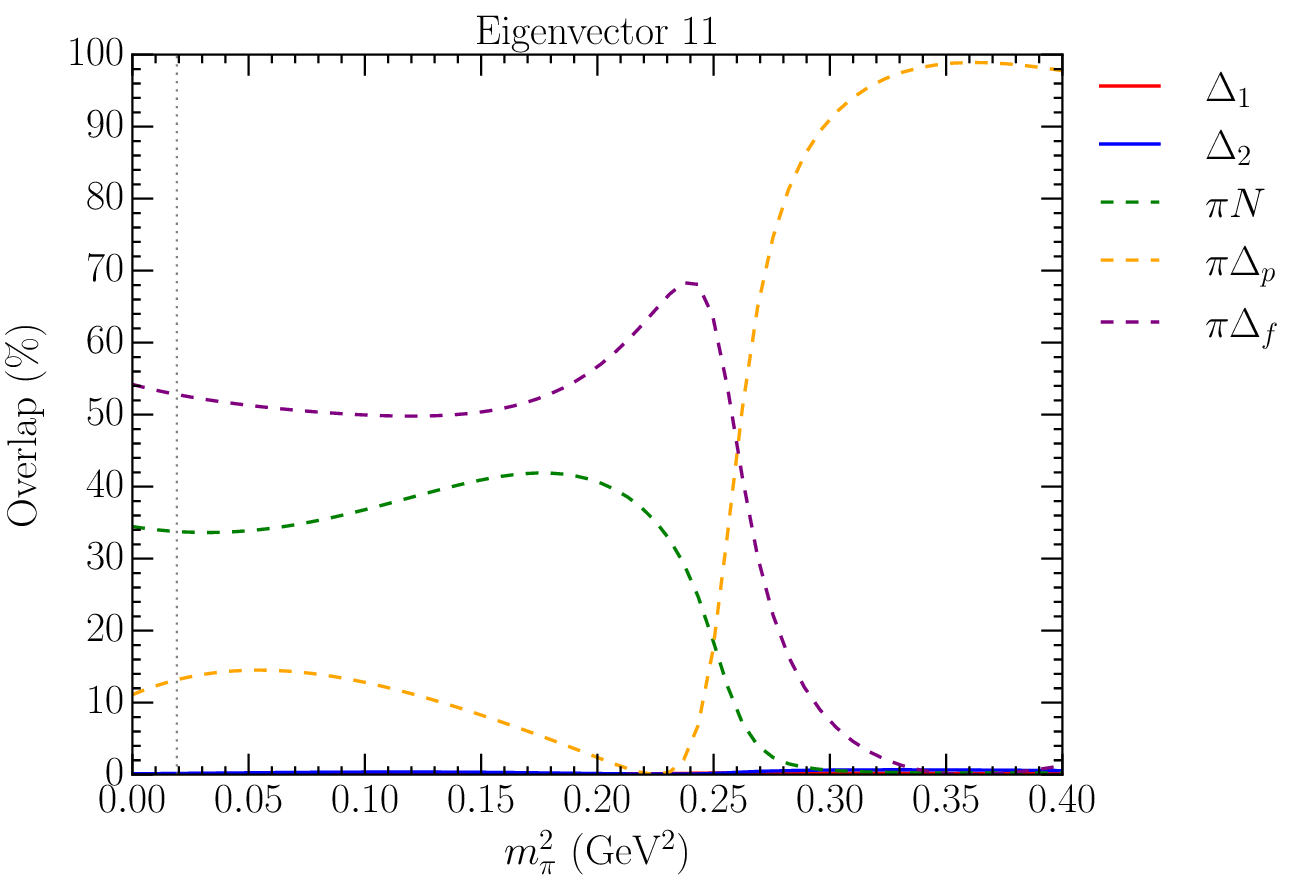}
		\caption{Eigenvector components for the 4th-11th fully interacting HEFT eigenstates in the  $ L = 4.16 $ fm calculation.}
		\label{fig:eigvec_comps_cls3}
	\end{center}
\end{figure*}

\subsection{Cyprus Collaboration Comparison}
\begin{figure*}
	\begin{center}
		\includegraphics[width=0.7\linewidth]{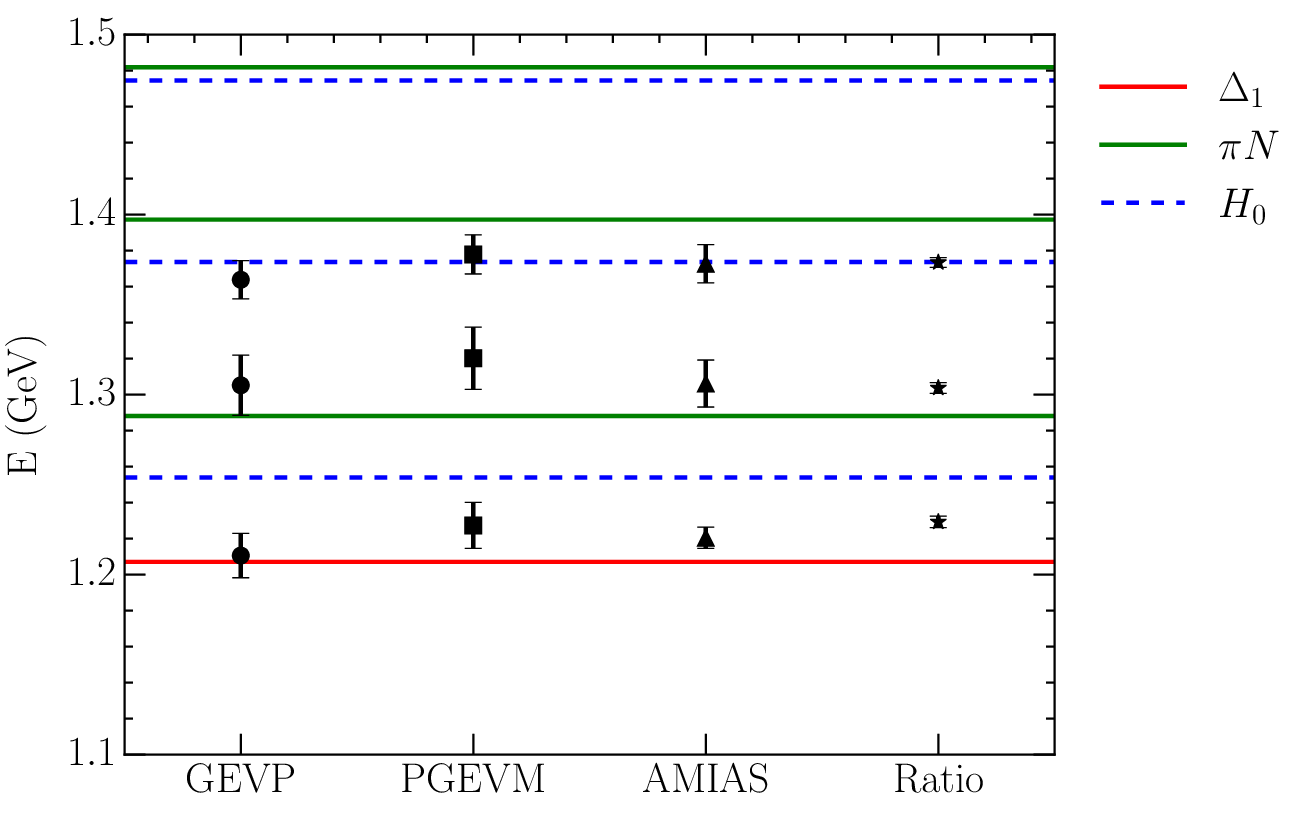}
	\caption{Plot comparing the energy levels between our HEFT model and those reported by the Cyprus group. Note that this comparison is made at the single value of $ m_\pi = 139.43 $ used in Ref.~\cite{Alexandrou:2023elk}. Non-interacting energies are shown as blue dashed lines, while fully interacting HEFT levels are shown as solid red and green lines indicating the predominant basis state component. The lattice QCD results from the Cyprus group are shown by the various symbols for each of the different methods used in extracting energies; GEVP (filled circles), PGEVM (squares), AMIAS (triangles) and Ratio (stars).}
	\label{fig:cyp_comparison}
	\end{center}
\end{figure*}

Recently, Ref.~\cite{Alexandrou:2023elk} provided lattice QCD results for $ \pi N $ energies at the physical quark mass point. To make contact with these results via HEFT, we use $ m_\pi = 139.43 $ MeV, $ L = 5.1 $ fm and $ m_N = 944 $ MeV. Assuming a similar increase from the finite volume for the $ \Delta $, we take $ m_\Delta = 1238 $ MeV. The resulting HEFT energy level predictions at the physical pion mass point are given in Fig.~\ref{fig:cyp_comparison} along with the results from the four methods used in Ref.~\cite{Alexandrou:2023elk} for determining the energy levels of $ \pi N $ scattering states, those being the Generalised Eigenvalue Problem (GEVP), Prony Generalised Eigenvalue Method (PGEVM), Athens Model Independent Analysis Scheme (AMIAS) and the Ratio method (RATIO). 

There is fair agreement between the HEFT energy levels and the lattice QCD results for the standard GEVP analysis. The largest discrepancy occurs for the highest energy results reported by the Cyprus group. All four analysis methods produce results consistent with the nearby non-interacting state. Future calculations should explore additional interpolating fields and more extensive Euclidean time evolution to examine the possible evolution of the eigenstate energy away from the noninteracting basis-state energy.

\clearpage

\subsection{HSC Comparison}\label{subsec:HSC}
\begin{figure*}
	\begin{center}
		\includegraphics[width=0.85\linewidth]{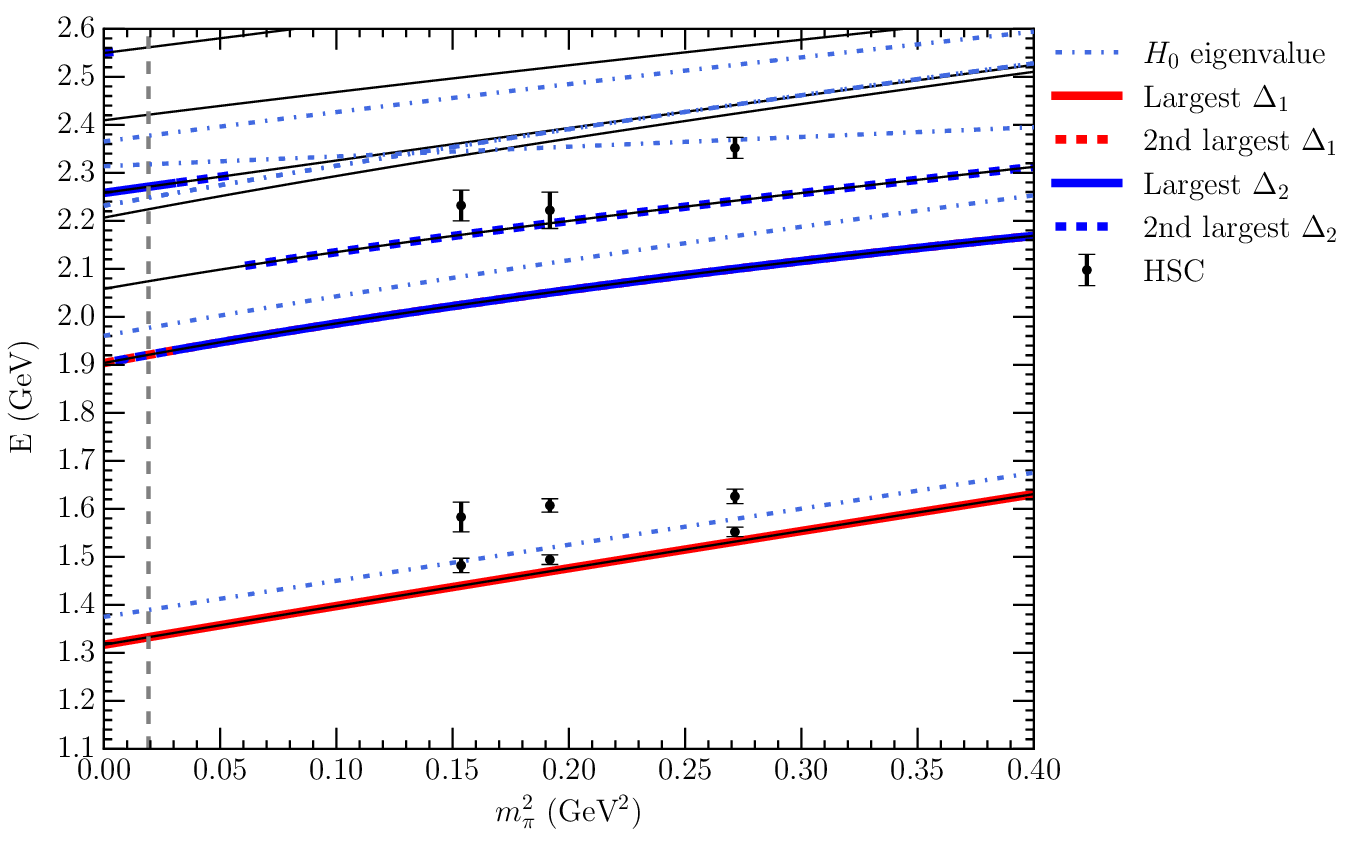}
		\caption{Finite-volume spectrum at a lattice size of $ L = 1.96 $ fm. Solid lines show interacting energy levels, while dashed lines show non-interacting levels. The lattice data from Ref. \cite{Bulava:2010yg} are shown as data points with errors.}
		\label{fig:finVol_HSC}
	\end{center}
\end{figure*}

\begin{figure*}
	\begin{center}
		\includegraphics[width=0.45\linewidth]{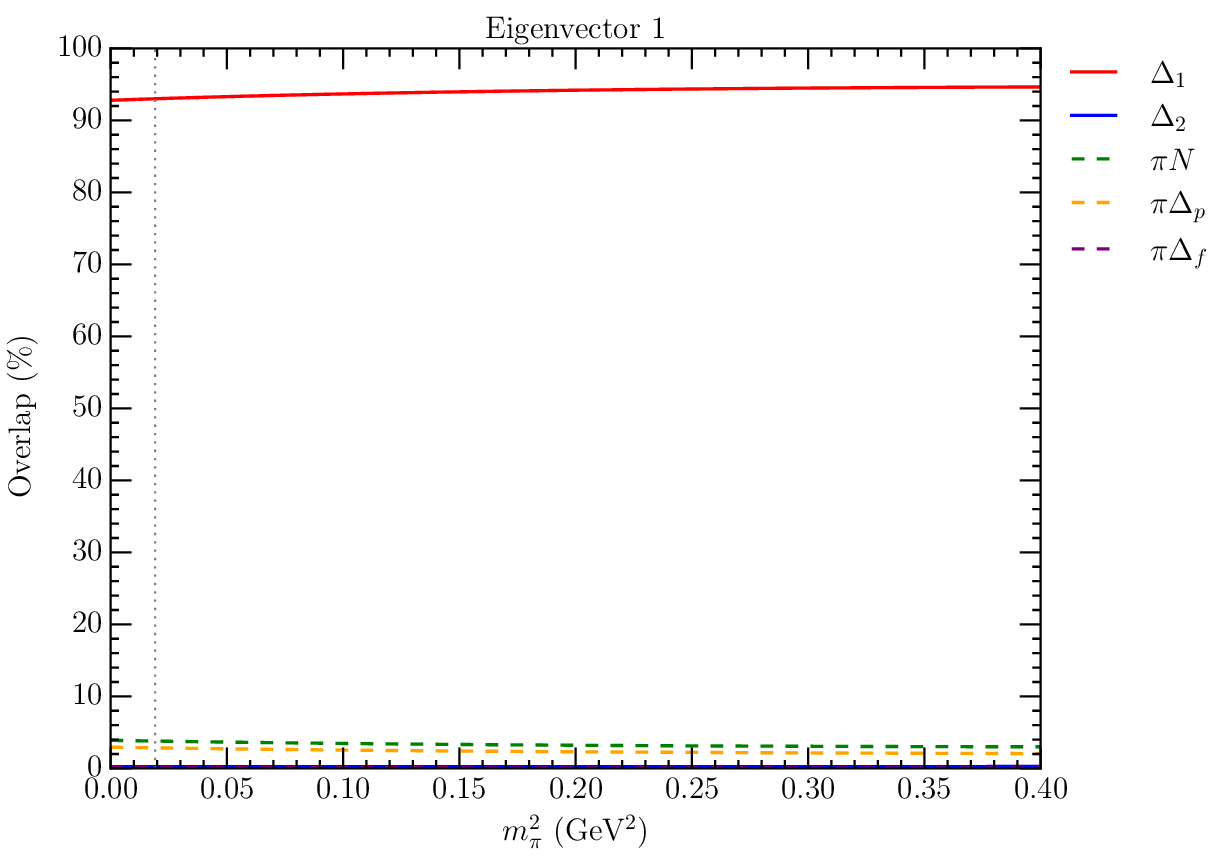}
		\includegraphics[width=0.45\linewidth]{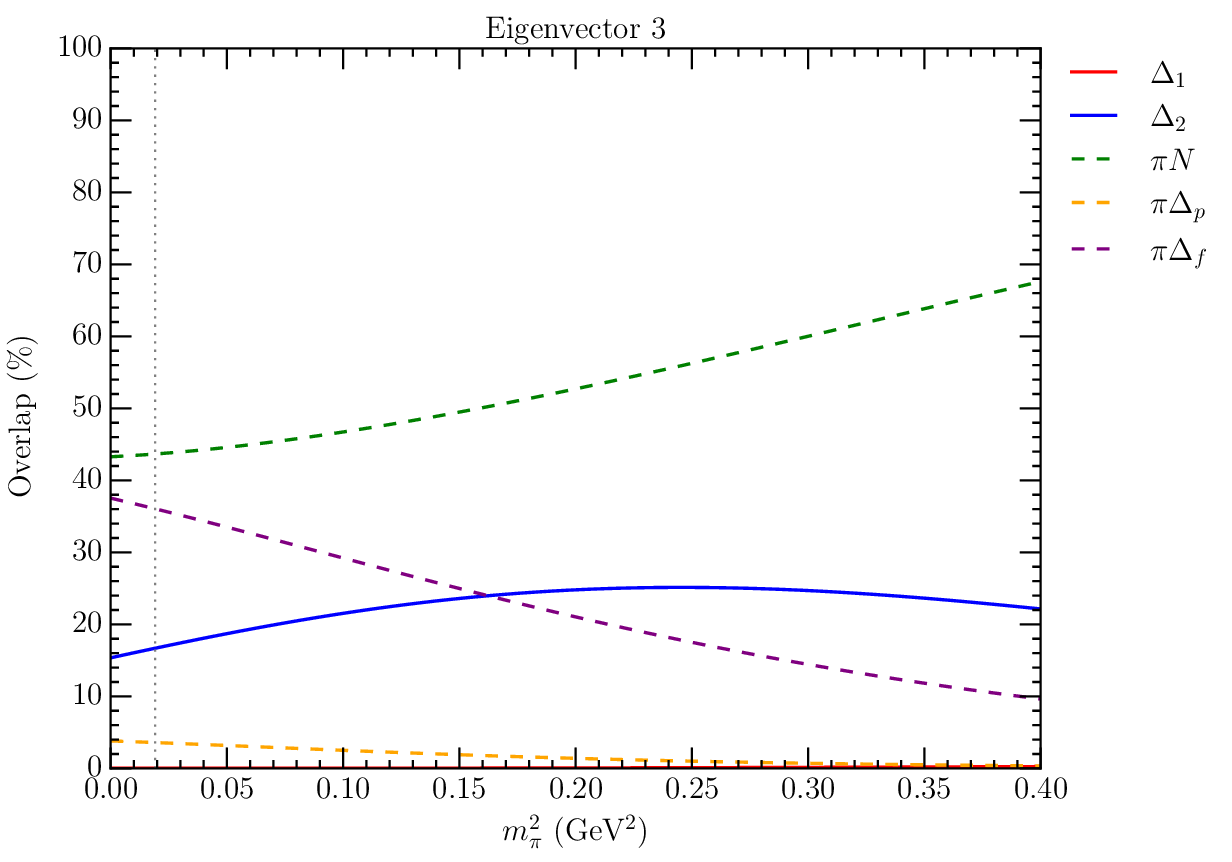}
		\includegraphics[width=0.45\linewidth]{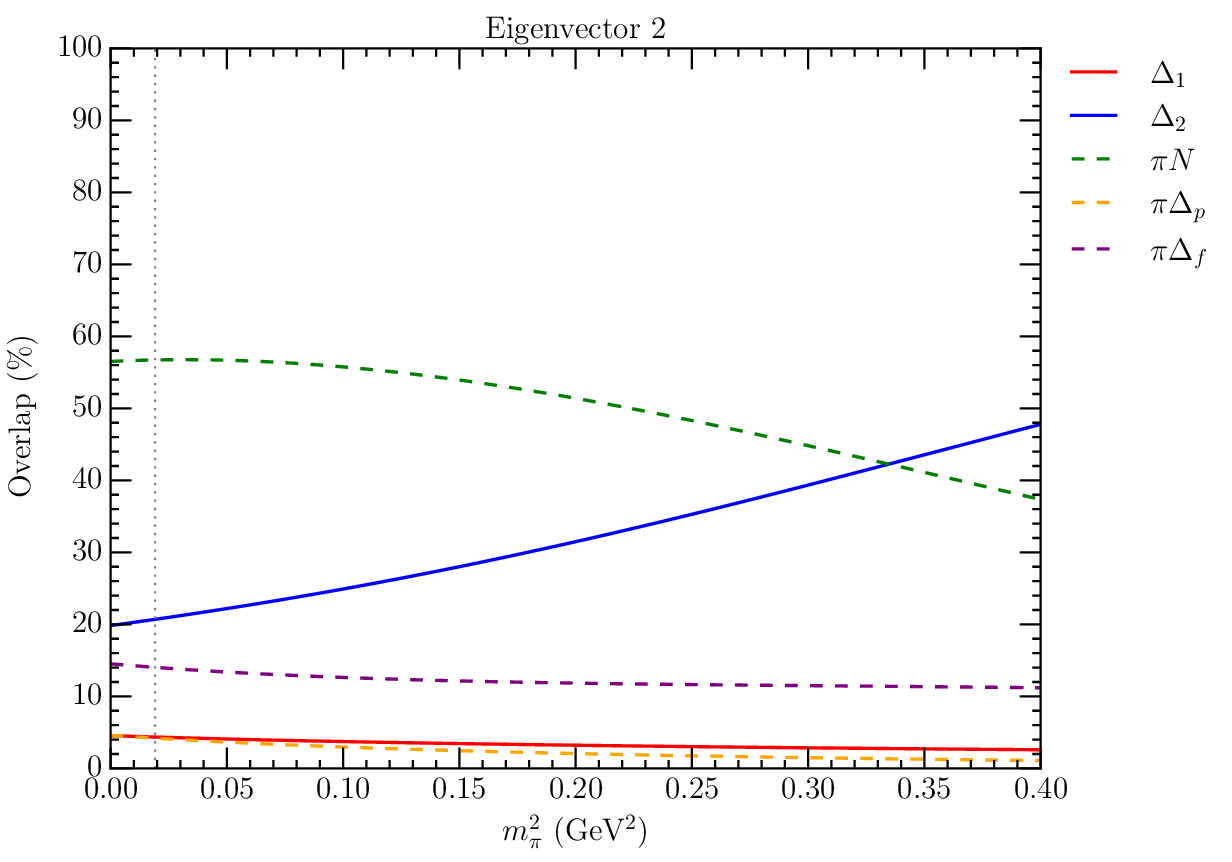}
		\includegraphics[width=0.45\linewidth]{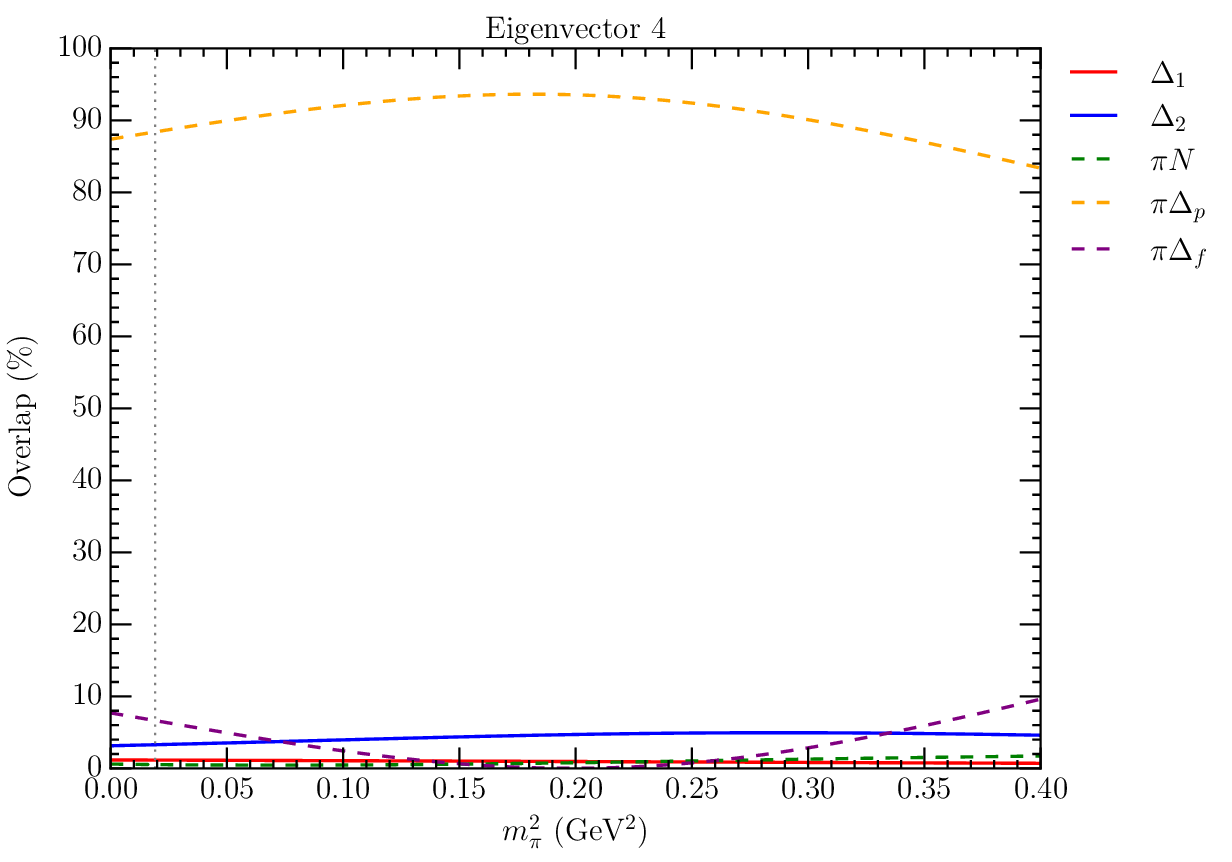}
		\caption{Eigenvector components for the first four fully interacting HEFT eigenstates in the $ L = 1.96 $ fm calculation.}
		\label{fig:eigvec_comps_hsc}
	\end{center}
\end{figure*}

Next, we compute the HEFT spectrum using a lattice size of $ L = 1.96 $ fm, as in Ref. \cite{Bulava:2010yg}. Taking their values of $ m_\pi = 392,\, 438 $ and $ 521 $ MeV, and taking $ m^*_N $, $ \alpha_N $ and $ m^*_\Delta, \alpha_\Delta $ so as to reproduce the non-interacting thresholds in Table VI of Ref.~\cite{Bulava:2010yg} yields the finite-volume spectrum given in Fig.~\ref{fig:finVol_HSC}. Additionally, we provide the eigenvector component plots for the four lowest energy levels in Fig.~\ref{fig:eigvec_comps_hsc}. 

Notably, we see clear evidence that while the HSC results for the lowest- and highest-energy states show fair agreement with the HEFT predictions, the first excited state has no nearby interacting energy level with which it can be associated. Indeed, on a lattice of around just 2 fm, the finite-volume spectrum levels are very sparse at low energies. The first $ \pi N $ dominated state is at $ \sim 2 $ GeV. In this light, the intermediate states remain unexplained. They cannot be associated with the $ \Delta(1600) $ (as had been suggested in Ref.~\cite{Bulava:2010yg}).

Disregarding the first excitation then, the remaining results are in fair agreement with HEFT; the ground state is approximately where HEFT predicts it to be, while the second excitation is likely a superposition of energy eigenstates having overlap with the bare basis state $ \Delta_2 $.

\subsection{Khan \textit{et al.} Comparison}\label{subsec:Khan}
\begin{figure*}
		\begin{center}
			\includegraphics[width=0.9\linewidth]{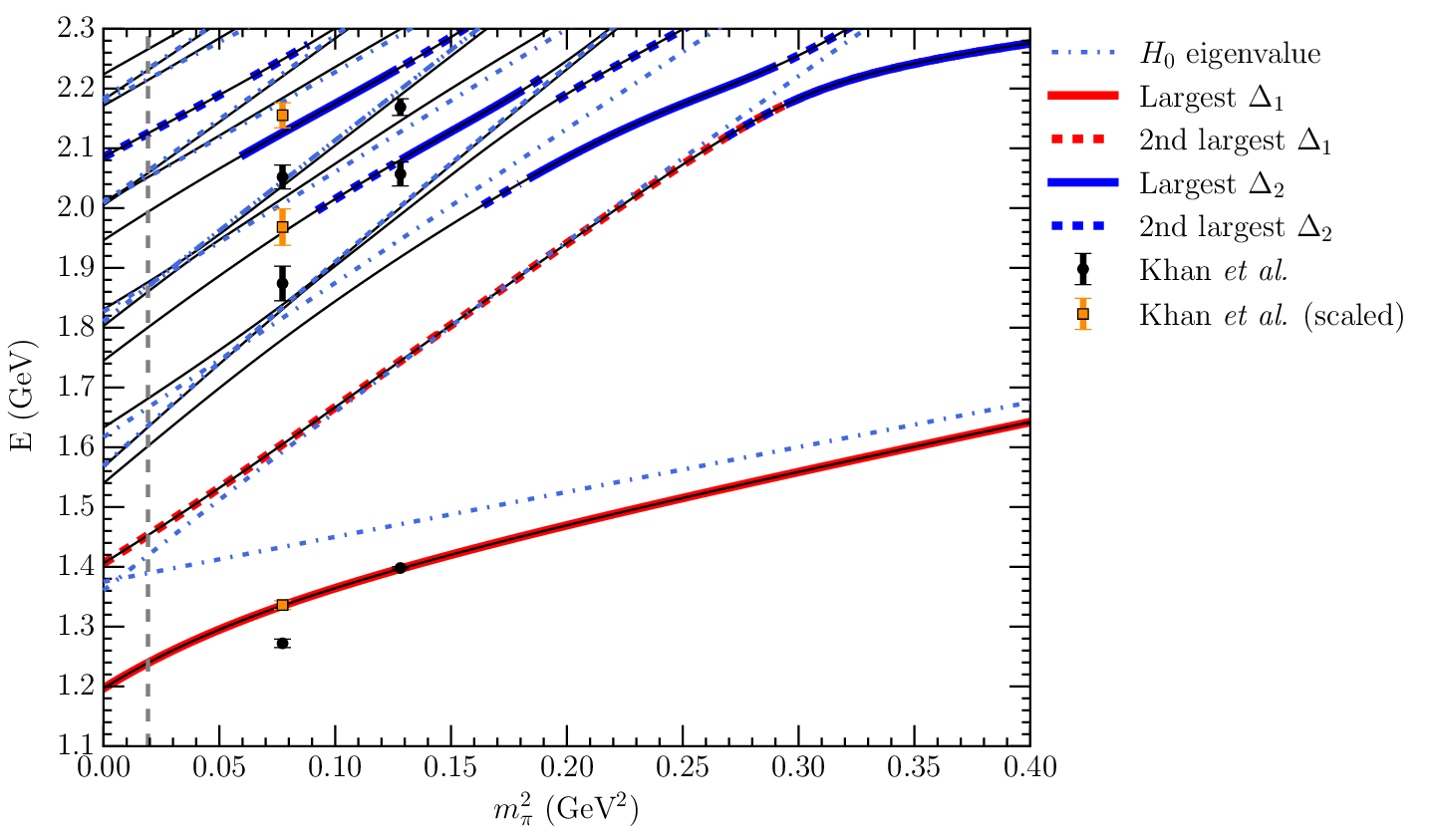}
			\caption{Finite-volume spectrum at a lattice size of $ L = 3.01 $ fm. Solid lines show interacting energy levels, while dashed lines show non-interacting levels. The lattice QCD results from Ref.~\cite{Khan:2020ahz} are shown as black circles with errors. The orange squares show the points from the lighter ensemble scaled up such that the ground state matches the prediction from HEFT. }
			\label{fig:finVol_Khan}
		\end{center}
	\end{figure*}

\begin{figure*}
		\begin{center}
			\includegraphics[width=0.45\linewidth]{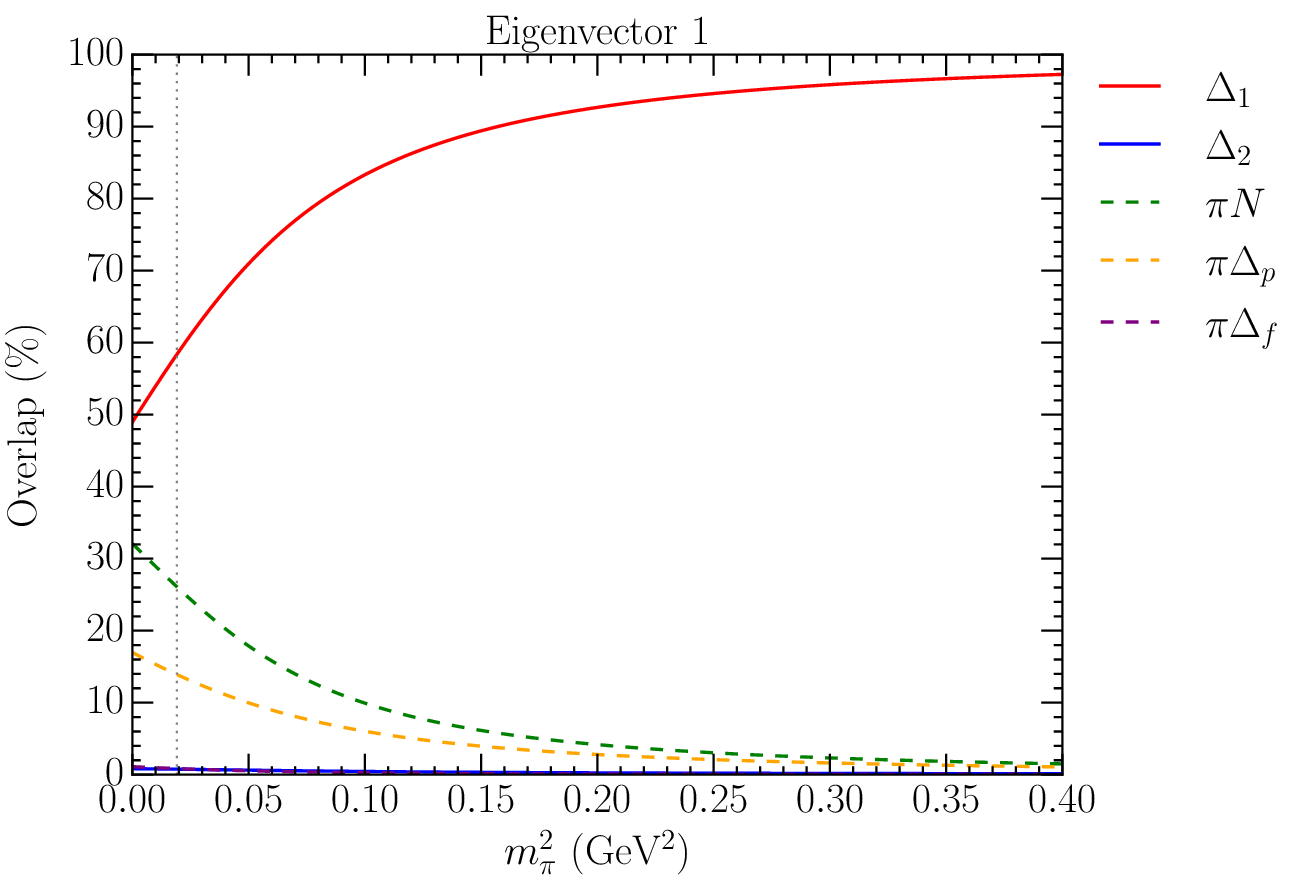}
			\includegraphics[width=0.45\linewidth]{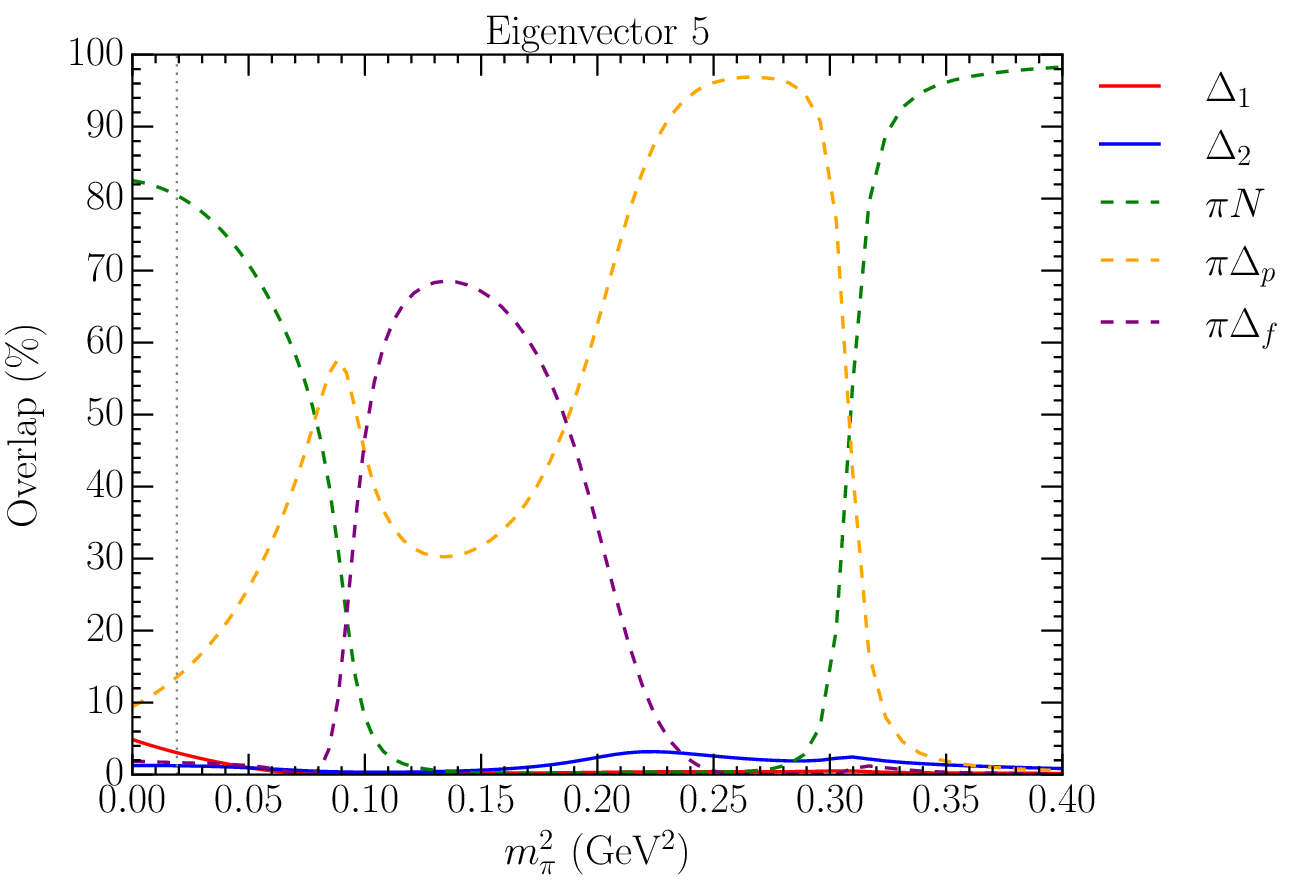}
			\includegraphics[width=0.45\linewidth]{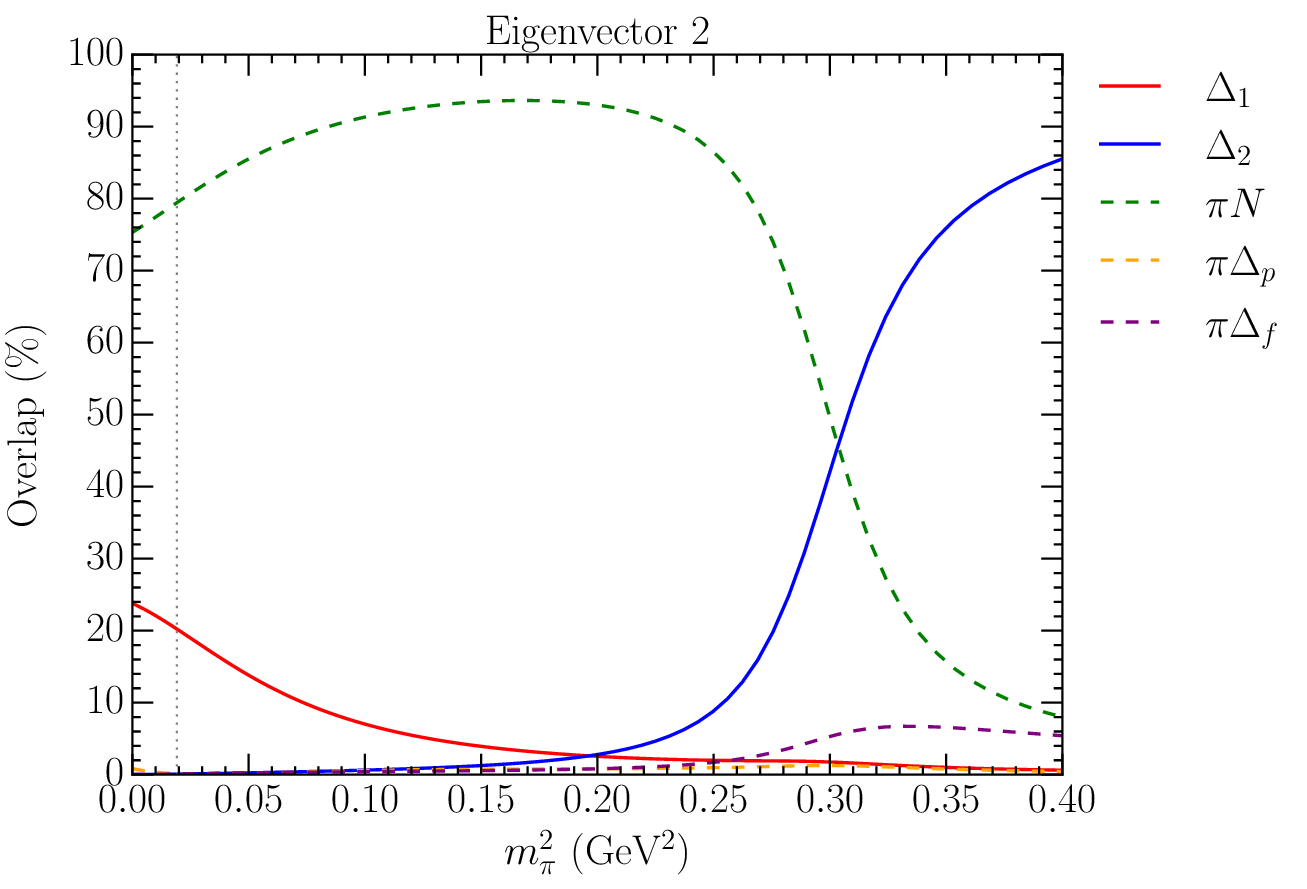}
			\includegraphics[width=0.45\linewidth]{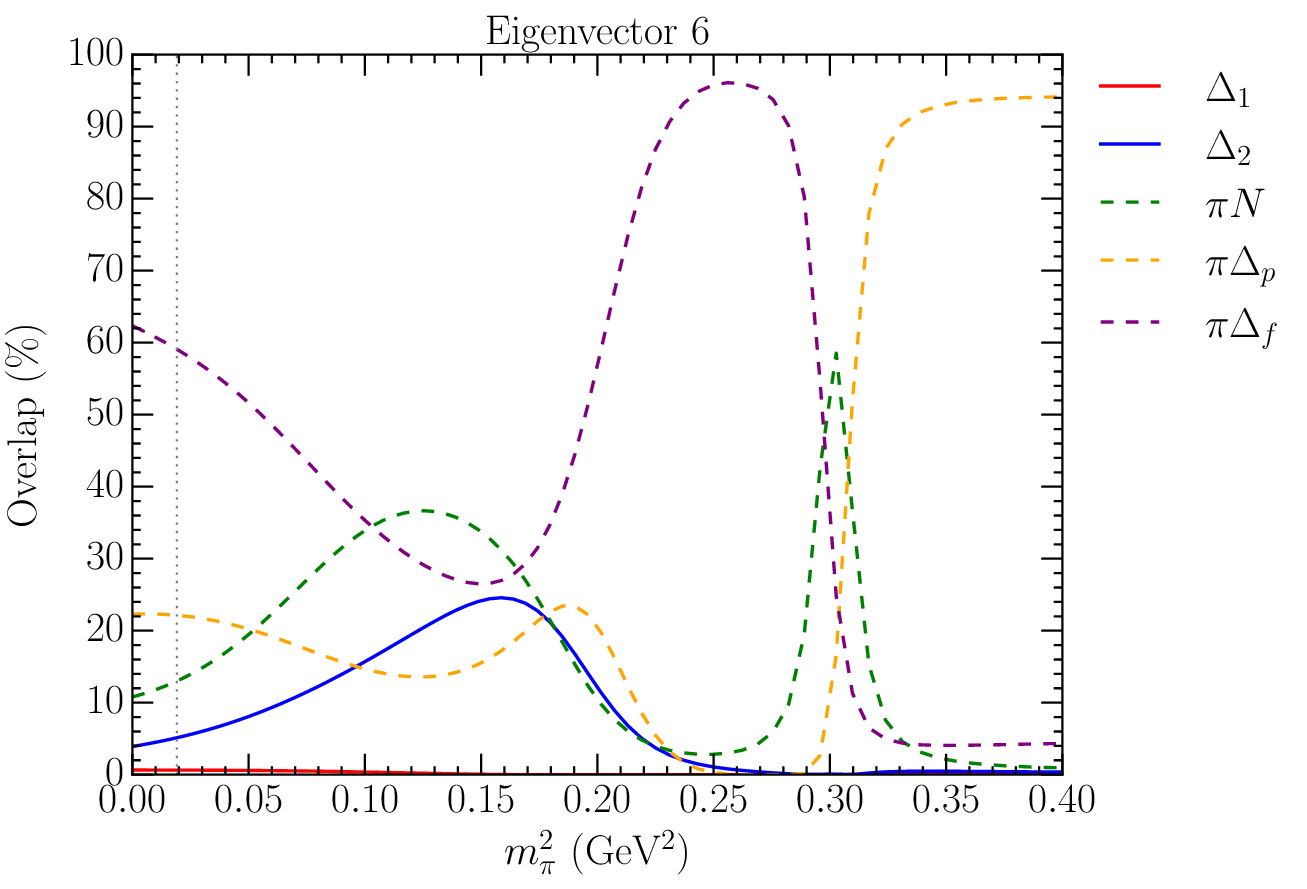}
			\includegraphics[width=0.45\linewidth]{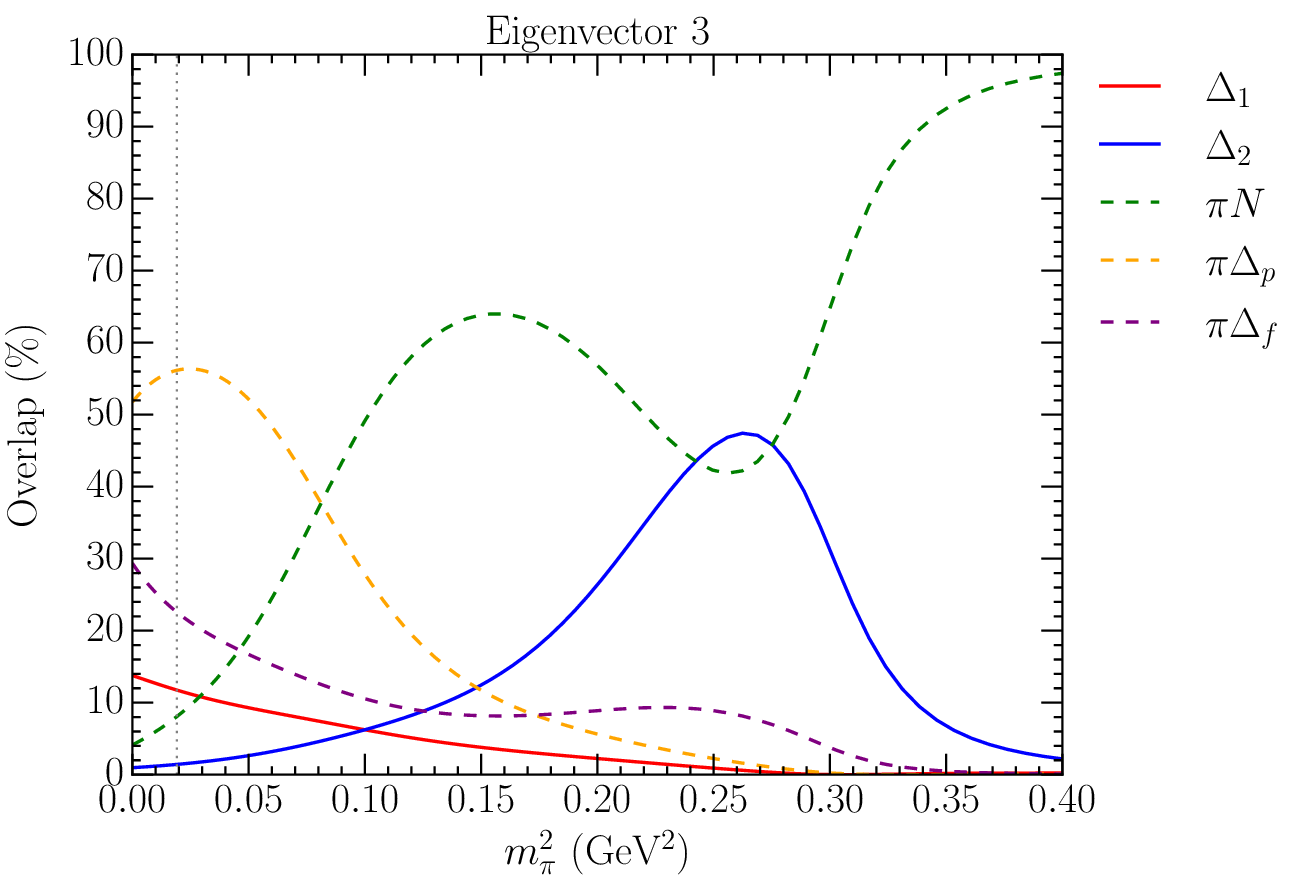}
			\includegraphics[width=0.45\linewidth]{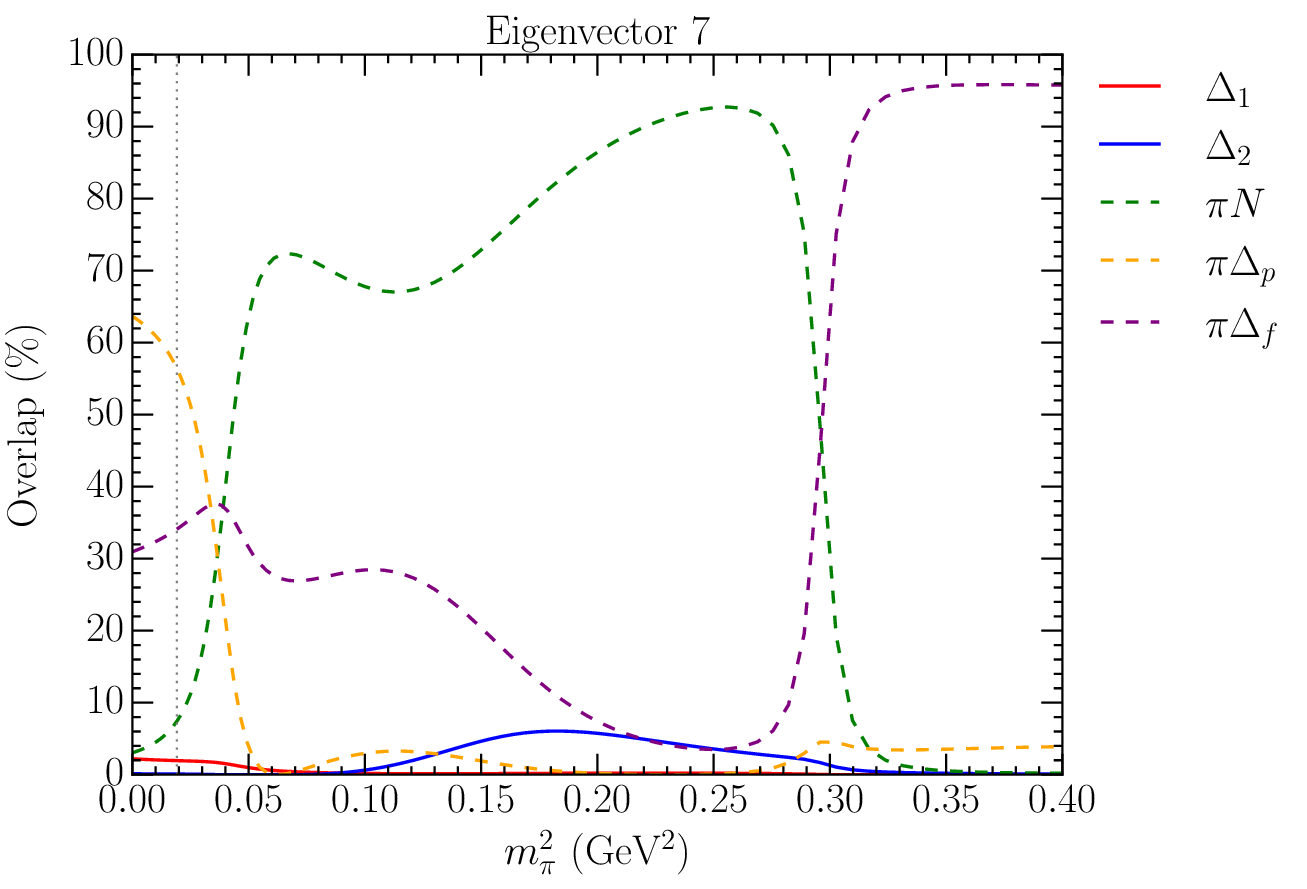}
			\includegraphics[width=0.45\linewidth]{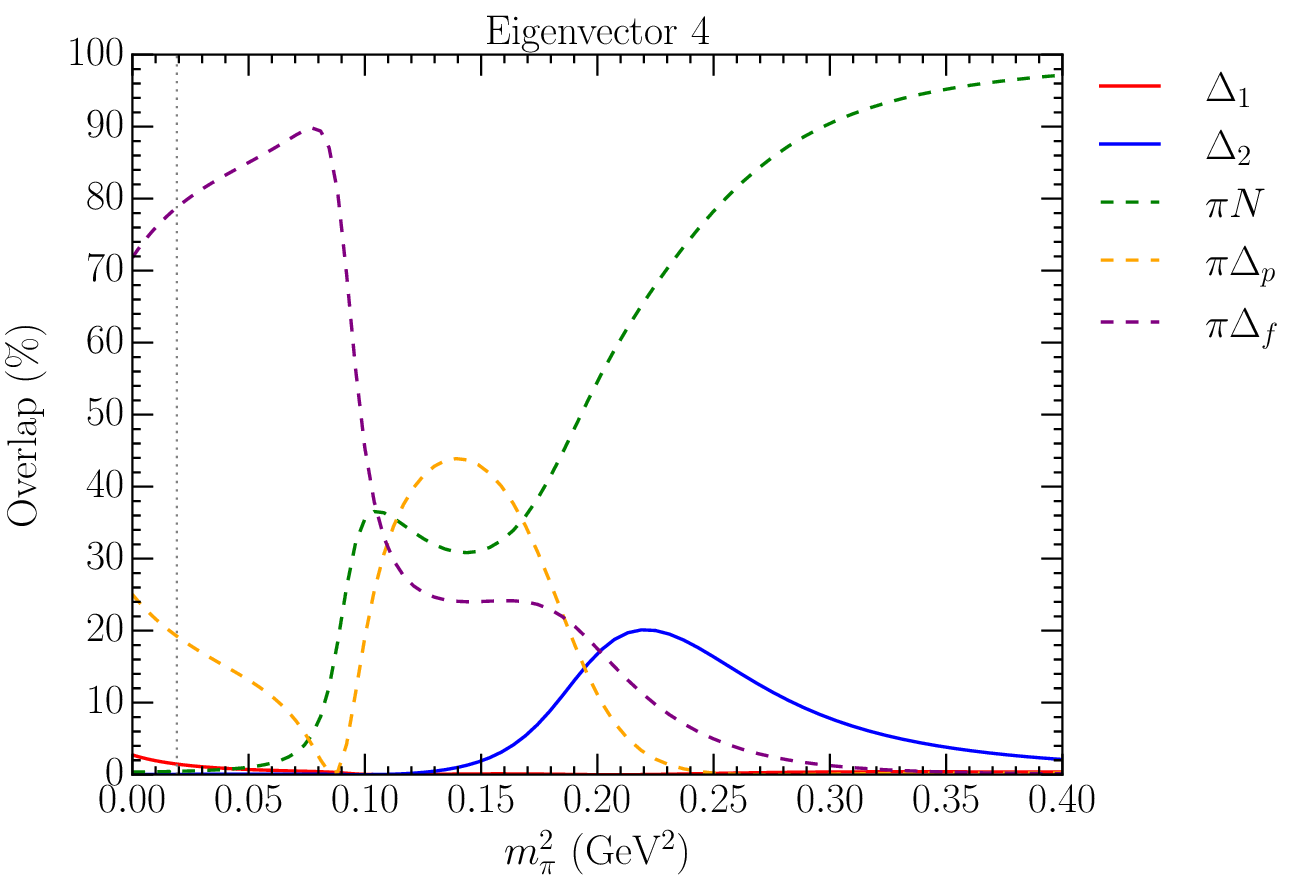}
			\includegraphics[width=0.45\linewidth]{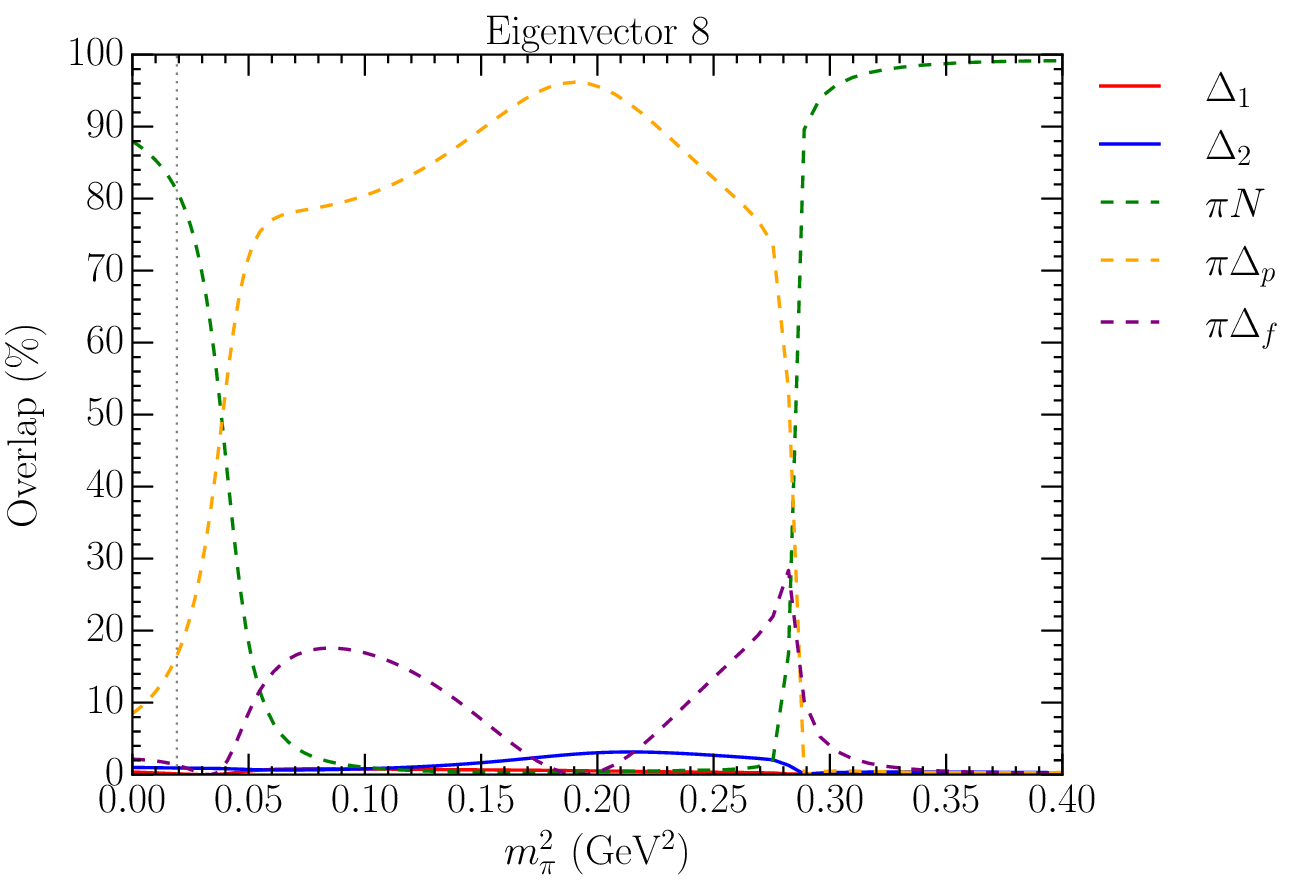}
			\caption{Eigenvector components for the first 8 fully interacting HEFT eigenstates in the $ L = 3.01 $ calculation.}
			\label{fig:eigvec_comps_khan}
		\end{center}
	\end{figure*}

As our final comparison with contemporary lattice QCD results, we compute the HEFT spectrum using a lattice size of $ L = 3.01 $ fm, as in Ref. \cite{Khan:2020ahz}. The resulting finite-volume spectrum is given in Fig.~\ref{fig:finVol_Khan}. Additionally, we provide the eigenvector component plots for the eight lowest energy levels in Fig.~\ref{fig:eigvec_comps_khan}.

Here, the reported lattice points align fairly well with the finite-volume spectrum generated via HEFT, with every point lying on or nearby a bare-basis state dominated interacting energy level with the exception of the lowest-lying state at the lighter quark mass. If we assume a systematic error in the lower quark mass results associated with scale setting, then all these masses move into positions which are in better agreement with the HEFT model (given by the ``scaled" points in Fig.~\ref{fig:finVol_Khan}).

The heavier quark-mass ensemble results are interesting as well; remarkably, the lowest-lying energy at the heavier quark mass sits on the lowest-lying HEFT energy level, despite the miniscule uncertainty. Additionally, the highest energy level reported is reminiscent of the $ m_\pi^2 \sim 0.15 $ and $ 0.27 $ GeV$ ^2 $ from Fig.~\ref{fig:finVol_3fm} in that it lies in a superposition of two energy levels dominated by the second bare state. Thus this result is also supported by HEFT.

Turning our attention to the eigenvector component plots, we consider the plots for eigenvectors 4, 5 and 6 as these correspond to the eigenvalues near the first excitation results of Ref. \cite{Khan:2020ahz}, and are approximately in the region associated with the $ \Delta(1600) $ resonance. Notably, these lattice QCD results were identified as being consistent with hybrid states excited by local lattice operators. The quantum numbers of the operators used are not exotic so these states can also be described by mixtures of two-particle channels.

Eigenvectors 4 and 5 possess very little bare state content at the pion masses used, instead being mostly comprised of two-particle basis states. On the other hand, eigenvector 6 shows a greater mixture of bare and two-particle states at the lighter quark mass. Importantly though, at the physical point, the bare basis state content of eigenvector 6 is greatly diminished. This provides further support for the interpretation of the $ \Delta(1600) $ as being a dynamically generated resonance, rather than a resonance with a dominant 3-quark core component.

\section{Conclusion}\label{sec:Conclusion}
We have presented an analysis of the low-lying $ \Delta $-baryon spectrum through a combination of lattice QCD and Hamiltonian Effective Field Theory (HEFT).

Using HEFT, we are able to simultaneously describe experimental data from Ref.~\cite{GWU:2023ex} and lattice QCD results obtained in Ref.~\cite{Hockley:2023yzn}. Starting with a simple description of the low-energy region around the $ \Delta(1232) $ resonance, we then extended this analysis to higher energies seeking to confront the $ \Delta(1600) $ and the $ 2s $ excitation observed in lattice QCD.

We have also validated our results through comparison with spectrum results obtained using lattice QCD on an array of volumes, at several pion masses. Excellent agreement was found with the results of the CLS Consortium and we saw agreement with a comparison to state-of-the-art results from the Cyprus Collaboration for the first two states. Our analysis was in fair agreement with the results from Khan \textit{et al.}, with the main discrepancies accounted for by a possible systematic error in the scale of their lightest-ensemble results. We find that the legacy results of the HSC collaboration are not fully supported by HEFT, with some of these results unable to be associated with any nearby finite-volume energy levels.

Based on our HEFT analysis, the ground state $ \Delta(1232) $ corresponds to a $ 1s $ state, and is well described by a quark-model-like core dressed by $ \pi N $ and $ \pi \Delta $ channels. The crucial finding from our analysis is that the corresponding $ 2s $ excitation is \emph{not} the $ \Delta(1600) $. Using the lattice QCD calculation of the first excited state in the $ \Delta(3/2^+)$ spectrum in Ref.~\cite{Hockley:2023yzn}, we saw that the $ 2s $ state sits at $ \sim 2.15 $~GeV, and can be explained in HEFT through the inclusion of a second bare state. This bare state couples to many states in the spectrum, but only weakly to one of the states close to the $ \Delta(1600) $ on the volumes used in Refs.~\cite{Hockley:2023yzn} and \cite{Morningstar:2021ewk}. This means that the states which are associated with the $ \Delta(1600) $ rely on highly non-trivial mixing in the two-particle basis states. We conclude then that, for a complete description of the $ \Delta(1600) $, it is insufficient to consider it as a simple $ 2s $ excitation of the $ \Delta(1232) $. One needs to include the rich resonance properties obtained through coupling to $ \pi N $ and $ \pi \Delta $ channels.

This work, along with previous studies of the Roper resonance \cite{Mahbub:2010rm,Roberts:2013ipa,Liu:2016uzk,Wu:2017qve} and the $ \Delta $-spectrum \cite{Hockley:2023yzn} show how LQCD and HEFT suggest new interpretations of the first excitations observed in the $ J^P = 1/2^+\, (3/2^+) $ spectrum for nucleons ($ \Delta $-baryons). Rather than seeing the first excitations as $ 2s $ radial excitations of the corresponding ground states, one finds that this role is filled by states at significantly higher energies. The Roper and the $ \Delta(1600) $ are then realised as dynamically generated resonances.

\begin{acknowledgments}
	LH would like to thank Zhan-Wei Liu for discussions during his visit to Adelaide. This work was supported by an Australian Government Research Training Program Scholarship and with supercomputing resources provided by the Phoenix High Performance Computing service at the University of Adelaide. This research was undertaken with the assistance of resources from the National Computational Infrastructure (NCI), provided through the National Computational Merit Allocation Scheme, and supported by the Australian Government through Grant No. LE190100021 via the University of Adelaide Partner Share. This research was supported by the University of Adelaide and the Australian Research Council through Grants No. DP190102215 and DP210103706 (DBL), and DP230101791 (AWT). 
\end{acknowledgments}

\clearpage

\appendix

\bibliography{bibliography}
		
\end{document}